\documentclass[aps,showpacs,showkeys,preprintnumbers,nofootinbib]{revtex4}
\usepackage{graphicx,epsfig,wrapfig}
\usepackage{amsmath,amssymb,bm}
\usepackage{slashed}
\usepackage[makeroom]{cancel}

\usepackage{color}

\definecolor{darkgreen}{rgb}{0,0.65,0}

\newcommand{\be}{\begin{equation}}
\newcommand{\ee}{\end{equation}}
\newcommand{\ba}{\begin{eqnarray}}
\newcommand{\ea}{\end{eqnarray}}
\newcommand{\sub}[1]{	\begin{subequations}
			#1
		     	\end{subequations} }

\newcommand{\di}{\!{\rm d}}
\newcommand{\la}{\langle}
\newcommand{\ra}{\rangle}

\newcommand{\fslash}[1] {{\not\! #1\,}}

\begin{document}

%============= TITLE, AUTHORS, AFFILIATION, PRE-PRINT NUMBER =======
\newcommand*{\UConn}{Departement of Physics, University of Connecticut,
Storrs, CT 06269, U.S.A.}\affiliation{\UConn}

%============= TITLE, AUTHORS, AFFILIATION, PRE-PRINT NUMBER =======
\title{\boldmath
	Energy momentum tensor and the $D$-term in the bag model}
\author{Matt J.~Neubelt, Andrew Sampino, Jonathan Hudson, Kemal Tezgin, 
Peter Schweitzer}\affiliation{\UConn}
\date{December 2019}
\begin{abstract}
The energy-momentum tensor (EMT) form factors pave new ways for
exploring hadron structure. Especially the $D$-term related to the
EMT form factor $D(t)$ has received a lot of attention due to its 
attractive physical interpretation in terms of mechanical properties. 
We study the nucleon EMT form factors and the associated densities in 
the bag model which we formulate for an arbitrary number of colors $N_c$ 
and show that the EMT form factors are consistently described in this 
model in the large-$N_c$ limit.
The simplicity of the model allows us to test in a lucid way many 
theoretical concepts related to EMT form factors and densities
including recently introduced concepts like normal and tangential 
forces, or monopole and quadrupole contributions to the angular 
momentum distribution. We also study the $D$-terms of $\rho$-meson, 
Roper resonance, other $N^\ast$ states and $\Delta$-resonances. 
Among the most interesting outcomes is the lucid demonstration of 
the deeper connection of EMT conservation, stability, the virial 
theorem and the negative sign of the $D$-term. 
\end{abstract}
%\preprint{\red{\tt version 02a}}
% 
% 
% 03.50.-z classical field theory 
% 11.10.St field theory bound states
% 12.39.Fe Chiral Lagrangians
% 12.39.Dc Skyrmions
% 11.27.+d field theory aspects of cosmic strings
% 11.15.Pg Expansions for large numbers of components (e.g., 1/Nc expansions)
% 12.38.Gc Lattice QCD calculations
% 12.38.Lg Other nonperturbative calculations
% 13.60.Hb Total+inclusive cross sections (including deep-inelastic processes)
% 
\maketitle

%\tableofcontents

%====== SECTION 1: INTRODUCTION ====================================
\section{Introduction}
\label{Sec-1:introduction}

The perspective to access hadronic EMT form factors 
\cite{Kobzarev:1962wt} through studies of generalized parton distribution 
functions (GPDs) \cite{Mueller:1998fv} in hard exclusive reactions 
\cite{Ji:1996ek,Radyushkin:1996nd,Collins:1996fb,Goeke:2001tz,Diehl:2003ny}
and their attractive interpretation in terms of mechanical properties 
\cite{Polyakov:2002yz} have attracted lots of interest in recent
literature, see the review \cite{Polyakov:2018zvc}. 
EMT form factors were studied in models 
\cite{Ji:1997gm,Petrov:1998kf,Schweitzer:2002nm,Ossmann:2004bp,
Wakamatsu:2005vk,Wakamatsu:2006dy,Goeke:2007fp,Goeke:2007fq,
Wakamatsu:2007uc,Cebulla:2007ei,Kim:2012ts,Jung:2013bya,
Jung:2014jja,Perevalova:2016dln,Abidin:2008ku,Abidin:2008hn,Brodsky:2008pf,
Abidin:2009hr,Mamo:2019mka,Megias:2004uj,Megias:2005fj,Broniowski:2008hx,
Kumar:2017dbf,Chakrabarti:2015lba,Mai:2012yc,Mai:2012cx,Cantara:2015sna,
Gulamov:2015fya,Nugaev:2019vru,Hudson:2017xug,Freese:2019bhb,Freese:2019eww},
chiral perturbation theory \cite{Belitsky:2002jp,Ando:2006sk,Diehl:2006ya},
% NEW
meson-dominance approach \cite{Masjuan:2012sk}, 
dispersion relations \cite{Pasquini:2014vua}, lattice QCD 
\cite{Mathur:1999uf,Hagler:2003jd,Hagler:2007xi,Shanahan:2018pib}, 
QCD lightcone sum rules \cite{Anikin:2019kwi} and for photons 
\cite{Gabdrakhmanov:2012aa,Friot:2006mm}. Especially the form 
factor $D(t)$ \cite{Polyakov:1999gs,Teryaev:2001qm} gained 
increased attention due to its interpretation in terms of internal 
forces \cite{Polyakov:2002yz} spurred by recent attempts to 
extract phenomenological information on $D(t)$ 
\cite{Kumano:2017lhr,Nature,Kumericki:2019ddg}.

In this work we present a study of EMT properties
in one of the simplest hadronic models: the bag model 
\cite{Chodos:1974je,Chodos:1974pn,DeGrand:1975cf}. 
This model was introduced more than 40 years ago, but is still in use 
and continues giving helpful contributions to the understanding of 
hadron structure. In fact, the bag model has been used as an exploratory 
theoretical framework in many instances, often being the first model 
(or one of the first models) where newly introduced hadronic properties 
were investigated, including studies of nucleon structure functions 
\cite{Jaffe:1974nj,Celenza:1982uk}, transversity and other chiral-odd 
parton distribution functions \cite{Jaffe:1991ra}, transverse momentum 
dependent parton distributions
\cite{Courtoy:2008vi,Avakian:2008dz,Avakian:2010br}, or double 
parton distribution functions \cite{Chang:2012nw,Rinaldi:2013vpa}.
The bag model was also the first model where GPDs and EMT form factors 
were studied \cite{Ji:1997gm}.

% NEW
% In this work we will extend the work of Ref.~\cite{Ji:1997gm} in 
In the present study we will extend the work of Ref.~\cite{Ji:1997gm} in 
multiple respects, and investigate within this model concepts which 
appeared only after Ref.~\cite{Ji:1997gm}. This includes the EMT 
densities introduced in \cite{Polyakov:2002yz} and further 
developed in \cite{Polyakov:2018zvc} and 
\cite{Lorce:2017wkb,Schweitzer:2019kkd,Polyakov:2018guq,Polyakov:2018exb,
Lorce:2018egm,Polyakov:2018rew,Cosyn:2019aio,Polyakov:2019lbq}.
The bag model provides an attractive theoretical framework for that. 
The version of the bag model used in this work is at variance with 
chiral symmetry which is a drawback. This model has, however,
also important advantages: it is a consistent theoretical framework.
Its simplicity allows one to obtain lucid insights which 
are more difficult to deduce from more complex models. 
Our results will help to improve the understanding 
of the nucleon structure and the EMT densities. 
% NEW The layout of this work is as follows.
The layout of our study is as follows.

After defining the EMT form factors and densities in 
Sec.~\ref{Sec-2:EMT-in-general}, we briefly introduce the bag model 
in Sec.~\ref{Sec-3:bag-model} and study the quark EMT form factors in 
Sec.~\ref{Sec-4:EMT-FFs-quarks} using a formulation of the model for a
large number of colors $N_c$.
The large-$N_c$ limit will allow us to avoid technical problems associated 
with the evaluation of form factors in so-called independent-particle models 
like the bag model. We will use the large-$N_c$ limit as a tool to derive 
consistent model expressions, and show that the $1/N_c$-corrections to 
the form factors are relatively small for small momentum transfers.
In addition, the large-$N_c$ limit provides a rigorous justification 
for the concept of 3D densities which are studied in detail in 
Sec.~\ref{Sec-5:densities}. We will evaluate the ``gluonic'' form 
factor $\bar{c}^G(t)$ due to the bag which can only be computed by 
taking advantage of the EMT density formalism, and we will rigorously
prove the internal consistency of the description.
The Sec.~\ref{Sec-6:D-term} presents an extensive study
of the $D$-term for the nucleon and other hadronic states
including $N^\star$ states, $\rho$-mesons, and the 
$\Delta$-resonances. We also include an insightful study of
hypothetical highly excited bag model states. This is the only study 
of EMT properties of excited states available in literature
besides $Q$-balls \cite{Mai:2012cx} and we make the interesting 
observation that in both systems asymptotically the $D$-term grows as 
$D=-\,{\rm const}\times M^{8/3}$ with the mass $M$ of the excitation,
even though the excited states have much different internal 
structures in the two frameworks. The Sec.~\ref{Sec-7:D-limiting-cases}
is dedicated to studies of limiting cases like the heavy quark limit,
the large bag-radius limit, and the non-relativistic limit of the nucleon,
and discuss the behavior of the $D$-term in these limits.
Our study is complemented by an instructive discussion in 
Sec.~\ref{Sec-8:Bogoliubov} of the $D$-term in a predecessor of the 
bag model \cite{Thomas:2001kw}, the Bogoliubov model \cite{Bogo:1967}, 
which is a counter-example where the nucleon is not fully consistently
described. As a consequence one finds an unphysical (positive) $D$-term
in this model. This example also illustrates the necessity to study the
complete EMT structure. 
The conclusions are presented in Sec.~\ref{Sec-9:conlusions} and 
technical details can be found in Appendix~\ref{App}.
Some of our results where previously mentioned in 
\cite{Hudson:2017oul,Neubelt:2019wnn}.

%====== SECTION 2: FORM FACTORS IN GENERAL =========================
\section{EMT form factors}
\label{Sec-2:EMT-in-general}

The EMT form factors \cite{Kobzarev:1962wt} can be defined
in QCD in the following way
\ba
    \la p^\prime| \hat T^a_{\mu\nu}(0) |p\rangle
    = \bar u(p^\prime)\biggl[A^a(t)\,\frac{P_\mu P_\nu}{M_N}+
    J^a(t)\ \frac{i(P_{\mu}\sigma_{\nu\rho}+P_{\nu}\sigma_{\mu\rho})
    \Delta^\rho}{2M_N} 
    + D^a(t)\,
    \frac{\Delta_\mu\Delta_\nu-g_{\mu\nu}\Delta^2}{4M_N}+
    \bar{c}^a(t)\,M_N\,g_{\mu\nu}\biggr]u(p)\, ,
    \label{Eq:ff-of-EMT} 
\ea
where the kinematic variables are defined as
\be
	P=\frac12(p+p'), 	\quad
	\Delta=(p'-p),		\quad
	t=\Delta^2.
    	\label{Eq:kin-variables}
\ee
The EMT form factors for different partons $a=g,\,u,\,d,\,\dots\,$
depend on renormalization scale $\mu$, e.g.\ $A^a(t)=A^a(t,\mu^2)$,
which we not always indicate for brevity. 
The total EMT form factors $A(t)=\sum_a A^a(t,\mu^2)$, and analogous for 
$J(t)$ and $D(t)$, are renormalization scale independent.
The appearance of the form factors $\bar{c}^a(t,\mu^2)$ signals that
the separate quark and gluon EMTs are not conserved.
Only the total EMT is conserved and consequently $\sum_a \bar{c}^a(t,\mu^2)=0$.

The form factors of the EMT in Eq.~(\ref{Eq:ff-of-EMT}) can be interpreted
\cite{Polyakov:2002yz} in analogy to the electromagnetic form factors
\cite{Sachs} in the Breit frame where $\Delta^0=0$.
In the Breit frame one can define the static energy-momentum tensor as
\be\label{Def:static-EMT}
    T_{\mu\nu}(\vec{ r},\vec{ s}) =
    \int\frac{\di^3\Delta}{2E(2\pi)^3}\;\exp(-i\vec{\Delta}\vec{ r})\;
    \la p^\prime,S^\prime|\hat{T}_{\mu\nu}(0)|p,S\ra
\ee
with initial and final nucleon polarizations $S$ and $S^\prime$ 
defined such that they are equal to $(0,\vec{s})$ in the respective 
rest frames, where the unit vector $\vec{s}$ denotes the quantization 
axis for the nucleon spin. This interpretation is subject to 
``relativistic corrections'' as in the case of electromagnetic 
form factors \cite{Polyakov:2002yz,Sachs} and is exact 
in the large-$N_c$ limit \cite{Polyakov:2018zvc}. 

The component $T_{00}(\vec{ r})$ describes the energy density,
and the components $T_{ik}(\vec{ r})$ characterize the spatial 
distributions of forces experienced by the partons \cite{Polyakov:2002yz}.
Both are independent of the polarization vector. The components 
$T_{0k}(\vec{ r},\vec{ s})$ are related to the distributions of 
angular momentum. At $t=0$ the form factors satisfy the constraints
\ba
    A(0)&=&
    \frac{1}{M_N}\,\int\di^3r\;T_{00}(\vec{ r})=1
    \;,\nonumber\\
    J(0)&=&
    \int\di^3r \;\epsilon^{ijk}\,s_i\,r_j\,
    T_{0k}(\vec{ r},\vec{ s})=\frac12\, , \nonumber\\
    D(0)&=&
    -\frac{2M_N}{5}\, \int\di^3r\;T_{ij}(\vec{ r})\,
    \left(r^i r^j-\frac{\vec{ r}^{\,2}}3\,\delta^{ij}\right)\equiv D\,.
    \label{Eq:M2-J-d1}
\ea
The constraints on $A(0)$ and $J(0)$ can be traced back to the fact
that the EMT matrix elements contain information on the particle's  mass 
and spin and are dictated by the transformation properties of the states
\cite{Lowdon:2017idv,Cotogno:2019xcl}.
The value of the form factor  $D(t)$ at $t=0$ is not constrained by
any general principle.
The components $T_{ij}(\vec{ r})$ of the static stress tensor encode 
the information on the distribution of pressure and shear forces 
\cite{Polyakov:2002yz}
\be\label{Eq:T_ij-pressure-and-shear}
    T_{ij}(\vec{ r})
    = s(r)\left(\frac{r_ir_j}{r^2}-\frac 13\,\delta_{ij}\right)
        + p(r)\,\delta_{ij}\, . \ee
Here $p(r)$ describes the radial distribution of the pressure
inside the hadron, and $s(r)$ is the distribution of shear forces 
\cite{Polyakov:2002yz}.  Both functions are related to each other 
due to the EMT conservation by the differential equation
\be\label{Eq:p(r)+s(r)}
    \frac23\;\frac{\partial s(r)}{\partial r\;}+
    \frac{2s(r)}{r} + \frac{\partial p(r)}{\partial r\;} = 0\;.
\ee
The conservation of the EMT also provides two equivalent 
expressions for the $D$-term  in terms of $p(r)$ or $s(r)$ as 
\ba\label{Eq:d1-from-s(r)-and-p(r)}
        D &=& -\,\frac{4}{15}\;M_N \int\di^3r\;r^2\, s(r)
         =     M_N \int\di^3r\;r^2\, p(r)\;.
\ea
Further properties of EMT densities will be discussed below.

%====== SECTION 3: BAG MODEL =======================================
\section{\boldmath The bag model}
\label{Sec-3:bag-model}

In the bag model one describes baryons (mesons) by placing $N_c=3$
non-interacting quarks (a $\bar qq$ pair) in a color-singlet state
inside a ``bag.'' In its rest frame the bag is a spherical region of 
radius $R$ carrying the energy density $B>0$ \cite{Chodos:1974je}.
The Lagrangian of the bag model can be written as \cite{Thomas:2001kw}
\be\label{Eq:Lagrangian-bag}
   	{\cal L} = {\cal L}_Q + {\cal L}_{\rm surf} + {\cal L}_G,\quad
        {\cal L}_Q = \sum\limits_q\biggl[\bar\psi_q\,\biggl(
	-\frac{i}{2}\overleftarrow{\fslash{\partial}}
	+\frac{i}{2}\overrightarrow{\fslash{\partial}}
	-m\biggl)\psi_q\biggr]\Theta_V, \quad
	{\cal L}_{\rm surf} = \frac12\,\sum\limits_q
	\bar\psi_q\,\psi_q\:\eta^\mu\partial_\mu\Theta_V,\quad
	{\cal L}_{\rm G} = -B\,\Theta_V
\ee
with the following definitions referring to the rest frame of the bag
\be
	\Theta_V=\Theta(R-r),	\quad
	\delta_S=\delta(R-r),	\quad
	\eta^\mu=(0,\vec{e}_r),	\quad
	\vec{e}_r = \vec{r}/r,	\quad
	r=|\vec{r}\,|.
\ee
In Eq.~(\ref{Eq:Lagrangian-bag}) we defined for later convenience 
the contributions of quarks ${\cal L}_Q$, ``gluons'' ${\cal L}_G$, and 
the interaction ${\cal L}_{\rm surf}$ with the bag surface. We deal with 
a very crude model of confinement, so the contribution of ``gluons'' 
should not to be understood literally.
It ``resembles'' the QCD gluon contribution remotely in the sense that 
{\sl(i)} it cannot be expressed in terms of fermionic degrees of freedom, 
and {\sl(ii)} is crucial for the formation of bound states in this model.
In fact, if we let $R\to\infty$ then $\Theta_V\to1$, $\partial_\mu\Theta_V\to0$
and we recover free and unbound quarks.
The Euler-Lagrange equation of the theory (\ref{Eq:Lagrangian-bag}) 
are given by
\begin{subequations}
\label{Eq:eom-all}
\ba
    (i\fslash{\partial}-m)\psi_q	
	&=& \,0 \,\;\;\;\mbox{for}\;\;\; r<R
	\;\;\;\,\,\mbox{(free quarks)},\phantom{\frac12}\label{Eq:eom-q}\\
    i\fslash{\eta}\,\psi_q      	
	&=& \psi_q \;\;\mbox{for}\;\;\; \vec{r}\in S  
    	\;\;\;\,\mbox{(linear boundary condition)}, 	\label{Eq:eom-lin}\\
    -\,\frac12\,\sum_q\eta_\mu\partial^\mu\bar\psi_q\psi_q
	&=& B \;\;\;\mbox{for}\;\;\; \vec{r}\in S\;\;\;
	\mbox{(non-linear boundary condition)}.		\label{Eq:eom-non}
\ea
\end{subequations}
% The Euler-Lagrange equation of the theory (\ref{Eq:Lagrangian-bag}) for 
% $\bar{\psi}_q$ yields: {\sl(i)} Eq.~(\ref{Eq:eom-q}) when evaluated for 
% $r<R$, {\sl(ii)} the linear boundary condition (\ref{Eq:eom-lin}) when 
% evaluated in the limit $r\to R$ from below. The non-linear boundary 
% condition (\ref{Eq:eom-non}) can be derived by treating $\Theta_V$
% as a ``field'' at the surface of the bag where this ``field'' varies. 
The boundary conditions (\ref{Eq:eom-lin},~\ref{Eq:eom-non}) are
equivalent to the statement that there is no energy-momentum flow 
out of the bag, i.e.\ $\eta_\mu T^{\mu\nu}(t,\vec{r})=0$ for 
$\vec{r}\in S$ \cite{Chodos:1974je}, which provides a simple model
of confinement.

In the positive parity sector, which contains the ground state, 
the wave-functions are given by
\be\label{Eq:bag-wave-function}
    \psi_{s}(t,\vec{r}) = e^{-i\varepsilon_it}
    \,\phi_s(\vec{r}) 
	\, , \;\;\;
    \phi_s(\vec{r})   = \frac{A}{\sqrt{4\pi}}\,
    \left(\begin{array}{l}
        \alpha_+j_0(\omega_ir/R)\,\chi_s \phantom{\displaystyle\frac11}\\
        \alpha_-j_1(\omega_ir/R)\,i\vec{\sigma}\vec{e}_r\chi_s 
    \end{array}\right)
	\, , \;\;\;
    	A = \biggl(\frac{\Omega_i(\Omega_i- m R)}
	{R^3j_0^2(\omega_i)(2\Omega_i(\Omega_i-1)+mR)}\biggr)^{\!1/2}
\ee
where $\alpha_\pm=\sqrt{1\pm mR/\Omega_i}$ with 
$\Omega_i=\sqrt{\omega_i^2+m^2R^2}$. The $\sigma^i$ are $2\times2$ 
Pauli matrices, and $\chi_s$ are two-component Pauli spinors. The 
spherical Bessel functions are defined in App.~\ref{App}.
The single-quark energies are given by $\varepsilon_i=\Omega_i/R$ 
where the $\omega_i$ denote solutions of the transcendental equation
\be\label{Eq:omega-transcendental-eq}
	\omega_i = (1-mR-\Omega_i)\,\tan\omega_i \,,
\ee
whose lowest (ground state) solution is % $\omega_0\approx 2.0427869\dots$ 
$\omega_0\approx 2.04$ for massless quarks. If $mR$ is varied from 0 to 
infinity, the ground state solution $\omega_0=\omega_0(mR)$ covers the interval
\be
	2.04 \lesssim \omega_0(mR) \le \pi \,.
\ee 
The momentum space wave functions are defined by the Fourier transform 
$\varphi_s(\vec{k}) = \int d^3r\,e^{-i\vec{k}\vec{r}}\,\phi_s(\vec{r})$
and given by
\be
    \varphi_{s}(\vec{k})=\sqrt{4\pi}\,A\, R^3
    \left (\begin{array}{l} 
      \alpha_+t_0(k)\,\chi_s\\
      \alpha_-t_1(k)\,\vec{\sigma}\vec{e}_k\,\chi_s
    \end{array} \right ) \ ,
    \label{Eq:bag-wave-function-mom}
\ee
where $\vec{e}_k = \vec{k}/k$ with $k=|\vec{k}|$.
The functions $t_l(k)$ for $l=0,1$ are given by
\begin{equation}
    \label{Eq:t0-t1}
    t_l(k)=\int\limits_0^1du \,u^2 j_l(ukR)j_l(u\omega_i) \ .
\end{equation}
The constant $A$ in 
Eqs.~(\ref{Eq:bag-wave-function},~\ref{Eq:bag-wave-function-mom})
ensures the normalization
\be
	\int\di^3r\:
	\phi^\dag_{s^\prime}(\vec{r}^{\,})\,\phi^{ }_{s^{ }}(\vec{r}^{\,})
	=
	\int\frac{\di^3k}{(2\pi)^3}\:
	\varphi^\dag_{s^\prime}(\vec{k}^{\,})\,\varphi^{ }_{s^{ }}(\vec{k}^{\,})
	=
	\delta_{s^\prime\!s}\,.
\ee
The nucleon wave-functions with definite spin-isospin quantum numbers are 
constructed from the single-quark wave functions (\ref{Eq:bag-wave-function}) 
assuming SU(4) spin-flavor symmetry. We will not need the explicit expressions
here, and only quote the resulting SU(4) spin-flavor factors which appear
in respectively spin-independent ($N_q$) and spin-dependent ($P_q$) 
matrix elements for a proton made of $N_c$ quarks 
(for neutron interchange $u\leftrightarrow d$) \cite{Karl:1984cz}
\begin{subequations}
\label{Eq:Nq+Pq}
\ba\label{Eq:Nq} N_u = \frac{N_c+1}{2}\,, &&  N_d = \frac{\;N_c-1}{2}\,, \\
   \label{Eq:Pq} P_u = \frac{N_c+5}{6}\,, &&  P_d = \frac{-N_c+1}{6}\,. 
\ea
\end{subequations}
For the proton and $N_c=3$ the familiar values $N_u=2$, $N_d=1$, $P_u=\frac43$,
$P_d=-\,\frac13$ are reproduced. 

% But working with the general
% values in Eqs.~(\ref{Eq:Nq},~\ref{Eq:Pq}) will allow us to test 
% the model against predictions from general large-$N_c$ counting rules.

%====== SECTION 4: FORM FACTORS ====================================
\section{The EMT form factors of quarks}
\label{Sec-4:EMT-FFs-quarks}

In this section we compute the matrix elements of the quark EMT 
$T^{\mu\nu}_Q$ in the limit of a large number of colors $N_c$, check 
the consistency of the results, discuss the role of $1/N_c$ corrections,
and compare to results from literature.

\subsection{\boldmath 
Kinematics and scaling of EMT form factors in large $N_c$ limit}

In this limit the nucleon mass behaves as $M_N={\cal O}(N_c)$. This means 
the  nucleon is a heavy particle, and its motion is non-relativistic, 
i.e.\ the nucleon energies $E$ and $E^\prime$ are given by
$M_N+{\cal O}(N_c^{-1})$, while the nucleon momenta $\vec{p}^{ }$ 
and $\vec{p}^{\:\prime}$ are of the order ${\cal O}(N_c^0)$.
For the kinematic variables (\ref{Eq:kin-variables}) this implies
\be\label{Eq:kin-large-Nc}
	P^0={\cal O}(N_c), 		\quad
	\vec{P}={\cal O}(N_c^0),	\quad
	\vec{\Delta}={\cal O}(N_c^0),	\quad
	\Delta^0={\cal O}(N_c^{-1}).
\ee
Thus $P^\mu = (M_N,0,0,0)$ and 
$\Delta^\mu=(0,\vec{\Delta})$ and $t=-\vec{\Delta}^{\:2}$
modulo $1/N_c$ corrections. Notice that the non-relativistic 
motion concerns only the nucleon. The motion of the quarks inside 
the nucleon can still be ultra-relativistic for light or massless quarks. 
In the large-$N_c$ limit the bag model is still a relativistic model. 
Only if in addition to the large-$N_c$ limit one also would choose 
to make the quarks heavy, would one recover the picture of a 
non-relativistic quark model
(which we shall explore in Sec.~\ref{Sec-7:D-limiting-cases}).

In order to evaluate the expressions for the EMT form factors 
(\ref{Eq:ff-of-EMT}) we also have to take into account the
large-$N_c$ behavior of the quark EMT form factors \cite{Goeke:2001tz}
\be\label{Eq:EMT-ff-large-Nc}
	A^Q(t) 	={\cal O}(N_c^0), 	\quad 
	J^Q(t)	={\cal O}(N_c^0), 	\quad
	D^Q(t)	={\cal O}(N_c^2), 	\quad
  \bar{c}^Q(t)	={\cal O}(N_c^0).
\ee
Notice that the index $Q$ denotes the isoscalar $(u+d)$ flavor 
combinations. The isovector $(u-d)$ flavor combinations have
different $N_c$ scalings:
$A^{u\!-\!d}(t)={\cal O}(N_c^{-1})$,
$J^{u\!-\!d}(t)={\cal O}(N_c)$,
$D^{u\!-\!d}(t)={\cal O}(N_c)$,
$\bar{c}^{u\!-\!d}(t)={\cal O}(N_c^{-1})$ \cite{Goeke:2001tz}.

\subsection{\boldmath Form factors of the symmetric quark EMT in bag model}

In the large-$N_c$ limit, i.e.\ considering 
Eqs.~(\ref{Eq:kin-large-Nc},~\ref{Eq:EMT-ff-large-Nc}), 
the expressions for the EMT form factors (\ref{Eq:ff-of-EMT}) become
\sub{\begin{align}
\langle p^\prime,s^\prime| \hat T^{00}_Q(0) |p,s\rangle
      	= 2\,M_N^2\biggl[
        & A^Q(t) - \frac{t}{4M_N^2}\,D^Q(t)
      	+ \bar{c}^Q(t)\biggr]\delta_{ss^\prime}  
	\label{Eq:EMT-Breit-T00}\\
\langle p^\prime,s^\prime| \hat T^{ik}_Q(0) |p,s\rangle
    	= 2\,M_N^2\biggl[
	& D^Q(t)\,\frac{\Delta^i\Delta^k-\delta^{ik}{\vec{\Delta}}^2}{4M_N^2}
    	- \bar{c}^Q(t)\,\delta^{ik} \biggr] \delta_{ss^\prime}  
	\label{Eq:EMT-Breit-Tik}\\
\langle p^\prime,s^\prime| \hat T^{0k}_Q(0) |p,s\rangle
	= 2\,M_N^2\biggl[
	& J^Q(t)\;\frac{(-i\,\vec{\Delta}\times\vec{\sigma}_{s^\prime s})^k}{2\,M_N}
	\biggr]
	\label{Eq:EMT-Breit-T0k}
\end{align}}
where we used $\chi^\dag_{s'}\chi^{ }_s=\delta_{ss'}$ and defined
$\vec{\sigma}_{s^\prime s}=\chi^\dag_{s'}\vec{\sigma}\chi^{ }_s$.
The generic expression to evaluate nucleon matrix elements of 
quark bilinear operators of the type $\bar{\Psi}_q\hat{O}\Psi_q$ 
in the bag model in the large-$N_c$ limit is given by
\be
	\la N(p^\prime\!,s^\prime)|\,
	\overline{\Psi}_q\:\hat{O}\,\Psi_q\,|N(p,s)\ra
	=2M _N
	\int\frac{\di^3k}{(2\pi)^3}\;
	\overline{\varphi}_{s^\prime}(\vec{k}^\prime)\,
	\hat{O}\,\varphi_{s^{ }}(\vec{k}\,),
	\quad
	\vec{k}^\prime=\vec{k}+\vec{\Delta}.	
\ee
The prefactor $2M_N$ originates in the large $N_c$ limit
from the factor $2P^0$ in the covariant normalization of the
nucleon states. The symmetric quark EMT is given by
(the arrows indicate which wave functions are differentiated)
\be\label{Eq:EMT-symm}
	T^{\mu\nu}_q = \frac{1}{4}\overline{\psi}_q\biggl(
	-i\overset{ \leftarrow}\partial{ }^\mu\gamma^\nu
	-i\overset{ \leftarrow}\partial{ }^\nu\gamma^\mu
	+i\overset{\rightarrow}\partial{ }^\mu\gamma^\nu
	+i\overset{\rightarrow}\partial{ }^\nu\gamma^\mu\biggr)\psi_q\,.
\ee
In order to perform the calculations we choose 
$\vec{\Delta}=(0,0,\Delta^3)$ and the nucleon polarization
along the $z$-axis. We define $k_\perp^2 = k_1^2+k_2^2$,
$k=|\vec{k}^{\,}|$, $k^\prime=|\vec{k}^{\,\prime}|$ with
$\vec{k}^{\,\prime}=\vec{k}+\vec{\Delta}=(k^1,k^2,k^3+\Delta^3)$ 
in our frame. The results read
\sub{\begin{align}
	A^Q(t)- \frac{t}{4M_N^2}\,D^Q(t) + {\bar c}^Q(t)  =
	\frac{4\pi A^2R^6N_c}{M_N}
	\int\!\frac{\di^3k}{(2\pi)^3\!}\,\varepsilon_0
	\biggl[\alpha_+^2t_0(k)t_0(k^\prime)+
	\alpha_-^2\vec{e}_k\vec{e}_{k'}t_1(k)t_1(k^\prime)\biggr],
	&	\label{Eq:EMT-bag-T00}\\
	\frac{t}{4M_N^2}\,D^Q(t) - {\bar c}^Q(t)	 =
	\frac{4\pi A^2R^6N_c}{M_N}\int\!\frac{\di^3k}{(2\pi)^3\!}\,
	\alpha_+\alpha_-\frac{k_\perp^2}{2}
	\biggl[	t_0(k 	   )t_1(k^\prime)\,\frac{1}{k^\prime}\,+
		t_0(k^\prime)t_1(k      )\frac{1}{k      }\,\biggr],
	&	\label{Eq:EMT-bag-T11}\\
	- {\bar c}^Q(t)	=
	\frac{4\pi A^2R^6N_c}{M_N}\int\!\frac{\di^3k}{(2\pi)^3\!}\,
	\alpha_+\alpha_- \frac{(k^{\prime3}+k^3)}{2}\,
	\biggl[	t_0(k 	   )t_1(k^\prime)\frac{k^{\prime3}}{k^\prime}+
		t_0(k^\prime)t_1(k      )\frac{k^3       }{k     }\biggr],
	&	 \label{Eq:EMT-bag-T33}\\
	J^Q(t) =
	4\pi A^2R^6\int\!\frac{\di^3k}{(2\pi)^3\!}\,
	\biggl[\alpha_+\alpha_-\frac{\varepsilon_0}{\Delta^3} \biggl(
		- t_0(k^\prime)t_1(k)\,\frac{k^3}{k}
		+ t_0(k)t_1(k^\prime)\,\frac{k^{\prime3}}{k^\prime}\biggr)
	+ \frac{k_\perp^2}{2} \alpha_-^2
	\frac{t_1(k)}{k}\,\frac{t_1(k^\prime)}{k^\prime}\biggr].
	&	 \label{Eq:EMT-bag-T0k}
\end{align}}
Hereby 
Eq.~(\ref{Eq:EMT-bag-T00}) follows from $T^{00}_Q$ in 
    (\ref{Eq:EMT-Breit-T00}), 
Eq.~(\ref{Eq:EMT-bag-T11}) follows from $T^{11}_Q$ or $T^{22}_Q$ in
    (\ref{Eq:EMT-Breit-Tik}),
Eq.~(\ref{Eq:EMT-bag-T33}) is obtained from $T^{33}_Q$ in
    (\ref{Eq:EMT-Breit-Tik}), and
Eq.~(\ref{Eq:EMT-bag-T0k}) follows from $T^{01}_Q$ or $T^{02}_Q$ in
    (\ref{Eq:EMT-Breit-T0k}), 
while $T^{03}_Q$ vanishes. 
The canonical EMT has a symmetric part which coincides with
what we discussed above, and an anti-symmetric part which is 
discussed in App.~\ref{App}.

\subsection{\boldmath Numerical results}
\label{Sec:num-res-FFs}

\begin{figure}[b!]
\centering
\includegraphics[width=4.5cm]{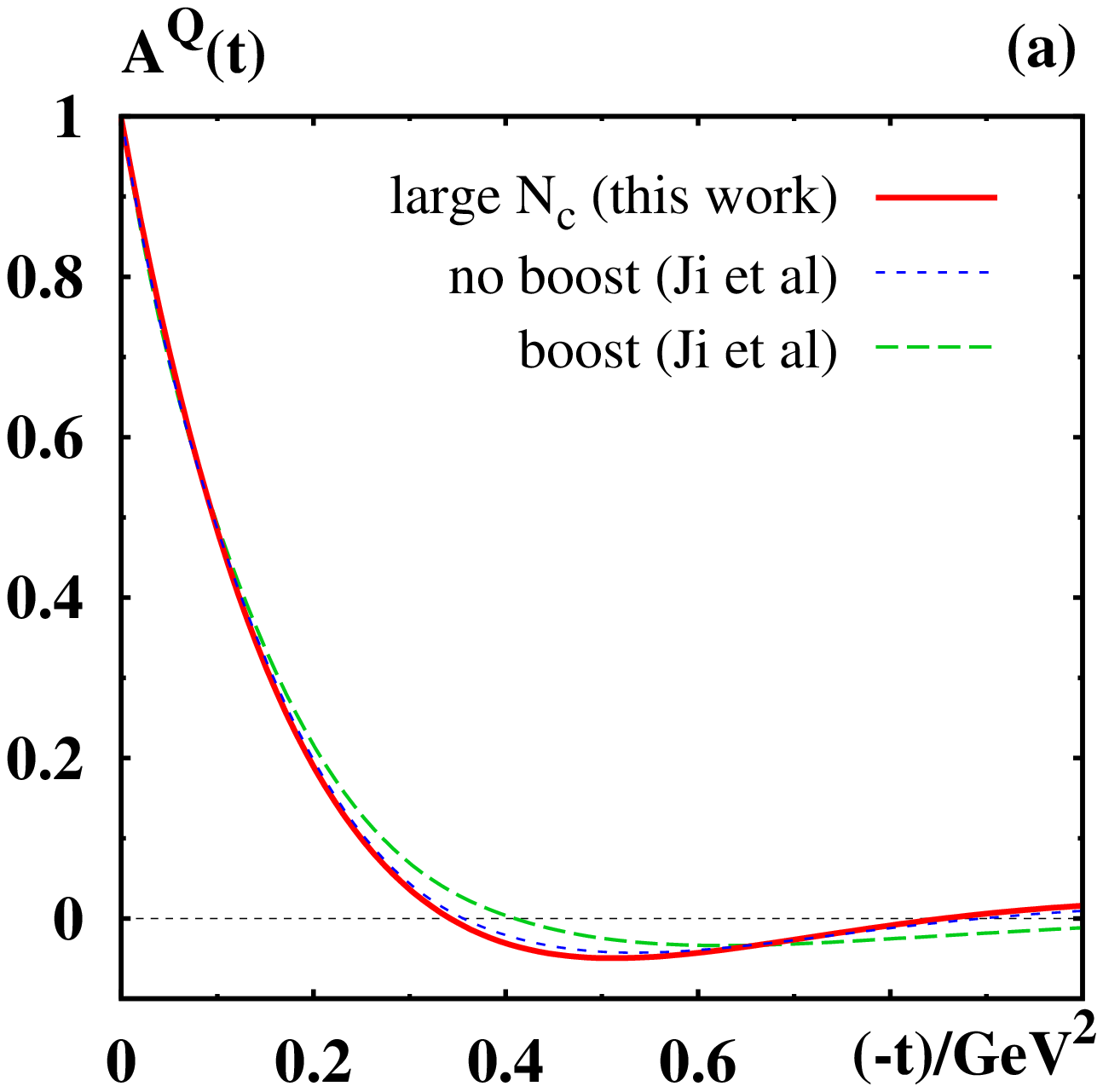}%
\includegraphics[width=4.5cm]{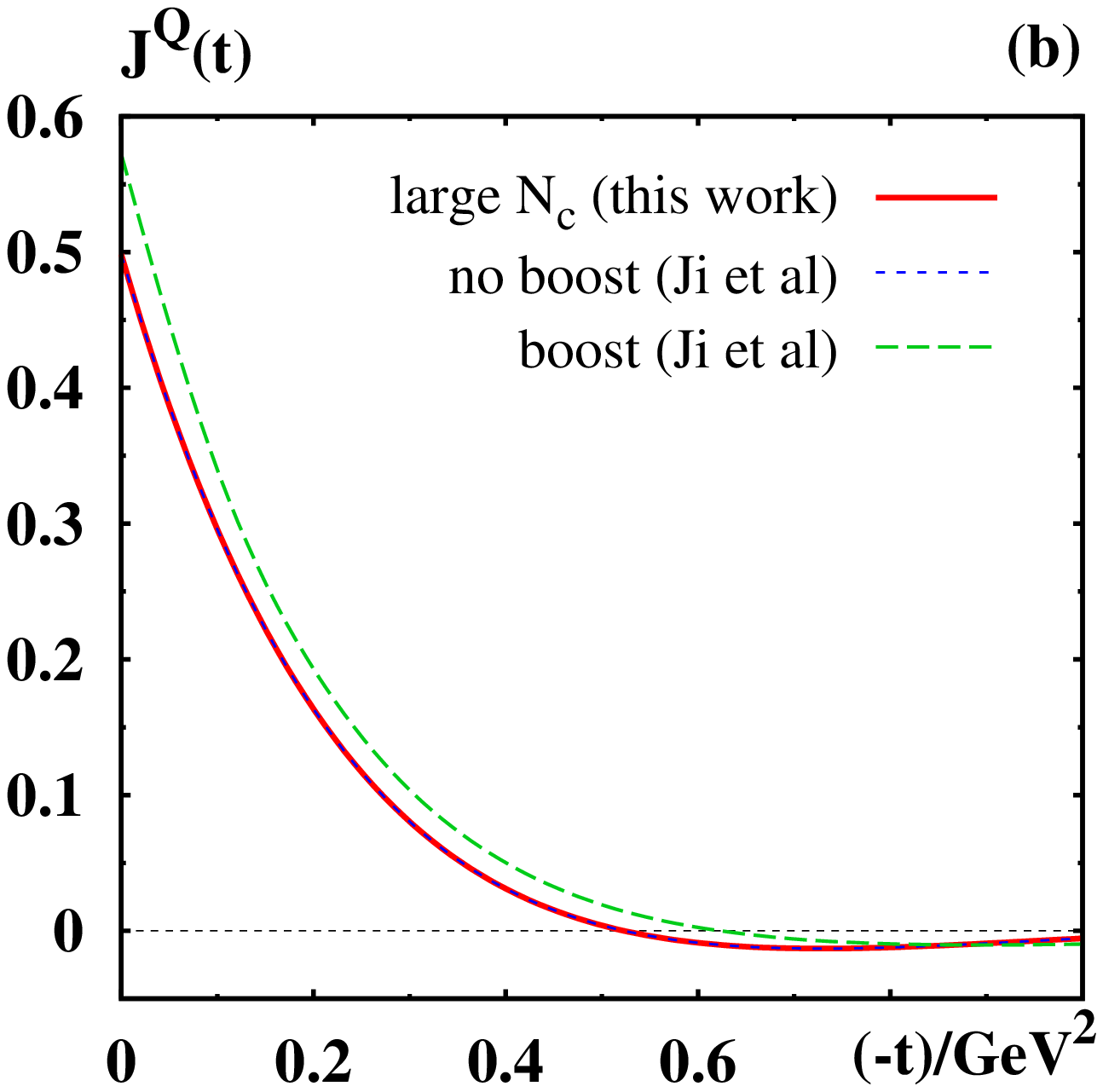}%
\includegraphics[width=4.5cm]{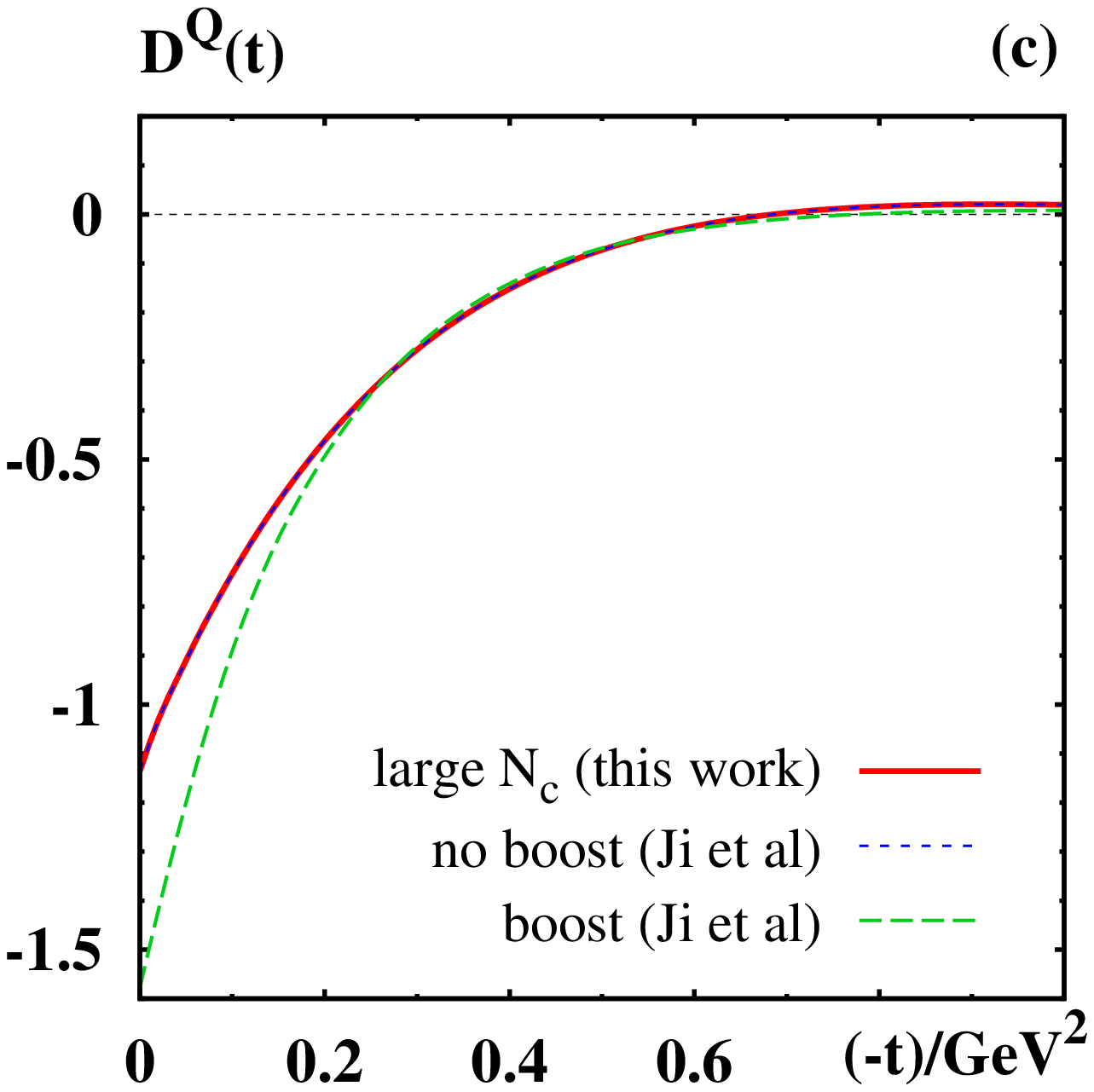}%
\includegraphics[width=4.5cm]{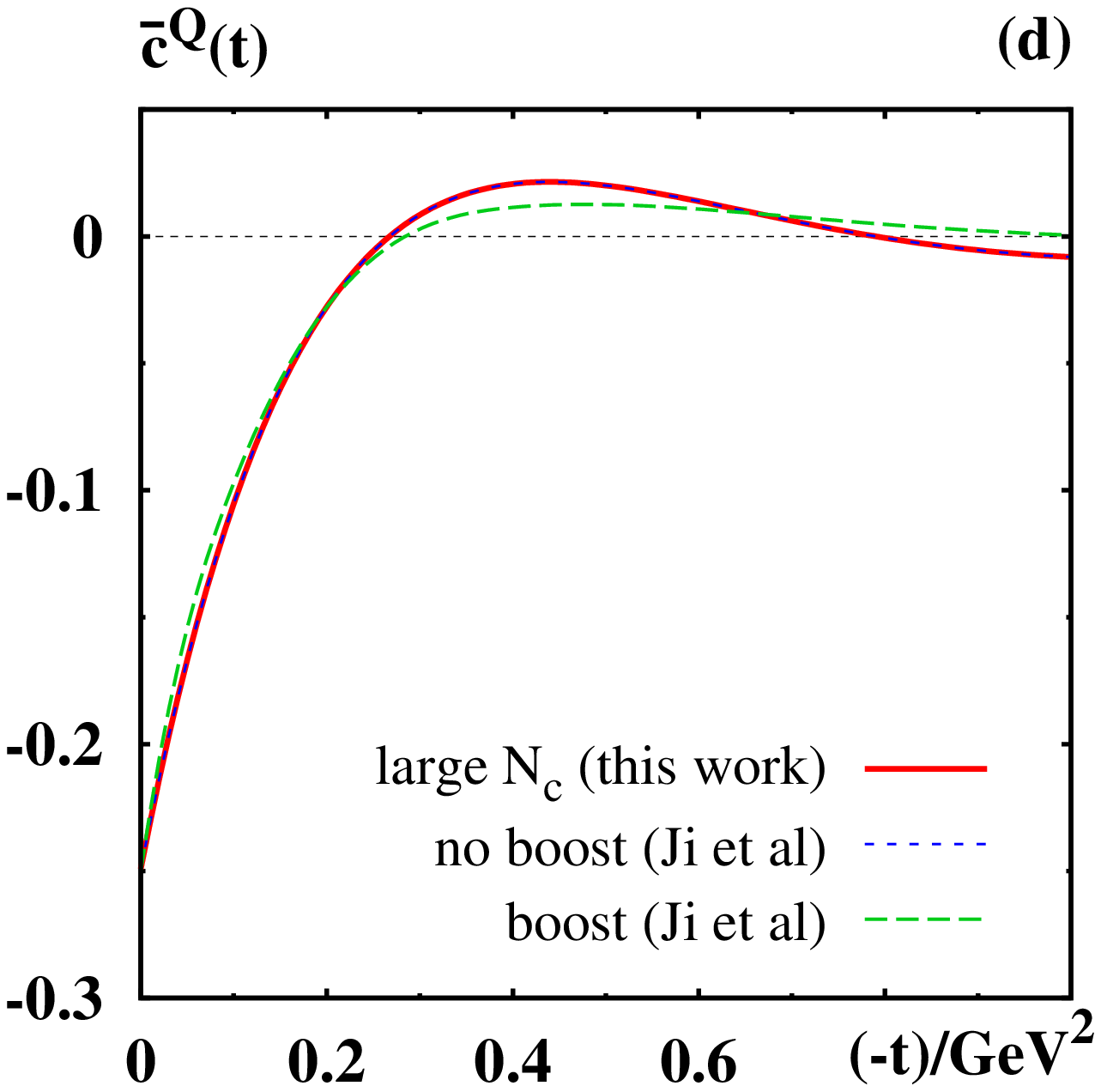}

\caption{\label{Fig-01:FFs} EMT form factors of quarks in the bag model
	in the large $N_c$ limit (solid lines, this work). For comparison
	we also show results by Ji et al, Ref.~\cite{Ji:1997gm}, computed
	in the bag model without (dotted lines) and with (dashed lines)
	considering boosts. This comparison shows the effects of 
	relativistic and $1/N_c$--corrections within the 
	``independent particle model treatment'' in the bag model.
	For finite $N_c$ there are further corrections associated
	with the independent particle model treatment, see text.}
\end{figure}

Evaluating Eqs.~(\ref{Eq:EMT-bag-T00}--\ref{Eq:EMT-bag-T0k}) for
massless quarks yields the curves shown in Fig.~\ref{Fig-01:FFs} 
as solid lines. These results refer to the leading order in the
large $N_c$ limit and are consequently valid for $|t|\ll M_N^2$.
The obtained form factors satisfy the general requirements at
$t=0$ namely $A^Q(0)=1$ and $J^Q(0)=\frac12$. Furthermore it is
$\bar{c}^Q(0)=-\,\frac14$ which is a bag model specific result 
\cite{Ji:1997gm}. All three constraints can be proven analytically,
but the proofs are lengthy, not enlightening and we do not show them.
The $D$-term is not fixed by any general constraint. It assumes
the value $D^Q(0)=-1.145$ for massless quarks. We will discuss
the $D$-term in more detail below in Sec.~\ref{Sec-6:D-term}.

The results $A^Q(0)=1$ and $J^Q(0)=\frac12$ mean that quarks
carry $100\,\%$ of the momentum and spin of the nucleon.
The appearance of the form factor $\bar{c}^Q(t)\neq0$ means,
however, that the quark part of the EMT, $T^{Q}_{\mu\nu}$, is
not conserved. To have a conserved total EMT one must include
also non-fermionic contributions associated with the bag,
i.e.\ ``gluonic contributions'' in the sense explained in 
Sec.~\ref{Sec-3:bag-model}.
At this point it is not clear how to formulate a wave-function of the 
bag and compute the ``gluonic'' EMT form factors in the bag model, but 
in Sec.~\ref{Sec-5:densities} we will see that this can be naturally 
achieved by taking advantage of the concept of 3D spatial EMT densities.

%=============== BEGIN FIGURE 1: FORM FACTORS ======================
%================= END FIGURE 1 ====================================

\subsection{\boldmath $1/N_c$ corrections}

The large-$N_c$ results are theoretically consistent which is crucial
for our study. However, it is instructive to get insights on the size 
of $1/N_c$--corrections by comparing our results with those of 
Ref.~\cite{Ji:1997gm} obtained for finite $N_c$. We can distinguish 
different types of $1/N_c$-corrections.
If we do not implement the kinematic effects (\ref{Eq:kin-large-Nc}) 
and large-$N_c$ counting rules (\ref{Eq:EMT-ff-large-Nc}) we recover
the ``no-boost results'' by Ji et al from \cite{Ji:1997gm}. This type of 
$1/N_c$ correction only affects the form factor $A(t)$ where it has a small 
effect for $|t|$ below $1\,{\rm GeV}^2$, see the curve depicted by the dotted 
line in comparison to the solid line in Fig.~\ref{Fig-01:FFs}a. The form 
factors $J^Q(t)$, $D^Q(t)$, $\bar{c}^Q(t)$ are not affected by these 
corrections, so the no-boost results from \cite{Ji:1997gm} 
(dotted lines) coincide with our large-$N_c$ results (solid lines)
in Figs.~\ref{Fig-01:FFs}b--d. 

A conceptually different type of corrections arises because for finite 
$N_c$ it is necessary to take into account relativistic corrections 
associated with boosting the quark wave function
(\ref{Eq:bag-wave-function-mom}) to a frame where the nucleon 
moves with velocity $\vec{v}$:
$\psi(t,\vec{x})\to S(\Lambda_{\vec{v}})\psi(t^\prime,\vec{x}{ }^\prime)$ 
with $S(\Lambda) = \exp(w \gamma^0\gamma^3)$ where $\Lambda_{\vec{v}}$ 
is the Lorentz transformation for a boost along $z$-axis with 
$\vec{v}=(0,0,\tanh(w))$ where $\sinh(w)=|\vec{\Delta}|/(2M_N)$ 
\cite{Ji:1997gm}. The results obtained in this way are depicted 
as dashed lines in Fig.~\ref{Fig-01:FFs}.
The constraint $J^Q(0)=\frac12$ is no longer satisfied, see 
Fig.~\ref{Fig-01:FFs}b, because ``the boosted bag wave function 
does not have the correct Lorentz symmetry'' \cite{Ji:1997gm}.
This artifact can in principle be avoided using 
Peierls-Yoccoz projections \cite{Peierls:1957er} or 
center-of-mass freedom separation methods \cite{Wang:1982tz}
which were not performed in \cite{Ji:1997gm}. For our purposes it is 
completely sufficient to observe that in practice such boost effects --- 
even if they were not entirely consistently estimated in \cite{Ji:1997gm} --- 
constitute a small correction. It is important to stress that in the
large-$N_c$ limit $|\vec{\Delta}|/(2M_N)\to 0$ and this type of 
relativistic corrections is negligible.

The third type of $1/N_c$ corrections is due to the fact that the bag 
model belongs to a class of so-called independent particle models in 
which the form factors of one-body operators are strictly speaking zero: 
the transferred momentum is absorbed by only one ``active'' quark'' while 
the motion of the remaining ``spectator'' quarks is not affected. 
The nucleon wave function of such a configuration is strictly speaking zero.
In a more realistic description the nucleon wave-function would contain 
``correlations'' between the constituents through which the momentum 
transferred to the active quark would be redistributed among all 
constituents of the system such that the nucleon as a whole would recoil 
\cite{Ji:1997gm}. But the bag model quark wave functions are 
independent of each other, and lack explicit correlations.

At least in principle the bag model could provide correlations: the 
elastic scattering process could be thought of as consisting of two steps. 
In the first step the active quark absorbs the transferred momentum. 
In the second step the active quark ``bounces off'' the bag boundary, 
which subsequently transfers momentum to the spectator quarks, etc. 
Through such back-and-forth bouncing the transferred momentum would 
be redistributed among all constituents. For larger $|t|$ inelastic 
processes (bag deformation, creation of $\bar{q}q$-pairs) may become 
possible. Even though this simple mechanism cannot be expected to be 
realistic, at least in principle one could estimate correlation effects in 
this way. In practice this is too complex to consider, and a different way 
to heuristically estimate correlation effects was chosen in \cite{Ji:1997gm}:
a free parameter $\eta$ was introduced such that the momentum transfer to 
the active quark is $\vec{\Delta}\to\eta\vec{\Delta}/\cosh(w)$. It is 
intuitively expected that $\eta\sim 1/3$ to redistribute the momentum 
transfer among 3 quarks in a recoiled nucleon.
A reasonable description of the proton electromagnetic form factors was 
obtained for $\eta$ in the range of $\eta=$0.35--0.55 with the lower (higher) 
value yielding a better description of the data at large (intermediate) 
values of $|t|$ \cite{Ji:1997gm}.
The correlations modeled in this way impact the EMT form factors
more strongly than the 2 above-discussed types of $1/N_c$ corrections.
However, the discrepancy with the general constraints at $t=0$ becomes 
also more pronounced: e.g.\ for $\eta = 0.35$ one finds 
$J^Q(0) \approx 0.25$ \cite{Ji:1997gm} instead of $J^Q(0)=\frac12$
indicating that this method to estimate correlation effects is not 
trustworthy at small $|t|$, even though it improves the phenomenological
description of electromagnetic form factors at $|t|\lesssim2\,{\rm GeV}^2$ 
\cite{Ji:1997gm}. 
As our large-$N_c$ results are valid for small $|t|\ll M_N^2$, while
the results for $\eta\neq 1$ from \cite{Ji:1997gm} are more appropriate
at larger $|t|$, a direct comparison is not meaningful and we 
refrain from it.

Notice that in the large-$N_c$ limit also this type 
of corrections vanish. Let us recall that correlations were introduced
to allow the active quark to redistribute the momentum transfer among all
constituents such that the entire system changes its direction and 
the nucleon as a whole is deflected. However, in the leading order of the 
large-$N_c$ limit the momentum transfer is small, $|t|\ll M_N^2$, and the 
recoil of the heavy nucleon ($M_N\sim N_c$) is negligible. 
Thus, one can consistently evaluate form factors in the bag model without the need to introduce 
correlations. 
(Notice that absence of correlations in the large-$N_c$ limit is a
peculiarity of the bag model. Other models formulated in large-$N_c$ 
limit like the chiral quark soliton or Skyrme models 
\cite{Goeke:2007fp,Goeke:2007fq,Wakamatsu:2007uc,Cebulla:2007ei,
Kim:2012ts,Jung:2013bya,Jung:2014jja,Perevalova:2016dln} 
exhibit strong correlations.)

To summarize, we may regard the results for the EMT form factors shown 
in Fig.~1 as valid for $|t|\ll M_N^2$ and theoretically consistent within 
the bag model in the large-$N_c$ limit. These results are subject to 
$1/N_c$ corrections which we may expect to be modest at smaller $|t|$ 
and more sizable especially at larger $|t|$. 
Our observations are in line with results from the Skyrme model 
of Ref.~\cite{Ji:1991ff} where relativistic recoil corrections 
(to electromagnetic form factors) 
were also found small for $|t|<1\,{\rm GeV}^2$.

%====== SECTION 5: DENSITIES =======================================
\section{The EMT densities in bag model}
\label{Sec-5:densities}

In order to compute the EMT densities one can perform the Fourier
transforms in Eq.~(\ref{Def:static-EMT}). In the large-$N_c$ limit
in the bag model one can also directly evaluate the EMT matrix elements
in coordinate space. Both ways yield the same result for quark EMT 
densities. But only the direct evaluation in coordinate space allows us 
to compute the contributions of the ``gluons'' in ${\cal L}_G$ and the 
``quark-gluon interaction'' in ${\cal L}_{\rm surf}$ as defined in 
(\ref{Eq:Lagrangian-bag}). We obtain 
\begin{subequations}
\ba
    T^{00}_q(r) &=&
    \frac{N_q\,A^2}{4\pi}\;\frac{\Omega}{R}\,
    \biggl(\alpha_+^2j_0^2+\alpha_-^2j_1^2\biggr)\,\Theta_V\,,
    \label{Eq:EMT-stat-T00q}\\
    T^{0k}_q(\vec{r}) &=& 
	-\,\frac{1}{2}\;\frac{P_q\,A^2}{4\pi}\,
	\biggl(2\,\alpha_+\alpha_-\,\frac{\Omega}{R}\,j_0\,j_1
	+\alpha_-^2\frac{j_1^2}{r}\biggr)
	\epsilon^{klm}\,e_r^l\,S^m \,\Theta_V \,,
    \label{Eq:EMT-stat-T0kq}\\
    T^{ik}_q(\vec{r}) &=& 
    \frac{N_q\,A^2}{4\pi}\;\alpha_+\alpha_-\Biggl(
    \biggl(j_0\,j_1^\prime-j_0^\prime\,j_1-\frac{j_0\,j_1}{r}\;\biggr)
    e_r^ie_r^k \; + \; \frac{j_0\,j_1}{r}\;\delta^{ik} \Biggr)\,\Theta_V \,,
    \label{Eq:EMT-stat-Tijq}\\
    T^{\mu\nu}_G(r) &=& g^{\mu\nu}\;B\;\Theta_V \,,
    \label{Eq:EMT-stat-G}\\
    T^{\mu\nu}_{\rm surf}(\vec{r}) &=& 0 \, .\phantom{\Biggl(\frac11\Biggr)}
    \label{Eq:EMT-stat-surf} 
\ea
\end{subequations}
For brevity we suppress the arguments of the Bessel functions
$j_i=j_i(\omega\, r/R)$, and primes denote differentiation with 
respect to $r$. The quark flavor dependence is encoded in the SU(4) 
spin-flavor factors (\ref{Eq:Nq+Pq}). The contribution of 
${\cal L}_{\rm surf}$ vanishes, but we obtain the contribution 
$T^{\mu\nu}_G(r)=g^{\mu\nu}\;B\;\Theta_V $ associated with
non-fermionic (``gluonic'') effects.

%=============== BEGIN FIGURE 2: DENSITIES =========================
\begin{figure}[b!]
\begin{centering}
\includegraphics[height=3.95cm]{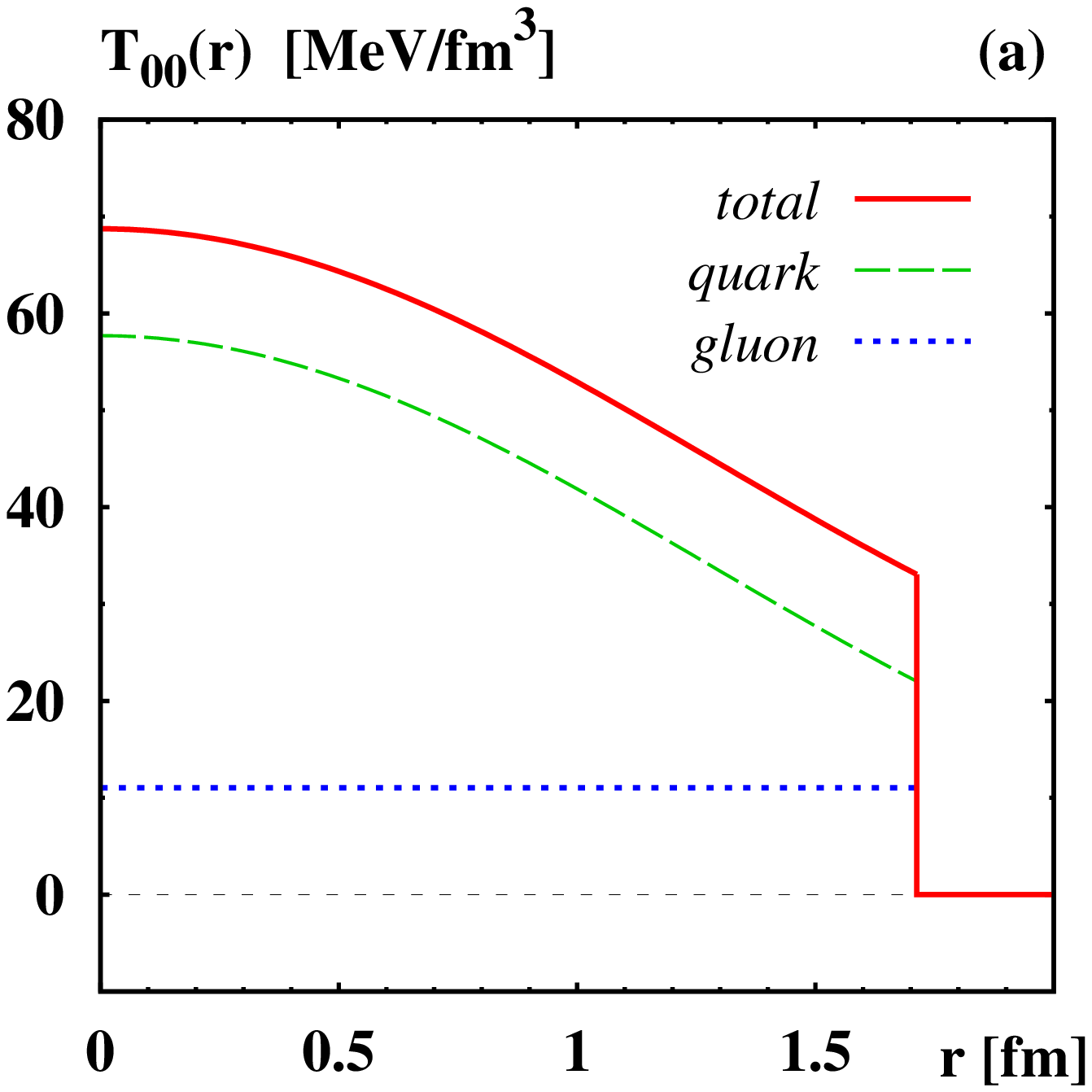} \ 
\includegraphics[height=3.95cm]{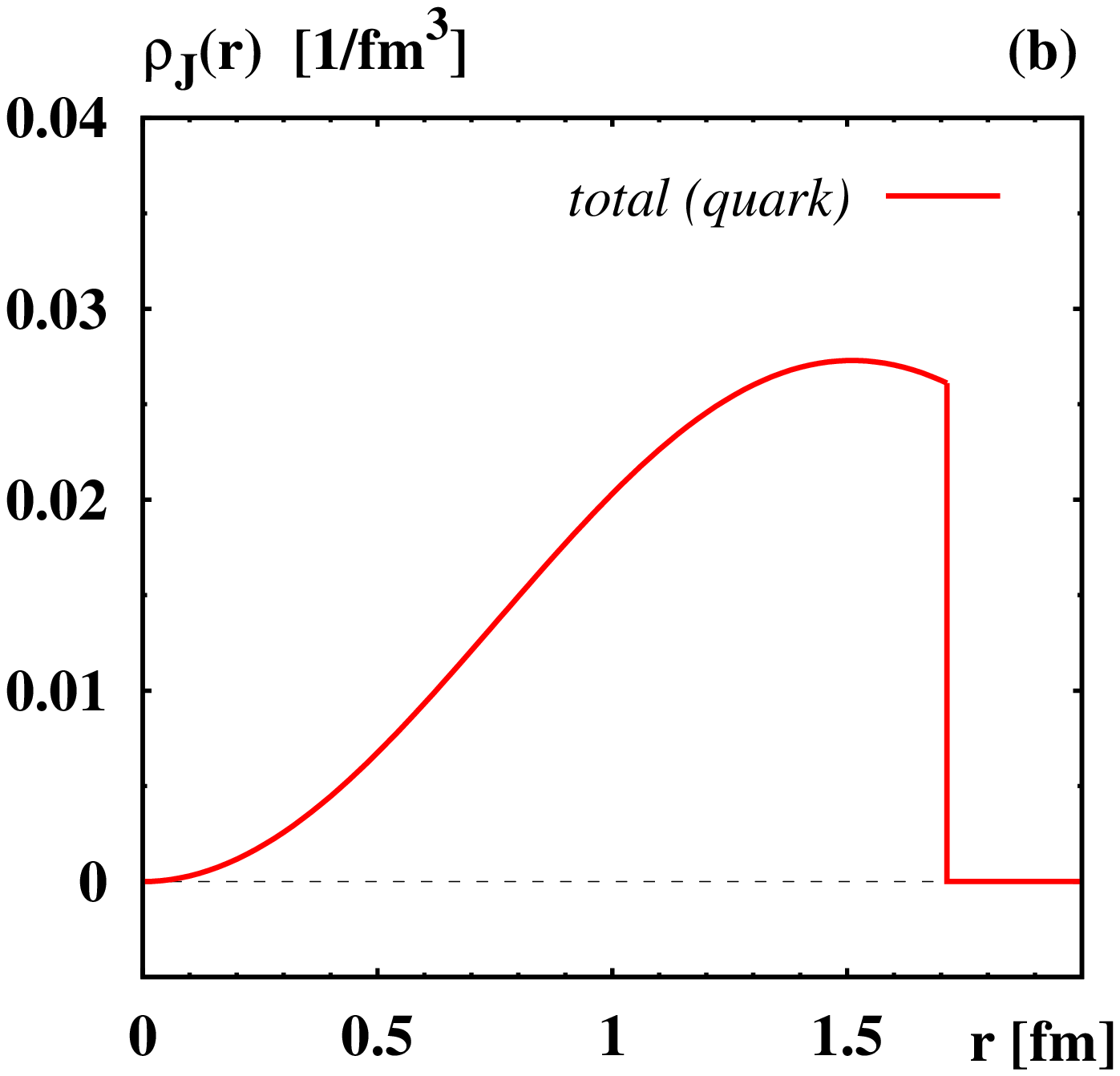} \ 
\includegraphics[height=3.95cm]{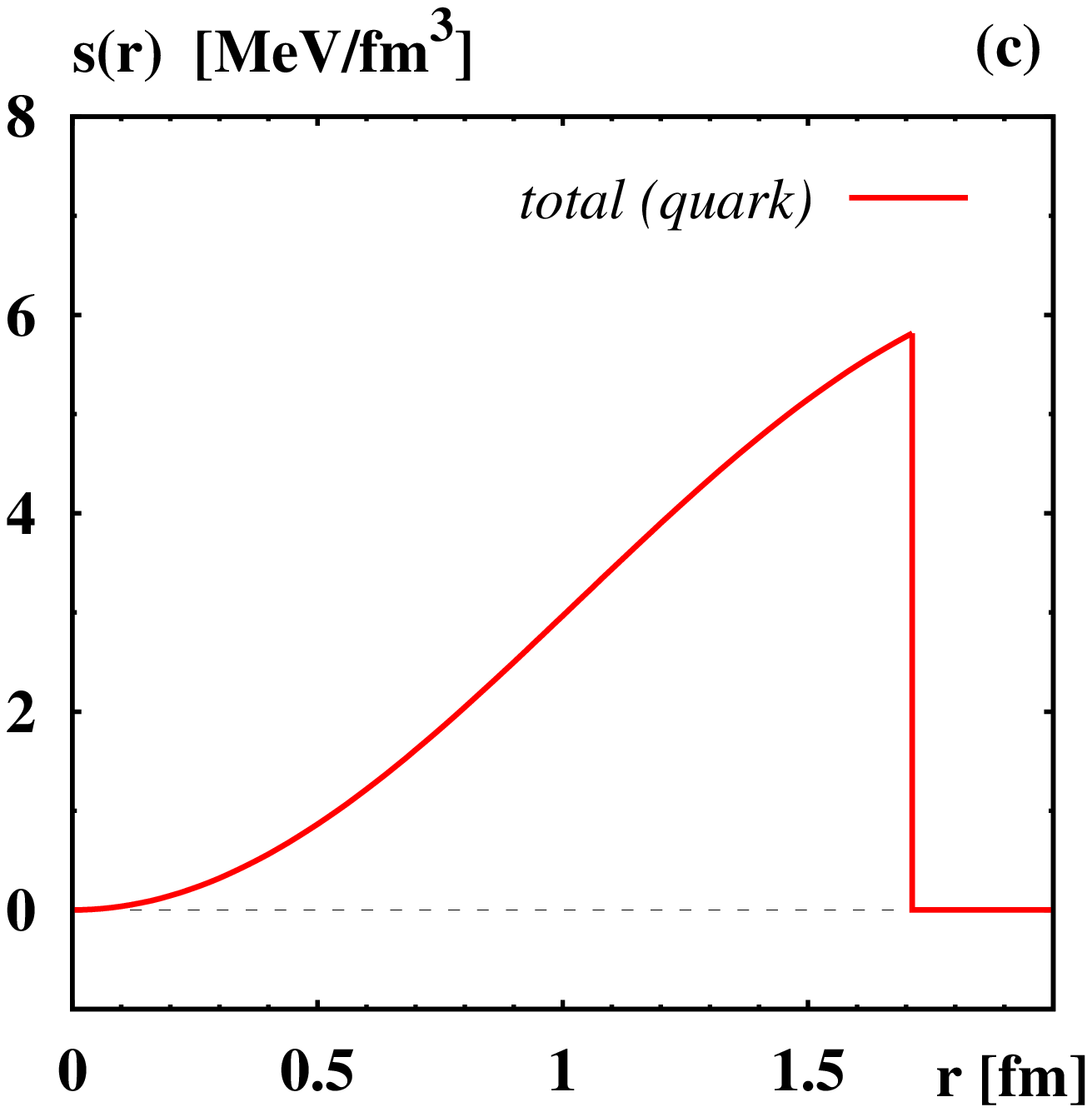} \
\includegraphics[height=3.95cm]{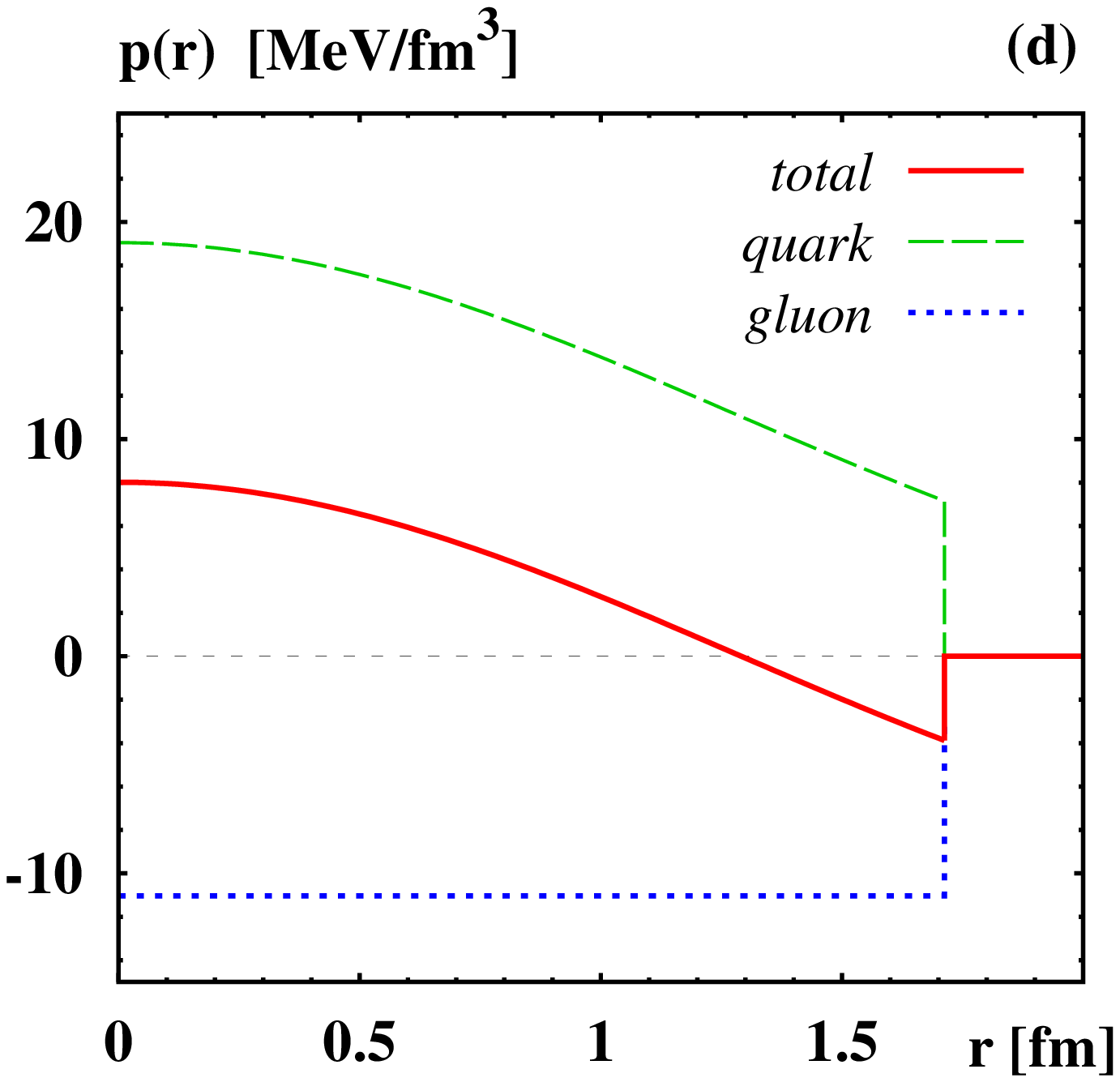} 
\par\end{centering}
\caption{\label{Fig-2:T00-p-s} (a) The energy density $T_{00}(r)$, 
(b) density $\rho_J(r)$ characterizing the angular momentum
density, (c) shear force distribution $s(r)$, and (d) 
pressure distribution $p(r)$ as functions of $r$ in the bag model 
for massless quarks. The vertical lines mark the position of the 
bag boundary (at $R=1.71\,$fm for massless quarks).
In the case of $T_{00}(r)$ and $p(r)$ the contributions from
quarks and ``gluons'' are shown in addition to the total result.
For $\rho_J(r)$ and $s(r)$ the total result is entirely due to quarks.}
\end{figure}
%================= END FIGURE 2 ====================================

\subsection{Energy density and mass}
\label{Sec-5a:T00}

The energy density $T_{00}(r)$ receives the contribution $T_{00}^Q(r)$ 
from quarks, Eq.~(\ref{Eq:EMT-stat-T00q}), and a contribution from 
``gluons'' $T_{00}^G(r)=B\,\Theta_V $ in Eq.~(\ref{Eq:EMT-stat-G}).
The quark and ``gluon'' contributions to the energy 
density are shown in Fig.~\ref{Fig-2:T00-p-s}a.
The integrated contributions are
\be\label{Eq:M-decompose}
	M_N^Q = N_c\,\varepsilon_0 \, ,\quad
	M_N^G = \frac{4\pi}{3}\,R^3 B\,.
\ee
For massless quarks the relative contributions of quarks and ``gluons'' 
to the nucleon mass are $M_N^Q \, : \, M_N^G = 3 \, : \, 1$.
This can be derived in 2 ways: (i) it follows from the
nonlinear bag boundary condition (\ref{Eq:eom-non}). Equivalently,
(ii) it can be derived from minimizing the nucleon mass understood
as a function of $R$ as follows. Since $\varepsilon_0=\omega_0/R$ we have 
\be\label{Eq:M-virial}
	M_N^\prime(R) = \frac{\partial}{\partial R} \biggl(N_c\,
	\frac{\omega_0}{R}+\frac{4\pi}{3}\,R^3 B\biggr)\stackrel{!}{=} 0 
	\quad \Leftrightarrow \quad 
	N_c\,\omega_0=4\pi\,R^4 B \,.
\ee
From Eqs.~(\ref{Eq:M-decompose},~\ref{Eq:M-virial}) we see that 
$M_N^Q = \frac34\,M_N$ and $M_N^G=\frac14\,M_N$ (for massless quarks).
This can be viewed as a bag-model version of the ``virial theorem.'' 
We recall that e.g.\ in soliton models virial theorems are derived 
by rescaling the coordinates $\vec{r}\to \lambda\vec{r}$ in the 
functional defining the nucleon mass. Considering infinitisemal
variations around $\lambda=1$ leaves the nucleon mass invariant,
i.e.\ $\delta M_N = 0$. This implies relations among different 
contributions to the nucleon mass \cite{Goeke:2007fp,Cebulla:2007ei}.
In the bag model the situation is simpler: the ``variation'' of the 
nucleon mass assumes the simple form stated in Eq.~(\ref{Eq:M-virial})
for massless quarks. For massive quarks $\omega_0=\omega_0(mR)$ depends 
also on $R$ and the virial theorem has a somewhat different form,
see App.~\ref{App:virial-massive}.
Notice that (\ref{Eq:M-virial}) shows that the constant $B={\cal O}(N_c)$ 
where one has to keep in mind that the bag radius $R={\cal O}(N_c^0)$ since
the size of baryons is of order $N_c^0$ in the large $N_c$ limit.

\subsection{Angular momentum density}
\label{Sec-5b:ang-mom}

The components $T_{0k}(\vec{r})$ depend on the 
nucleon polarization (which we do not indicate for brevity), and
receive only a contribution from quarks. The angular momentum density 
is given by
\ba
	J_q^i(\vec{r}) &=& \epsilon^{ijk} r^j T^{0k}_q(\vec{r})
%	= \epsilon^{ijk}	\epsilon^{klm}\,r^j\,e_r^l\,S^m \Biggl[
%	-\,\frac{1}{2}\;\frac{P_q\,A^2}{4\pi}\,
%	\biggl(\frac{2\omega}{R}\,\alpha_+\alpha_-j_0\,j_1
%	+\alpha_-^2\frac{j_1^2}{r}\biggr)
%	\,\Theta_V 	\Biggr] \nonumber\\
%	&=& 
%	(\delta^{il}\delta^{jm}-\delta^{im}\delta^{jl})
%	\,e_r^j\,e_r^l\,S^m\;r\Biggl[
%	-\,\frac{1}{2}\;\frac{P_q\,A^2}{4\pi}\,
%	\biggl(\frac{2\omega}{R}\,\alpha_+\alpha_-j_0\,j_1
%	+\alpha_-^2\frac{j_1^2}{r}\biggr)
%	\,\Theta_V \Biggr] \nonumber\\
%	&=& 
%	S^m(\delta^{im}-\,e_r^i\,e_r^m)
%	\;r\Biggl[
%	\frac{1}{2}\;\frac{P_q\,A^2}{4\pi}\,
%	\biggl(\frac{2\omega}{R}\,\alpha_+\alpha_-j_0\,j_1
%	+\alpha_-^2\frac{j_1^2}{r}\biggr)
%	\,\Theta_V \Biggr] \nonumber\\
%	&=& 
         =
	S^m\biggl[\delta^{im}\rho^q_{J}(r)_{\rm mono}
	+\,\biggl(e_r^i\,e_r^m-\frac13\delta^{im}\biggr)\rho^q_{J}(r)_{\rm quad}
	\biggr]\,,
\ea
with the monopole \cite{Goeke:2007fp} and quadrupole \cite{Lorce:2017wkb} 
contributions
\be\label{Eq:ang-mom-dens}
	\rho^q_{J}(r)_{\rm mono}
	=-\frac23\,\rho^q_{J}(r)_{\rm quad} \equiv \rho^q_{J}(r), \quad
	\rho^q_{J}(r) = 
	\frac13\;\frac{P_q\,A^2}{4\pi}\;r\,
	\biggl(\frac{2\Omega}{R}\,\alpha_+\alpha_-j_0\,j_1
	+\alpha_-^2\frac{j_1^2}{r}\biggr)
	\,\Theta_V \,.
\ee
The relation 
$\rho^q_{J}(r)_{\rm mono}=-\frac23\,\rho^q_{J}(r)_{\rm quad}$ is a general 
result \cite{Schweitzer:2019kkd} which the bag model respects. The 
total angular momentum density $\rho_J(r)=\sum_q\rho^q_J(r)$ 
is normalized as $\int\di^3r\,\rho_J(r)=\frac12$ and shown in 
Fig.~\ref{Fig-2:T00-p-s}b.

\subsection{\boldmath Shear forces and pressure}
\label{Sec-5c:s-and-p}

The pressure and shear forces encoded in the stress tensor 
(\ref{Eq:T_ij-pressure-and-shear}) are given by the expressions
\begin{eqnarray}
  p(r) & = & 
  \Biggl[\frac{N_{c}\,A^{2}}{12\pi}\;\alpha_{+}\alpha_{-}
  \biggl(j_{0}j_{1}^{\prime}-j_{0}^{\prime}j_{1}+\frac{2}{r}j_{0}j_{1}\biggr)-B\Biggr]
  \,\Theta_V ,\nonumber \\
  s(r) & = & 
  \Biggl[\frac{N_{c}\,A^{2}}{4\pi}\;\alpha_{+}\alpha_{-}
  \biggl(j_{0}j_{1}^{\prime}-j_{0}^{\prime}j_{1}-\frac{1}{r}j_{0}j_{1}\biggr)\Biggr]
  \,\Theta_V .\label{Eq:p+s-bag}
\end{eqnarray}
The numerical results for massless quarks are shown in 
Figs.~\ref{Fig-2:T00-p-s}c and \ref{Fig-2:T00-p-s}d. 

In the liquid drop model of a large nucleus which exhibits 
a ``sharp edge'' at the radius $R_{\rm nucl}$ the shear force 
is given by $s(r)=\gamma\,\delta(r-R_{\rm nucl})$ where $\gamma$ 
is the surface tension \cite{Polyakov:2002yz}. 
The nucleon is a far more diffuse object then a large nucleus 
and Fig.~\ref{Fig-2:T00-p-s}c shows that $s(r)$ is consequently
much more ``spread out'' than a $\delta$-function characterizing 
the shear force distribution of a large nucleus. 

In all model calculations so far the pressure was found positive
in the inner region and negative in the outer region. This is
also the case in the bag model, see Fig.~\ref{Fig-2:T00-p-s}d.
The positive pressure in the inner region is associated with
repulsive forces directed towards outside. The negative pressure 
in the outer region corresponds to attractive forces directed toward 
inside. The repulsive and attractive forces must compensate each 
other according to the von Laue condition which is a necessary
condition for stability and will be discussed below in 
Sec.~\ref{Sec-5e:stability}.

The pressure distribution and the shear forces in Eq.~(\ref{Eq:p+s-bag}) 
satisfy the differential equation (\ref{Eq:p(r)+s(r)}).
This relation is a consequence of the conservation of the EMT,
$\partial_{\mu}T^{\mu\nu}=0$, and hence reflects the fact that in the
bag model the EMT is conserved and the description is internally 
consistent. 
% Further tests of the consistency of the model based on the 
% EMT conservation will be presented in the following sections and in
% App.~\ref{App}.

\subsection{Normal and tangential forces}
\label{Sec-5d:normal-tangential-forces}

%=============== BEGIN FIGURE 3: NORMAL & TANGETIAL FORCES =========
\begin{figure}[t!]
\begin{centering}
\includegraphics[width=4cm]{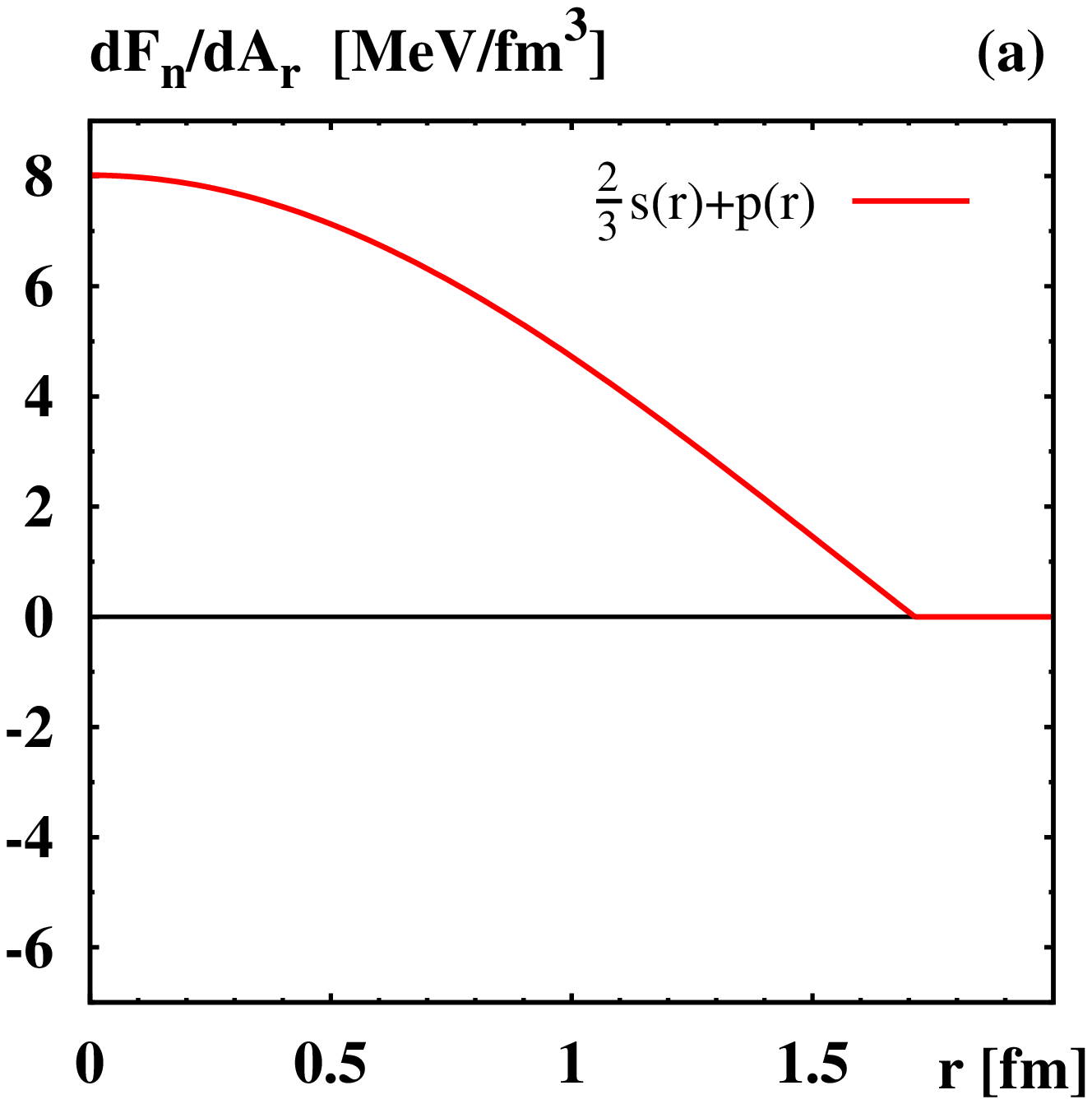}
\includegraphics[width=4cm]{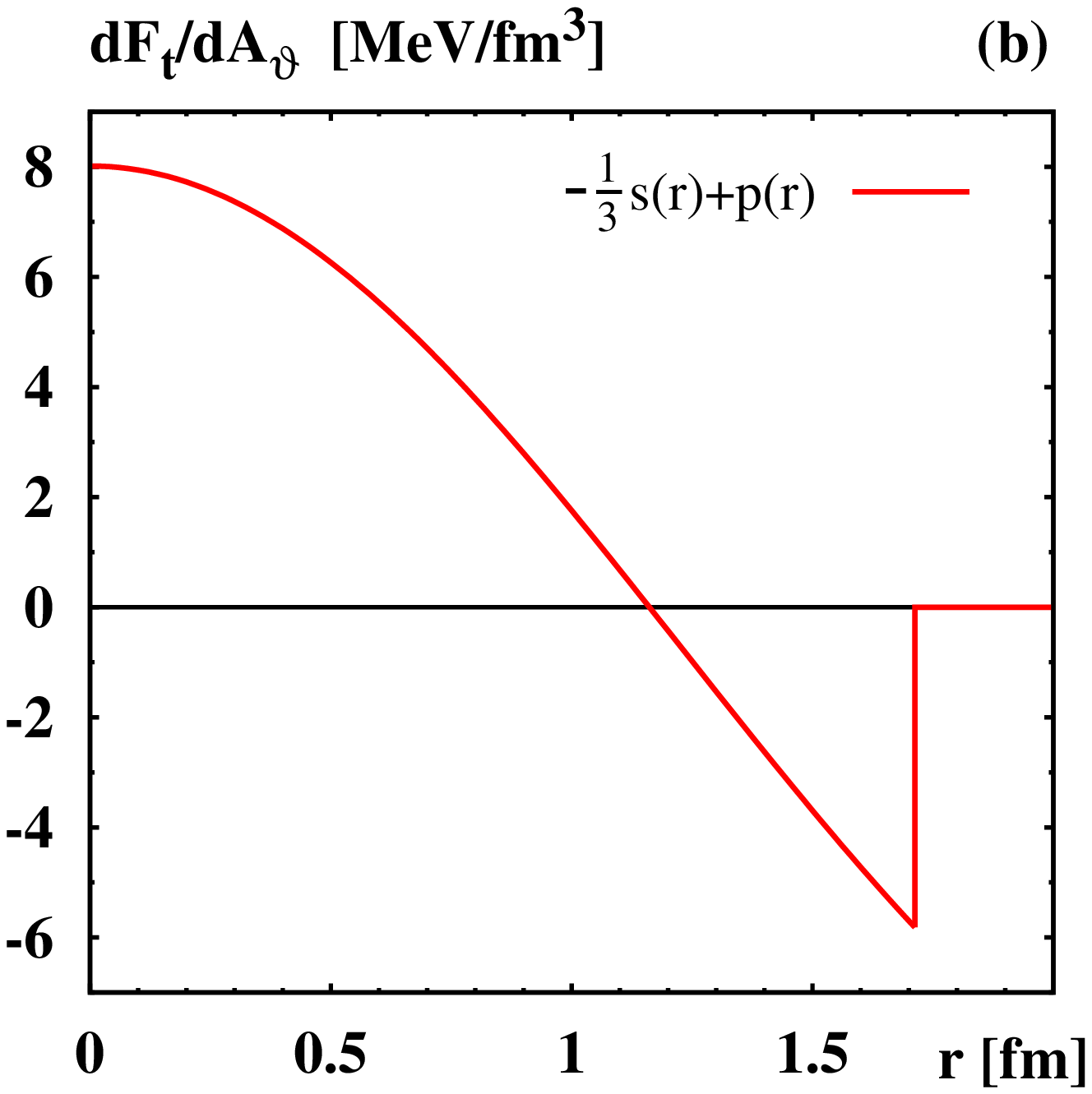}
\end{centering}
\caption{\label{Fig-4:forces} 
Densities of 
(a) normal $\di F_n/\di A_r = \frac23s(r)+p(r)$ and
(b) tangential 
$\di F_t/\di A_\vartheta=\di F_t/\di A_\varphi = -\frac13s(r)+p(r)$ 
forces per unit area
in bag model as functions of $r$. Mechanical stability requires 
$\frac23s(r)+p(r)>0$ inside the bag which is the case.}
\end{figure}
%================= END FIGURE 3 ====================================

The stress tensor (\ref{Eq:T_ij-pressure-and-shear}) is a 
symmetric $3\times3$ matrix whose eigenvectors are the unit vectors 
$\vec{e}_r$, $\vec{e}_\vartheta$, $\vec{e}_\varphi$ of the spherical 
coordinate system and eigenvalues are related to normal and tangential 
forces   \cite{Polyakov:2018zvc}. For spin-0 and 
spin-$\frac12$ particles the tangential eigenvalues (pertaining to
eigenvectors $\vec{e}_\vartheta$, $\vec{e}_\varphi$) are degenerate with the 
degeneracy being lifted only for higher spin $J\ge1$ particles. 
In our case the normal and tangetial forces per unit area 
are given by \cite{Polyakov:2018zvc}
\ba
     T^{ij}\;\di A_r^j 
     &=& \frac{\di F_n}{\di A_r}\;\,\di A_r \;e_{r}^i
      =  \biggl(\frac23s(r)+p(r)\biggr)\,\di A_r \;e_{r}^i\nonumber\\
     T^{ij}\;\di A_{\vartheta}^j 
     &=& \frac{\di F_t}{\di A_\vartheta}\;\,\di A_\vartheta \;e_{\vartheta}^i
      =  \biggl(-\frac13s(r)+p(r)\biggr)\,\di A_\vartheta \;e_{\vartheta}^i
\ea
where $\di\vec{A}_r = \di A_r \,\vec{e}_r$, etc denote the corresponding 
infinitesimal area elements. The results for normal forces 
$\di F_n/\di A_r$ and tangential forces 
$\di F_t/\di A_\vartheta=\di F_t/\di A_\varphi$ 
are shown in Fig.~\ref{Fig-4:forces}.

Mechanical stability requires that $\di F_n/\di A_r\ge 0$ with strictly 
$\di F_n/\di A_r > 0$ at all values of $r$ within the system 
\cite{Perevalova:2016dln}. The position where $\di F_n/\di A_r = 0$ marks 
the ``end'' of the system \cite{Polyakov:2018zvc}.
In the bag model it is consequently $\di F_n/\di A_r > 0$ for 
$0 \le r < R$ and the normal force vanishes at the finite radius $r=R$, 
as shown in Fig.~\ref{Fig-4:forces}a. This is a distinctly different 
situation than in soliton models where  $\di F_n/\di A_r > 0$ for all 
$0 \le r < \infty$ and the normal forces vanish only in the limit $r\to\infty$
\cite{Goeke:2007fp,Goeke:2007fq,Wakamatsu:2007uc,Cebulla:2007ei,
Kim:2012ts,Jung:2013bya,Jung:2014jja,Perevalova:2016dln}.
Other examples of finite size systems which are analogous in the sense
that $\di F_n/\di A_r$ vanishes at a finite radius are the 
liquid drop model \cite{Polyakov:2002yz} and neutron stars
whose radius is defined as that value of $r$ where the normal force 
per unit area (also called the hydrostatic pressure) vanishes 
\cite{Cedric-private}.

\subsection{Mechanical radius, surface tension, and diffusiveness}

The positivity of the normal forces allows one to introduce the
notion of a mechanical radius defined as \cite{Polyakov:2018zvc}
\be
        \la r^2\ra_{\rm mech}=
        \frac{\int\di^3r\:r^2[\frac23s(r)+p(r)]}
             {\int\di^3r\,   [\frac23s(r)+p(r)]}\,.
\ee
We obtain $\la r^2\ra_{\rm mech}^{1/2}=1.10\,{\rm fm}$ which is smaller 
than the proton charge radius $\la r^2\ra_{\rm el}^{1/2}=1.25\,{\rm fm}$ 
with our parameters. The values of the radii depend on how 
model parameters are fixed, e.g.\ a smaller proton charge radius 
of $1\,{\rm fm}$ was found in \cite{Chodos:1974pn} with a different 
parameter fixing. A more robust prediction might be the ratio 
$\la r^2\ra_{\rm mech}^{1/2}/\la r^2\ra_{\rm el}^{1/2}=0.88$
which is independent of how model parameters are fixed 
(for massless quarks). Interestingly, also the chiral 
quark soliton model predicts the mechanical radius 
to be smaller than the proton charge radius 
(by $25\,\%$ in that model) \cite{Polyakov:2018zvc}. 
Notice that the mechanical radius is the same for proton and neutron 
modulo small isospin violating effects, and hence constitutes a better 
concept for the nucleon ``size'' than the charge radius 
(which is negative for the neutron, giving insights on the 
distribution of charge inside neutron but not on its size).

One may define the property of ``surface tension'' for a hadron as
\be\label{Eq:gamma}
      \gamma = \int_0^\infty\di r\,s(r)\,,
\ee
if this integral exists.
In the bag model we find $\gamma = 4.26\,{\rm MeV\,/\,fm^2}$.
The concept of a surface tension is well-justifed in certain
situations, for instance for large nuclei \cite{Polyakov:2002yz}
or $Q$-balls \cite{Mai:2012yc}. The nucleon is much more
diffuse. In order to quantify the ``diffusiveness'' of a particle
we introduce the dimensionless measure $\Delta w^2$ for the 
``skin thickness'' of a particle defined in terms of the moments 
$\la r^n\ra_s$ of the shear force distribution as follows \cite{Mai:2012yc}
\be\label{Eq:diffusiness}
      \Delta w^2 
 = \frac{\bigl\la\, [r^2-\la r^2\ra_s^{ }]^2_{ }\bigr\ra_s^{1/2}}{\la r^2\ra_s^{ }}
 = \frac{\bigl[\la r^4\ra_s^{ }-\la r^2\ra_s^2\bigr]^{1/2}_{ }}{\la r^2\ra_s^{ }}
   \,, \quad
   \la r^n_{ }\ra_s^{ } = \frac{1}{\gamma}\,\int\di r\,r^n_{ }s(r)\,.
\ee
For a nucleus with a sharp edge in the liquid drop model the
shear force is given by $s(r)=\gamma\,\delta(r-R_A)$ where $R_A$
denotes the radius of the nucleus, and $\Delta w^2 =0$. One also
finds $\Delta w^2\to 0$ in the limit of very large $Q$-balls 
\cite{Mai:2012yc}. For realistic nuclei and finite-size $Q$-balls
the ``diffusiveness'' parameter $\Delta w^2$ is small.
For the nucleon $\Delta w^2 = 0.48$ in the bag model 
indicating that the nucleon is much more diffuse than a nucleus,
which is not unexpected.

%=============== BEGIN FIGURE 4: VON LAUE ==========================
\begin{figure}[b!]
%\begin{wrapfigure}[14]{r}{0.42\textwidth}
\begin{centering}
\includegraphics[width=5.3cm]{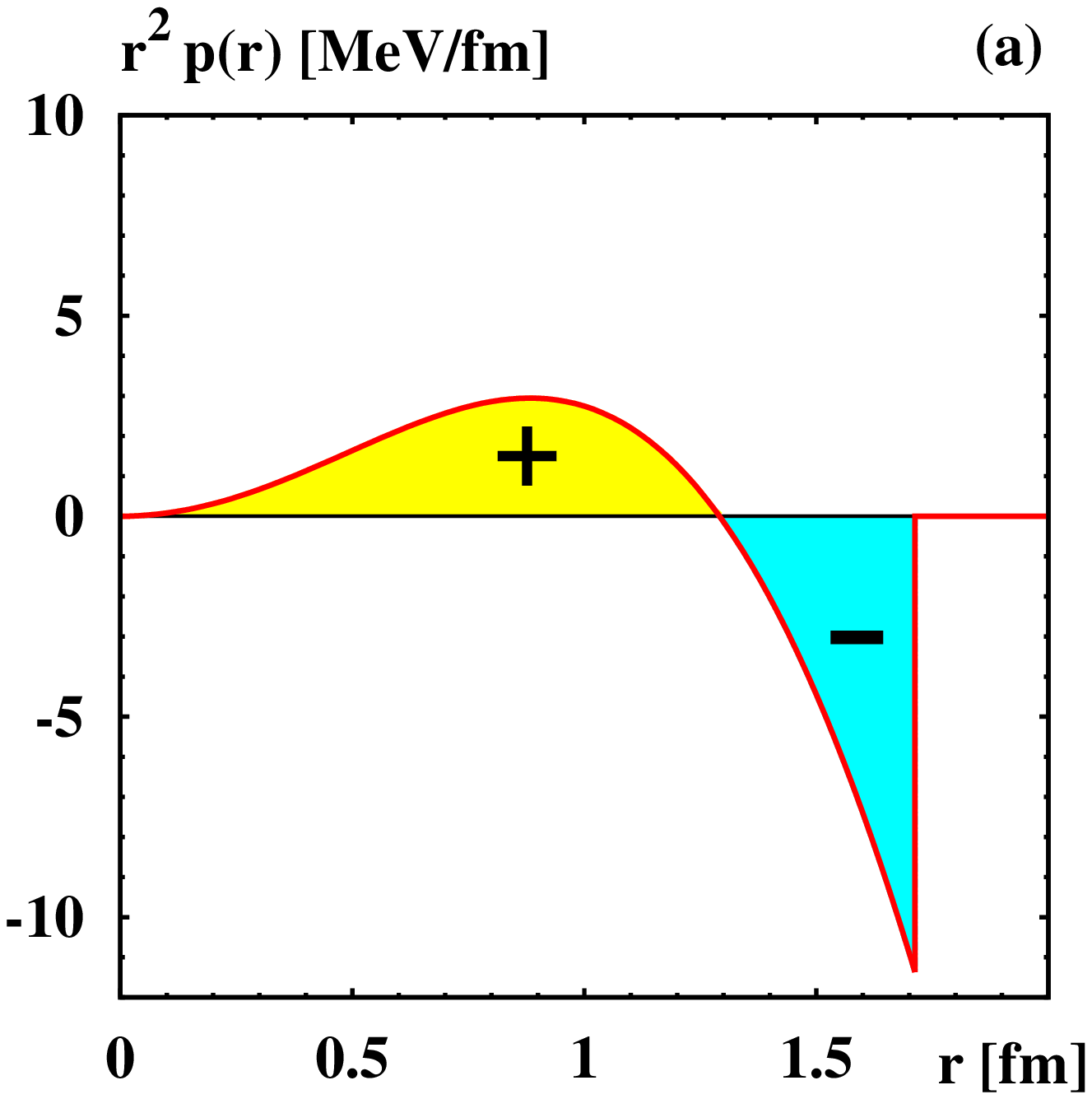}          \
\includegraphics[width=5.3cm]{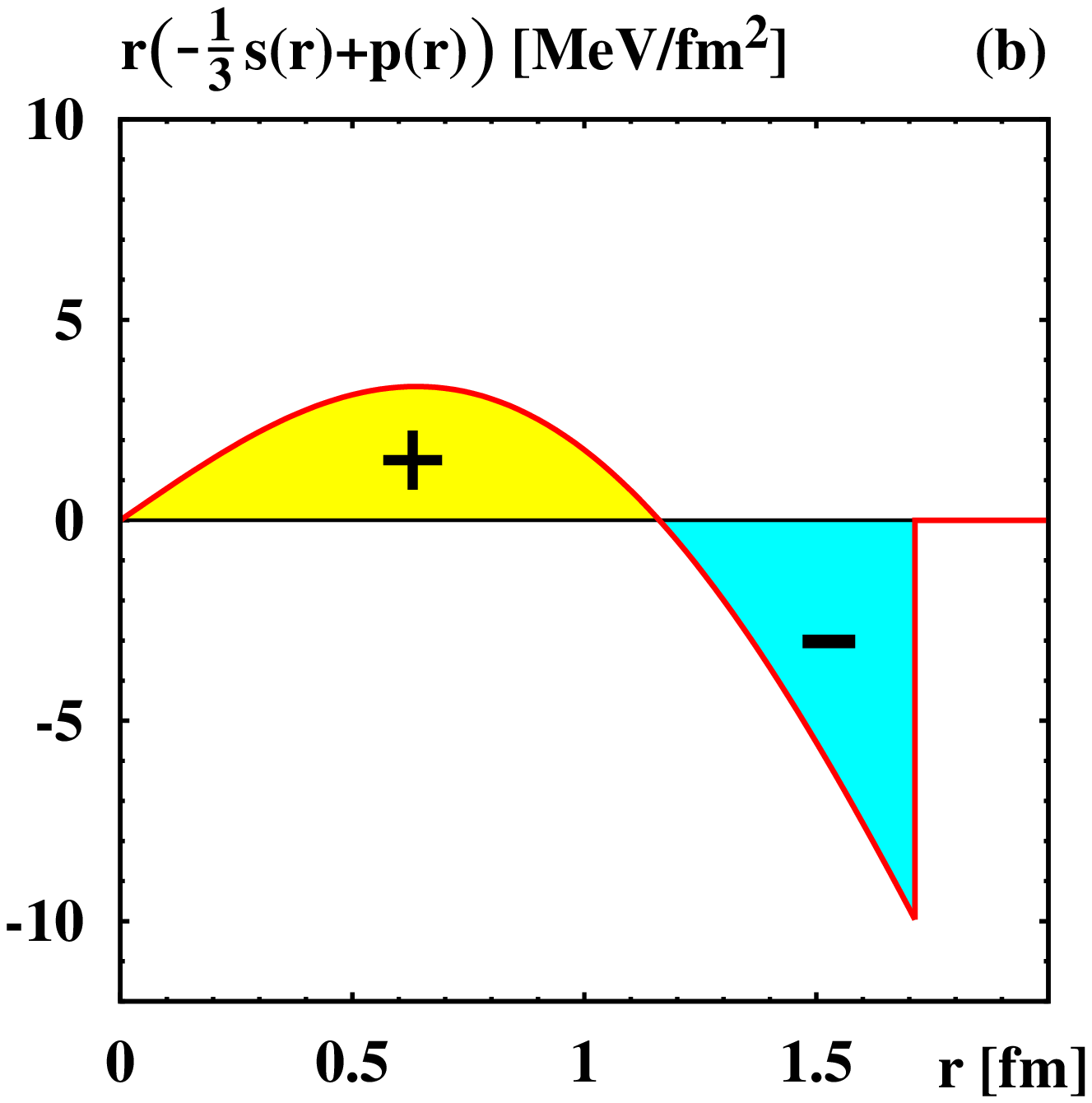}  \
\includegraphics[width=5.3cm]{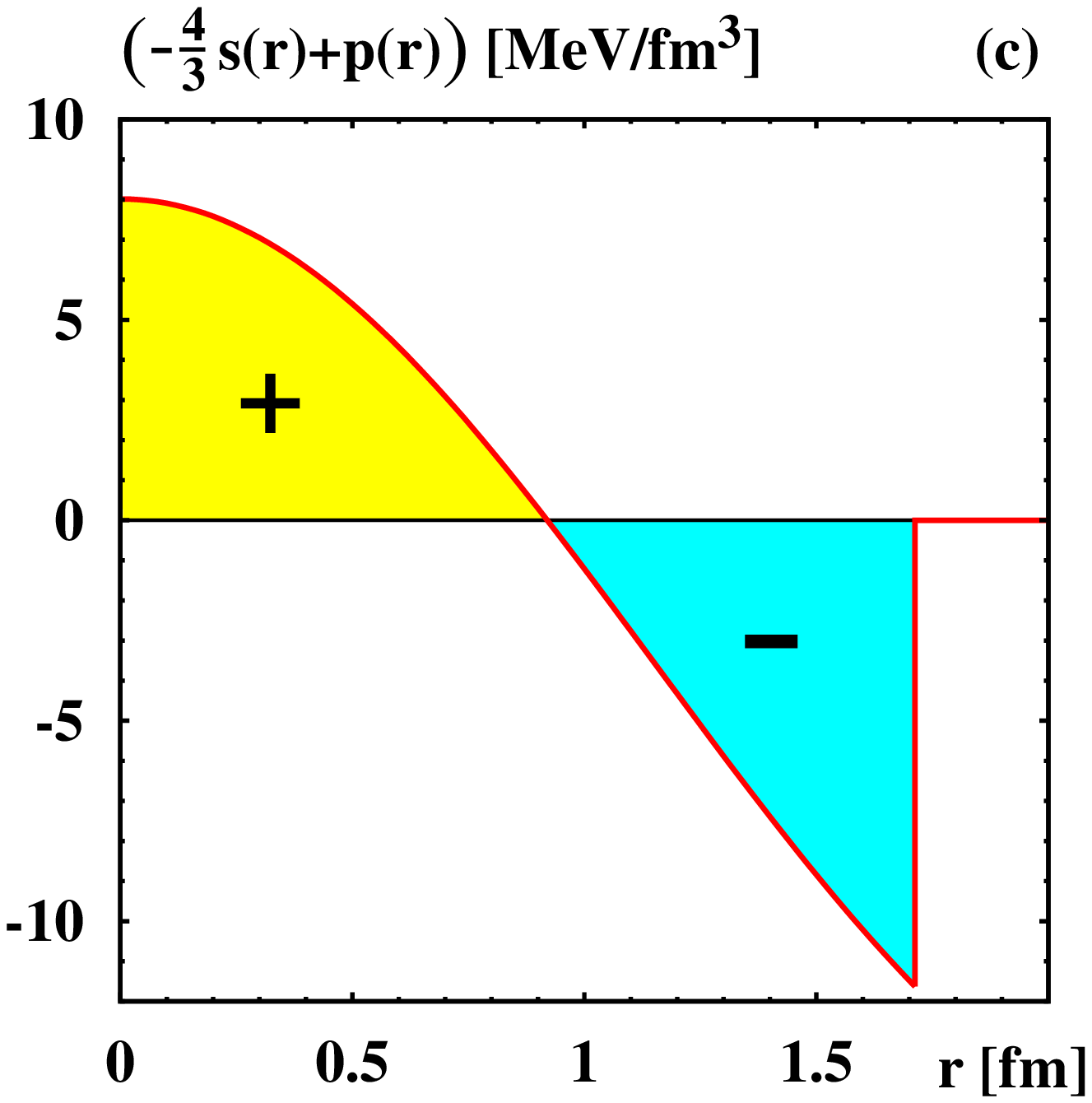} 
\end{centering}
\caption{\label{Fig-3:bag-model-r2-p} 
The 3D von Laue condition (a) and its lower-dimensional analogs 
in 2D (b) and 1D (c) in the bag model for massless quarks. 
The areas above and below the $r$-axis are equal and compensate 
each other according to the integrals in  Eq.~(\ref{Eq:stability}).}
%\end{wrapfigure}
\end{figure}
%================= END FIGURE 4 ====================================

\subsection{EMT conservation: von Laue condition 
and its lower-dimensional analogs}
\label{Sec-5e:stability}

The pressure and shear forces must obey the following integral relations 
\be   \label{Eq:stability}
      \int\limits_0^\infty \!\di r\;r^2p(r)=0 \;,\quad
      \int\limits_0^\infty \!\di r\;r\biggl(-\frac13\,s(r)+p(r)\biggr)=0\;,\quad
      \int\limits_0^\infty \!\di r\;\biggl(-\frac43\,s(r)+p(r)\biggr)=0\;.
\ee
The first of these relations was introduced by von Laue 
in Ref.~\cite{von-Laue} and holds in 3D, the other two hold in 
respectively 2D and 1D and were derived in \cite{Polyakov:2018zvc}.

The conditions in (\ref{Eq:stability}) are proven analytically in 
App.~\ref{App:von-Laue}.
The physical interpretation of the first condition in (\ref{Eq:stability}) 
is as follows. The positive pressure in the inner region corresponds to 
repulsion and the negative pressure in the outer region corresponds to 
attraction. Mechanical stability requires that the attractive and 
repulsive forces compensate each other in the 3D integral in
(\ref{Eq:stability}) which is satisfied in the bag model as shown 
in Fig.~\ref{Fig-3:bag-model-r2-p}a.
The tangential force per unit area, $-\frac13\,s(r)+p(r)$, must satisfy
the 2D relation in (\ref{Eq:stability}) which is the case in the bag model
as illustrated in Fig.~\ref{Fig-3:bag-model-r2-p}b. The interpretation 
of this condition is that the tangential forces within a 2D slice must 
compensate each other \cite{Polyakov:2018zvc}. Similarly also the 1D 
condition in (\ref{Eq:stability}) is satisfied in the bag model, 
which is illustrated in Fig.~\ref{Fig-3:bag-model-r2-p}c.
It is also instructive to discuss the ``finite-volume von Laue condition'' 
\cite{Polyakov:2018zvc}
\be\label{Eq:finite-von-Laue}
      \int\limits_{|\vec{r}{ }^{\,\prime}|\le r} \!\!\! \di^3r^\prime\;p(r^\prime) =
%      V(r)\,\frac{\di F_n}{\di A_r} = 
      V(r)\,\biggl(\frac23\,s(r)+p(r)\biggr) \,,
\ee
where the integration goes over the volume $V(r)=\frac43\,\pi\,r^3$.
The sum rule (\ref{Eq:finite-von-Laue}) is satisfied for all $0\le r < R$. 
However, in the bag model at $r=R$ one practically deals with the 3D 
relation in (\ref{Eq:stability}). Since $V(R)\neq0$ is non-zero, this
means that the normal force per unit area $\frac23\,s(r)+p(r)$ must vanish 
at the bag boundary, which is the case and emerges here as a necessary 
condition to comply with the von Laue condition in (\ref{Eq:stability}).
Notice that $\frac23\,s(r)+p(r)$ must vanish at the bag boundary
also in order to comply with (\ref{Eq:p(r)+s(r)}). The differentiation of 
$\Theta_V$-functions in the bag model expressions (\ref{Eq:p+s-bag}) for 
$p(r)$, $s(r)$ yields the contribution $\delta(r-R)\,[\frac23\,s(r)+p(r)]$
to (\ref{Eq:p(r)+s(r)}) which must and does vanish at $r=R$.
% The relations (\ref{Eq:stability},~\ref{Eq:finite-von-Laue}) 
% are consequences of the EMT conservation and constitute necessary 
% but not sufficient conditions for stability. 

The integrands $p^{(3D)}(r)=p(r)$, $p^{(2D)}(r)=-\frac13 s(r)+p(r)$, 
and $p^{(1D)}(r)=-\frac43 s(r)+p(r)$ in (\ref{Eq:stability}) are 
special cases of pressures in $n$-dimensional ($nD$) spherically 
symmetric mechanical systems. In general pressure and shear forces 
of a $kD$ system are related to those in $nD$ subsystems 
(if $k<n$ the roles of system and subsystem interchange) as 
 \cite{proceeding-new}
\begin{align}
  p^{(nD)}(r)=\frac kn\  p^{(kD)}(r)+\frac{k (n-k)}{n}\;\frac{1}{r^k}
  \int_0^r dr^\prime\ r^{\prime\,k-1} p^{(kD)}(r^\prime), & \nonumber\\
  s^{(nD)}(r)=-\frac {k}{n-1}\ p^{(kD)}(r)+\frac{k^2}{n-1}\;\frac{1}{r^k}
  \int_0^r dr^\prime\ r^{\prime\,k-1} p^{(kD)}(r^\prime). & \label{eq:nDvskD}
\end{align}
The $s^{(nD)}(r)$ and $p^{(nD)}(r)$ in (\ref{eq:nDvskD}) satisfy 
$\frac{n-1}{n}\frac{\partial\;}{\partial r}s^{(nD)}(r)
+\frac{n-1}{r}s^{(nD)}(r)
+\frac{\partial\;}{\partial r}p^{(nD)}(r)=0$ 
and
$\int_0^\infty\di r\;r^{n-1}p^{(nD)}(r)=0$
which are $nD$-versions of respectively (\ref{Eq:p(r)+s(r)}) and
(\ref{Eq:stability}). Such relations can be useful e.g.\ in holographic
 approaches to QCD or in fractal theories \cite{proceeding-new}.
In the bag model these relations are valid for all $n,\,k>0$ including 
non-integer and aribtrarily large values.
The practical verification of such relations can in practice be
numerically challenging especially for large $n$.
In the bag model, thanks to the finite range of the densities,
it is possible to test the validity and consistency of the relations 
(\ref{eq:nDvskD}) for any value of $n,\,k>0$ in a non-trivial model.

\subsection{\boldmath EMT conservation: equivalence of $D$-term expressions}
\label{Sec-5f:D-term-equivalence}

The $D$-term can be computed using the expressions in terms of (i) pressure 
and (ii) shear forces according to Eq.~(\ref{Eq:d1-from-s(r)-and-p(r)}).
From (\ref{Eq:p+s-bag}) we find that the two equivalent expressions in 
Eq.~(\ref{Eq:d1-from-s(r)-and-p(r)}) yields the same result which can 
be written as
\be\label{Eq:D-term-bag}
	D= 
        \frac13\,M_N\,N_c\,\frac{A^2 R^4\!}{\omega_0^4}\;\alpha_+\alpha_-
	\biggl(
	-\frac{4}{15}\,\omega_0 ^3 + \omega_0
        -\frac25\,\omega_0\,\sin^2\omega_0-\sin\omega_0 \,\cos\omega_0
	\biggr)\,,
\ee
see App.~\ref{App:D-equivalence} for a detailed proof.
The possibility to compute the $D$-term by means of two different 
equivalent expressions is also due to EMT conservation. We will 
discuss the $D$-term in Sec.~\ref{Sec-6:D-term} in more detail.

\subsection{\boldmath EMT conservation: form factor $\bar{c}^G(t)$}
\label{Sec-5f:cbar-gluon}

In Sec.~\ref{Sec-4:EMT-FFs-quarks} we found the form factor
$\bar{c}^Q(t)\neq0$ from the evaluation of the quark EMT, which 
means that $T_{\mu\nu}^Q$ by itself is not conserved. EMT conservation
requires $\sum_a\bar{c}^a(t)=0$ if one takes into account all contributions 
in a system, i.e.\ in the bag model also the contribution of the bag which 
simulates gluons in the sense discussed in Sec.~\ref{Sec-3:bag-model}.
However, while it was straight forward to compute the quark EMT form
factors in Sec.~\ref{Sec-4:EMT-FFs-quarks}, it is not clear how to
compute the bag contribution to the form factors. At this
point we can take advantage of the EMT density framework.
Instead of using EMT form factors as an input for an interpretation 
in terms of EMT densities \cite{Polyakov:2002yz}, we proceed in the 
opposite direction and invert Eq.~(\ref{Def:static-EMT}) for the ``gluon'' 
contribution $T_{\mu\nu}^G(r)$ in Eq.~(\ref{Eq:EMT-stat-G}).
We obtain for massless quarks the result
\be
	\bar{c}^G(t)=\frac{1}{M_N}
	\int\di^3r \exp(i\vec{\Delta}\vec{r})\,B\,\Theta_V 
	= \frac{3}{4}\;\frac{j_1(qR)}{qR}\;,\quad
	q=\sqrt{-t}
\ee
where we eliminated the bag constant $B$ by means of Eq.~(\ref{Eq:M-virial}).
From the behavior of spherical Bessel functions for small arguments
$j_l(z)=z^l/(2l+1)!!+{\cal O}(z^{l+2})$ we find $\bar{c}^G(0)=\frac14$ to 
be compared with $\bar{c}^Q(0)=-\,\frac14$, see Sec.~\ref{Sec:num-res-FFs} 
and Fig.~\ref{Fig-01:FFs}d. In App.~\ref{App} we show that we have
$\forall\;t$
\be\label{Eq:sum-bar-c-bag}
	\bar{c}^Q(t)+\bar{c}^G(t)=0\,
\ee
as it is required by the conservation of the total EMT.

%====== SECTION 6: D-TERM ==========================================
\section{\boldmath The $D$-term}
\label{Sec-6:D-term}

In this Section we discuss the $D$-term for the nucleon and other hadrons, 
and we consider then several instructive limiting cases in the bag model.
Here and throughout in Secs.~\ref{Sec-6:D-nucleon}--\ref{Sec-6:D-highly-excited}
we consider massless quarks.
In Sec.~\ref{Sec-7:D-limiting-cases} we will discuss also $m\neq0$. 
The expression for the $D$-term of the nucleon was already quoted in 
(\ref{Eq:D-term-bag}). Let us generalize this result to a general
state. 
Mesons (baryons) are constructed in the bag model by placing a 
$\bar{q}q$ ($qqq$) in the bag in a color singlet state. The mass 
and bag radius of a general bag model state (for massless quarks)
are given by
\be\label{Eq:M-R-general}
      M = \frac43\,\frac{\sum_i\omega_i}{R}\,, \quad
      R= \biggl(\frac{\sum_i\omega_i}{4\pi B}\biggr)^{1/4}
\ee
which follows from the virial theorem (\ref{Eq:M-virial}).
For baryons the summation goes over $N_c=3$ occupied 
bag levels $\omega_i$, for mesons over two levels.
The bag constant is fixed as $B=0.0559\,{\rm fm}^{-4}$ 
to reproduce the nucleon mass. Inserting the expressions
for the normalization constant $A$ and $\alpha_\pm$ defined 
in (\ref{Eq:bag-wave-function}) and mass (\ref{Eq:M-R-general})
into Eq.~(\ref{Eq:D-term-bag}) we obtain the result for the
$D$-term of a general bag model state (made of massless quarks) 
\be\label{Eq:D-term-hadron}
	D= -\,\frac45\times
        \biggl[\sum\limits_{i=1}^{N_{\rm cons}}\omega_i\biggr]\;
        \biggl[\sum\limits_{i=1}^{N_{\rm cons}}
        \frac{\omega_i(4\,\omega_i^2-15) + 6\,\omega_i\,\sin^2\omega_i+
        15\,\sin\omega_i\,\cos\omega_i}{54\,\omega_i(\omega_i-1)\sin^2\omega_i}
        \biggr]\,.
\ee
We make two important observations. First, since 
$\omega_i\ge \omega_0 \approx 2.04$ it is $D<0$ for all 
hadron states constructed in the bag model including unstable resonances. 
This is in line with results from all theoretical studies so far.
Second, the dependence on the model parameter bag radius $R$ or bag constant
$B$ cancels out in the $D$-term which therefore only depends on the 
dimensionless numbers $\omega_i$ (for massless quarks,  
cf.\ Sec.~\ref{Sec-7:D-limiting-cases} for the case of massive quarks).

\subsection{\boldmath The $D$-term of the nucleon}
\label{Sec-6:D-nucleon}

For the nucleon we obtain from (\ref{Eq:D-term-bag},~\ref{Eq:D-term-hadron}) 
in the case of massless quarks the result
\be\label{Eq:D-nucleon}
	D_{\rm nucleon} = -1.145 \,,
\ee 
which is in agreement with the numerical bag model calculation 
of nucleon GPDs and EMT form factors in Ref.~\cite{Ji:1997gm}.
The magnitude of the nucleon $D$-term in the bag model is smaller 
compared to soliton models 
\cite{Goeke:2007fp,Goeke:2007fq,Wakamatsu:2007uc,Cebulla:2007ei,
Kim:2012ts,Jung:2013bya,Jung:2014jja,Perevalova:2016dln}.
This is not surprizing considering that $D=M_N\int\di^3r\;r^2p(r)$ 
is sensitive to long distances. 
In fact, in chiral models $|D|$ is larger in the chiral limit where 
$p(r)$ and $s(r)\propto$ $1/r^6$. For finite pion masses $m_\pi$ the 
densities decay exponentially like $e^{-m_\pi r}$. The range of internal 
forces decreases in the soliton models, and $|D|$ diminishes 
\cite{Goeke:2007fp}. Since the bag model has a finite radius, the 
value for $|D|$ is small. 
We remark that through the SU(4) spin-flavor factors (\ref{Eq:Nq+Pq}) the 
bag model complies with the large-$N_c$ predictions \cite{Goeke:2001tz}
\be\label{Eq:D-term-in-large-Nc}
      (D^u+D^d)_{\rm nucleon}={\cal O}(N_c^2), \quad
      (D^u-D^d)_{\rm nucleon}={\cal O}(N_c).
\ee

\subsection{\boldmath $\rho$ meson}

Placing in the lowest level of the bag a $\bar{q}q$ pair with aligned 
spins yields a state with the quantum numbers of a $\rho$-meson.
With $B$ fixed to reproduce the nucleon mass yields a mass of 
692$\,$MeV which agrees with the experimental value of the 
$\rho$-meson mass 775 MeV within 10$\,\%$. Other ways to fix model 
parameters can also be considered \cite{Hasenfratz:1977dt}.
In contrast to this there is no ambiguity as to the bag model 
prediction for the $D$-term (\ref{Eq:D-term-hadron}) which
does not depend on the bag radius $R$ or bag constant $B$. 
The model prediction is
\be\label{Eq:D-rho}
       D_{\rho\mbox{-}\rm meson}=\frac{4}{N_c^2}\,D_{\rm nucleon} = -0.509\,.
\ee
Recalling that $D_{\rm nucleon}={\cal O}(N_c^2)$, 
cf.\ (\ref{Eq:D-term-in-large-Nc}), we see that the $D$-term 
of the $\rho$-meson (and all mesons) is of ${\cal O}(N_c^0)$.

As there is no spin-spin interaction a $\bar{q}q$ pair with 
anti-aligned spins corresponding to a state with pion quantum numbers 
has exactly the same mass (and $D$-term) as the $\rho$-meson. However,
since the bag boundary does not respect chiral symmetry, the description 
of the pion in the bag model is inadequate. This becomes evident here in 
two ways. First, in the chiral limit the pion is massless while here
it remains massive and is mass-degenerate with the $\rho$-meson. 
Second, soft pion theorems predict $D_{\rm pion}=-1$ 
\cite{Novikov:1980fa,Voloshin:1980zf,Voloshin:1982eb,Leutwyler:1989tn}
while in the bag model one would obtain the same value as in (\ref{Eq:D-rho}).
Ways to construct light pion states have been discussed \cite{Donoghue:1979ax}.
The cloudy bag model \cite{Thomas:1982kv} reconciles the bag concept and 
chiral symmetry. Both approaches are beyond the scope of this work.
In any case, since it is not a Goldstone boson one may apply the bag 
model to the description of the $\rho$-meson with less reservations.
It will be interesting to test the bag model prediction 
$D_{\rho\mbox{-}\rm meson}\,:\,D_{\rm nucleon} = 4\,:\,9$ 
in other models or lattice QCD.

Notice that a spin-1 hadron like the $\rho$-meson has 6 form factors 
of the total EMT \cite{Abidin:2008ku,Polyakov:2019lbq,Cosyn:2019aio}.
Our $D_{\rho\mbox{-}\rm meson}$ corresponds to the form factors  
$D_0(t)=-\,{\cal G}_3(t)$ in the notations of respectively
\cite{Polyakov:2019lbq} or \cite{Cosyn:2019aio}. Studies of other 
$\rho$-meson EMT form factors will be left to future investigations.

\subsection{\boldmath $\Delta$-resonance}

Let us briefly also comment on the $D$-term of the $\Delta$-resonance.
As discussed in the previous section, due to absence of spin-spin 
interactions states differing by the spin quantum numbers are degenerate.
In particular, also the $\Delta$-resonance and the nucleon are degenerate,
and the $D$-term of the $\Delta$ is simply predicted to be
\be
    D_{\Delta\mbox{-}\rm resonance}= D_{\rm nucleon} = -1.145\,.
\ee
Even though the absolute value might be underestimated, the bag model
result $D_{\Delta\mbox{-}\rm resonance}= D_{\rm nucleon}$ is correct in large-$N_c$ 
\cite{Perevalova:2016dln}. This is another consistency test of the large-$N_c$
description of baryons in the bag model.

\subsection{Roper resonance}

The state $N(1440)$ known as Roper resonance has the quantum numbers of 
the proton $J^P=\frac12{ }^+$ but a 50$\,\%$ larger mass and its structure
``has defied understanding'' since its discovery in the 1960s, see the 
review \cite{Burkert:2019bhp}.  In the bag model it is described by placing 
two quarks in the ground state with $\omega_0 = 2.04$ and one quark in the
first excited state with $\omega_1= 5.40$. If one would use the same bag 
radius for the nucleon and the Roper, then the physical value of the Roper 
mass would be reproduced. A more consistent parameter treatment 
may be to use the same bag constant $B$ for nucleon and Roper, which yields 
a Roper mass of 1302$\,$MeV and underestimates the physical value by 10$\,\%$. 
This is not 
unreasonable for such a simple model. While the mass increases by about 
50$\,\%$, the pressure in the center increases by a factor of 7.5 as 
one goes from the ground state nucleon to the first excited state in the 
$J^P=\frac12{ }^+$ sector. The increase of the internal forces is 
reflected by an increase of the $D$-term for which 
Eq.~(\ref{Eq:D-term-hadron}) yields the value 
\be\label{Eq:D-Roper}
      D_{\rm Roper} = 5.846\,D_{\rm nucleon} = -6.695\,.
\ee
It is interesting to observe how strongly the $D$-term is varied as one 
goes from a ground state to an excited state within a theory. This is
mainly due to an increase of internal forces, and in line with studies 
of excited states in $Q$-ball systems \cite{Mai:2012cx}. 
It is not known how the $D$-term of the Roper can be studied in experiment, 
but it would be interesting to confront the prediction (\ref{Eq:D-Roper}) 
with results from other models or lattice QCD.
We remark that $\Delta w^2$ defined in
(\ref{Eq:diffusiness}) is for Roper $\Delta w^2 = 0.72$
showing that this state is even more diffuse than the nucleon,
as intuitively expected.

\subsection{Negative parity baryons}
\label{Sec-6:D-negative-parity}

The lightest baryon with quantum numbers $J^P=\frac12{ }^-$
is $N(1535)$. Negative parity solutions to the bag equations 
(\ref{Eq:eom-all}) are given by the same expression as positive 
parity solutions (\ref{Eq:bag-wave-function}), but with upper 
and lower components exchanged and with the $\omega_i$ 
obtained from (for massless quarks) the transcendental equation 
$\omega=(1+\omega)\,\tan\omega$ whose lowest energy solution is 
$\omega = 3.81$. Keeping $B$ fixed at the value required for the 
nucleon yields 1498 MeV and reproduces the mass of $N(1535)$ 
within $3\,\%$. Also here is the $D$-term independent of 
parameter fixing and we obtain
\be\label{Eq:D-N1535}
      D_{N(1535)} = 11.32\,D_{\rm nucleon} = -12.97
\ee
which confirms the trend that the $D$-terms grow for heavier 
excited states in the spectrum of a theory. Also in this case we 
are not aware of a practical method to learn about the $D$-term 
of the state $N(1535)$ from experiment. However, the interesting
prediction $D_{N(1535)} = 11.32\,D_{\rm nucleon}$ could be compared
to theoretical studies in other models. With $\Delta w^2=0.59$
this state is somewhat less diffuse than the Roper.

\subsection{Highly excited states in baryonic spectrum}
\label{Sec-6:D-highly-excited}

In this section we consider higher excited states in the bag model. 
While we do not expect a realistic description of the hadronic spectrum,
the bag model provides a consistent theoretical framework and it
is instructive to explore it.
For simplicity we consider massless quarks and limit ourselves 
to the positive parity baryons sector. 

\begin{figure}[t!]
\vspace{-5mm}
\begin{centering}
\includegraphics[width=7cm]{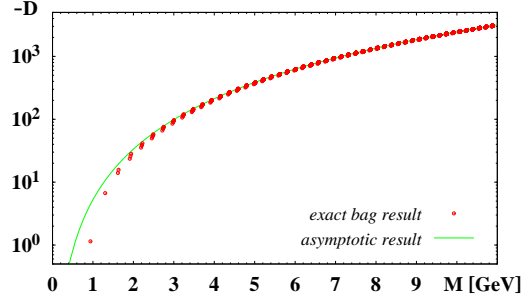}
\end{centering}
\caption{\label{Fig-5:D-vs-M}
$(-D)$ vs mass for the first 4000 states in the positive parity
sector for states made of massless up- and down-quarks. While the
masses increase by one order of magnitude, the $D$-terms grow by
4 orders of magnitude. The analytically derived asymptotic result
(\ref{Eq:D-term-asymp-vs-M}) for the $D$-term is shown as solid
line. The degeneracy pattern of the states is explained in the text.}
\end{figure}

As mentioned in the context of Eq.~(\ref{Eq:D-term-hadron}), 
we find $D<0$  for all excited states. Another important
observation is that the $D$-terms grow quickly with the mass of 
the hadron. To illustrate this point we plot the $D$-terms 
vs masses for $J^P=\frac12{ }^+$ baryons made of massless 
$u$- and $d$-quarks in Fig.~\ref{Fig-5:D-vs-M} which 
shows the first 4000 states: the first state is the nucleon
with $(M,\;D)=(938\,{\rm MeV},\;-1.145)$ and last state
has $(M,\;D)=(10.9\,{\rm GeV},\;-3068)$. All states 
above $2\,{\rm GeV}$ are hypothetical and practically 
in the continuum. Each state has a twofold degeneracy 
due to isospin quantum numbers.  (In our large-$N_c$ 
treatment the spectrum of $J^P=\frac32{ }^+$ baryons
looks exactly the same with a four-fold degeneracy
due to isospin $\frac32$ of $\Delta$-states.)
While the baryon masses increase by one order of 
magnitude in the range considered in Fig.~\ref{Fig-5:D-vs-M}, 
the $D$-terms grow by 4 orders of magnitude. This is in line 
with results from $Q$-balls \cite{Mai:2012cx}.

%=============== BEGIN FIGURE 5: D vs M ============================
%================= END FIGURE 5 ====================================

\begin{figure}[t!]
\vspace{-5mm}
\begin{centering}
\includegraphics[width=5.5cm]{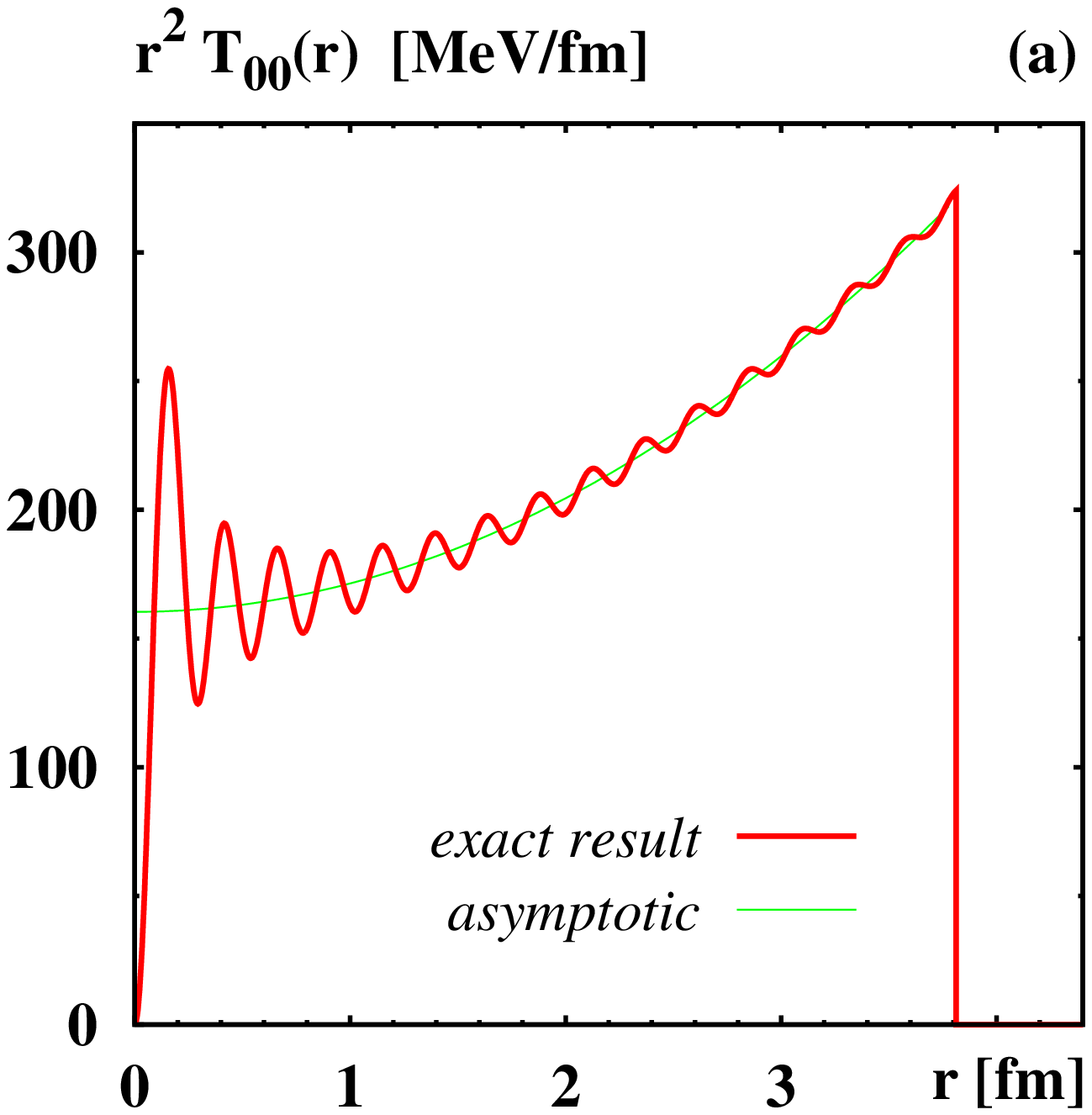}
\includegraphics[width=5.5cm]{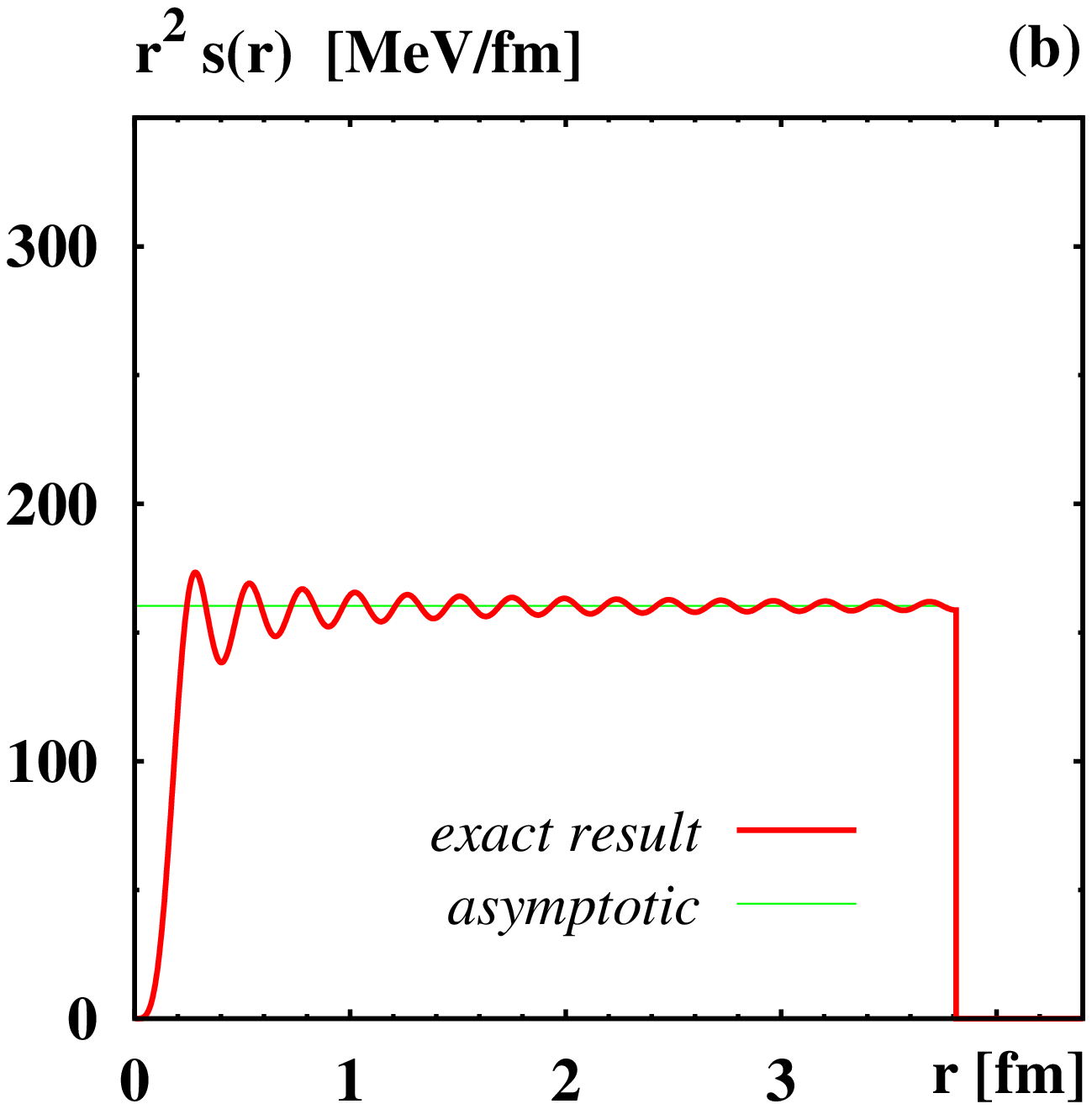}  
\includegraphics[width=5.5cm]{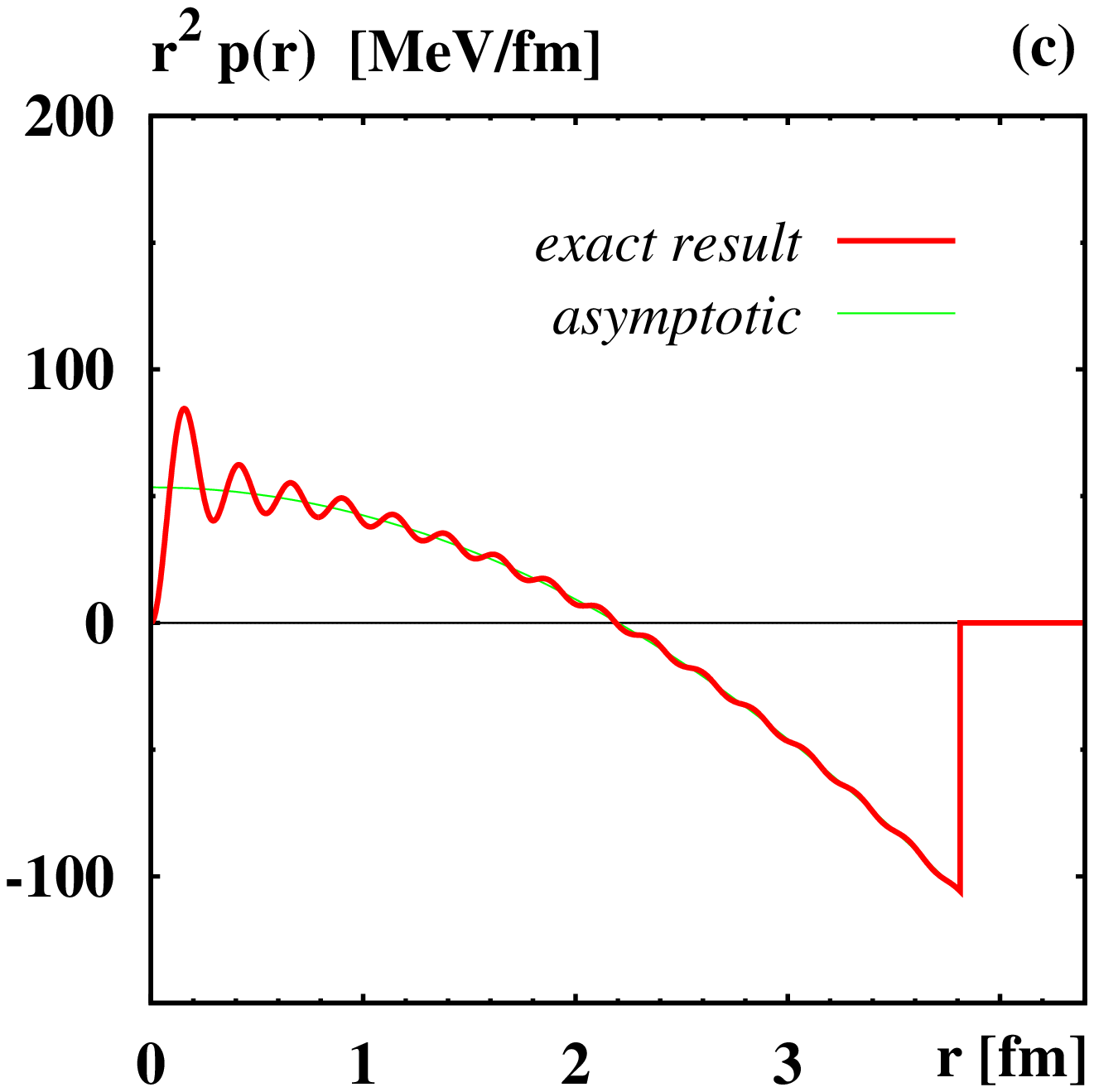}
\end{centering}
\caption{\label{Fig-6:dens-w15}
Solid lines: EMT densities $r^2T_{00}(r)$, $r^2s(r)$, $r^2p(r)$ 
for a (hypothetical) highly excited nucleon state with the tripply
occupied bag level $\omega_{15}=49.47\approx  (j+3/4)\,\pi$ with
$j=15$. For this state the bag radius is $R=3.8\,{\rm fm}$,
mass $M=10.24\,{\rm GeV}$,  $D$-term $D=-2607.7$ to be compared with the 
nucleon ground state where $\omega_0 = 2.04$, $R=1.71\,{\rm fm}$, 
$M=938\,{\rm MeV}$, $D=-1.145$.
Thin lines: asymptotic results for the bag model densities from 
Eq.~(\ref{Eq:EMT-densities-asymp}) for $\sum\omega_j=3\omega_{15}$
and $R$ as given by Eq.~(\ref{Eq:M-R-general}).}
\end{figure}

To get more insight we discuss the EMT densities of a
(hypothetical) highly excited nucleon state. For $Q$-balls
it was observed that $T_{00}(r)$ of the $N^{\rm th}$ (radial) 
excitation exhibits characteristic structures with $N$-shells
surrounding a ``core'' region, while $p(r)$ exhibits 
$(2N+1)$-nodes where $N=0$ refers to the ground state 
\cite{Mai:2012cx}. Is this also the case for excited states 
in the bag? The answer is no. 
For illustration we show in Fig.~\ref{Fig-6:dens-w15}
the EMT densities for the state with the level $\omega_{15}$ 
tripply occupied. This corresponds to a hypothetical $3163^{\rm th}$ 
excited state above the nucleon (ground) state with 
$(M,\,D)=(10.2\,{\rm GeV},\,-2608)$.
The EMT densities exhibit characteristic ``wiggles'' but 
$p(r)$ exhibits only one node. This is a general result:
no matter how highly excited a bag state is, $p(r)$ crosses
zero only once. Clearly, the spectrum of excitations in 
the bag model has a much different structure than the
$Q$-ball system \cite{Mai:2012cx}. One expects such
highly excited states to be very diffuse and the result
$\Delta w^2=2.9$ for this hypothetical state confirms it.

The solutions to the transcendental equation 
(\ref{Eq:omega-transcendental-eq}) are approximated by 
$\omega_j\to (j + 3/4)\,\pi$ for massless quarks to 
within an accuracy of better than $2\,\%$ already 
for $j\ge1$. For $\omega_{15}$ this asymptotic formula
has an accuracy of $2\times10^{-4}$.
Evaluating the expressions for $T_{00}(r)$, $p(r)$, $s(r)$ in 
Eqs.~(\ref{Eq:EMT-stat-T00q},~\ref{Eq:EMT-stat-G},~\ref{Eq:p+s-bag})
for asymptotically large $\omega_j$ yields
\ba
      \biggl[r^2T_{00}(r)\biggr]_{\rm asymp} 
      &=& \frac{\sum_j \omega_j}{4\pi R^2}
      \;\biggl(1+\frac{r^2}{R^2}\biggr)
      \,\Theta_V \,,\nonumber\\
      \biggl[\,r^2p(r)\,\biggr]_{\rm asymp} 
      &=& \frac{\sum_j \omega_j}{4\pi R^2}
      \;\biggl(\frac13-\frac{r^2}{R^2}\biggr) 
      \,\Theta_V \,,\nonumber\\
      \biggl[r^2s(r)\biggr]_{\rm asymp} 
      &=& \frac{\sum_j \omega_j}{4\pi R^2}
      \;\Theta_V 
      \quad \mbox{for} \quad \omega_j\to (j+3/4)\,\pi
      \label{Eq:EMT-densities-asymp}
\ea
where $R$ is defined in Eq.~(\ref{Eq:M-R-general}) and it is understood 
that all quantities actually depend on a set of 3 (or 2) values 
of $\omega_j$ for a higher baryonic (or mesonic) excitation. 
Except for the small-$r$ region the asymptotic expressions yield 
a good description of the gross features of the exact densities 
as shown in Fig.~\ref{Fig-6:dens-w15}.

Remarkably, the asymptotic expression for $T_{00}(r)$ 
integrates to the exact expression for the baryon mass
in Eq.~(\ref{Eq:M-R-general}).
The asymptotic expressions for pressure and shear forces 
satisfy the differential equation (\ref{Eq:p(r)+s(r)}), 
and $p(r)_{\rm asymp}$ complies with the von Laue condition
albeit not with its lower-dimensional analogs in (\ref{Eq:stability}) 
where the exact small-$r$ details are essential. 
The asymptotic normal force $r^2[\frac23\,s(r)+p(r)]_{\rm asymp}=
(\sum_j \omega_j)(1-r^2/R^2)/(4\pi R^2)\,\Theta_V >0$
for $r<R$ and vanishes at $r=R$. The two equivalent expressions in 
Eq.~(\ref{Eq:d1-from-s(r)-and-p(r)}) yield the same asymptotic result 
for the $D$-term
\be\label{Eq:D-term-asymp}
        D_{\rm asymp} = -\frac{16}{135}\,
        \biggl(\sum_j\omega_j\biggr)^2 \;.
\ee
The energy mean square radius and mechanical radius
are $\la r_E^2\ra_{\rm asymp} = \frac25\,R^2$ and
$\la r_{\rm mech}^2\ra_{\rm asymp} = \frac15\,R^2$.
Finally, by exploring Eq.~(\ref{Eq:M-R-general}), we may 
eliminate the sum $\sum_j\omega_j$ in (\ref{Eq:D-term-asymp}) 
in favor of $M$ which yields (for mesons and baryons)
\be\label{Eq:D-term-asymp-vs-M}
      D_{\rm asymp} = -\,A\,M^{\,8/3}
\ee
with $A=\frac{1}{5}(16\pi\sqrt{3}\, B)^{-2/3}$. 
This asymptotic expression explains the strong rise of $D$ 
with the mass observed in Fig.~\ref{Fig-5:D-vs-M} where 
Eq.~(\ref{Eq:D-term-asymp-vs-M}) is depicted as solid line. 
Interestingly the spectrum of radial $Q$-ball excitations exhibits 
the same asymptotic relation: for the $N^{\rm th}$ excitation the $Q$-ball 
mass grows like $M\propto N^3$ and $D$-term as $D\propto -\,N^8$,
such that $D\propto -\,M^{8/3}$ \cite{Mai:2012cx} like in bag model. 
But the internal structure of the excitations is much different:
e.g.\ the $p(r)$ of the $N^{\rm th}$ excited $Q$-ball state exhibits 
$(2N+1)$ \cite{Mai:2012cx}, while the $p(r)$ of excited bag states 
have always only one node.

To end this section we comment on the near-degeneracies visible
in Fig.~\ref{Fig-5:D-vs-M} where the first 4000 states in the 
$J^P=\frac12{ }^+$ sector appear to be organized in a far
smaller set of near-degenerate multiplets. To understand these 
near-degeneracies we notice that $\omega_j\approx (j+3/4)\,\pi$
for $j\gtrsim 1$ can be further simplified 
\footnote{
     When deriving the asymptotic expressions for EMT densities 
     (\ref{Eq:EMT-densities-asymp}) it is necessary to use the 
     asymptotic solutions $\omega_j\to (j+3/4)\,\pi$ of 
     Eq.~(\ref{Eq:omega-transcendental-eq}).
     Once we deal with the integrated quantities like $M$ and 
     $D$ in (\ref{Eq:M-R-general},~\ref{Eq:D-term-asymp}) one
     may go one step further and approximate $\omega_j\to j\,\pi$
     for $j\gg1$. But we stress that this further simplification 
     could not be used in the derivation of the asymptotic EMT 
     densities (\ref{Eq:EMT-densities-asymp}).}
as $\omega_j\approx j\,\pi$ for large enough $j \gg 1$.
If always all three occupied levels $j_1$, $j_2$, $j_3$ complied with
this condition then the mass would be determined by three integers 
as $M\approx {\rm const}\,(j_1+j_2+j_3)^{3/4}$ and the $n^{\rm th}$ 
energy level would be $\frac12n(n+1)(n+2)$--fold degenerated
(like 3D harmonic oscillator formulated in Cartesian coordinates). 
Since for lower bag levels $j\gg1$ is of course not valid, in practice 
a lesser degeneracy pattern is realized in Fig.~\ref{Fig-5:D-vs-M}.

%=============== BEGIN FIGURE 6: HIGHER EXCITATIONS ================
%================= END FIGURE 6 ====================================

%====== SECTION 7: LIMITING CASES ==================================
\section{Limiting cases}
\label{Sec-7:D-limiting-cases}

In this section we assume that $mR\neq0$.
The lowest solution $\omega_0$ of the transcendental 
equation (\ref{Eq:omega-transcendental-eq}) depends on the
product $mR$. We will be especially interested in the limit
$\varepsilon = 1/(mR)\to 0$ where we have $\omega_0\to\pi$.
For our calculation the $\varepsilon$-corrections to $\omega_0$ are essential.
They can be determined analytically, and are given by
\be\label{Eq:w-mR}
       \omega_0 = \pi - \frac{\pi}{2}\,\varepsilon
       + \frac{\pi^3}{6}\,\varepsilon^3 
       - \frac{7\pi^4}{48}\,\varepsilon^4
       + \frac{\pi^3}{2}\biggl(\frac{1}{16}-\frac{\pi^2}{5}\biggr)
       \,\varepsilon^5
       + \frac{109\,\pi^5}{480}\,\,\varepsilon^6
       + {\cal O}(\varepsilon^7),
       \quad\mbox{for}\quad
       \varepsilon = \frac{1}{mR}\ll 1.
\ee
After exploring the virial theorem for $m\neq0$ in 
Eq.~(\ref{App:virial-massive-II}) of App.~\ref{App} the
bag constant becomes
\be
      B = \frac{N_c}{4R^3}\;\kappa\,\biggl(\varepsilon +
      {\cal O}(\varepsilon^2)\biggr), \quad
      \kappa = \frac{\pi}{R}, \quad c_0 = \frac{\pi}{2R^3}\,
\ee
where we also define a constant $c_0$ which will be used in the subsequent 
equations. For the EMT densities we obtain in the region $0\le r\le R$ 
for $\varepsilon\ll 1$ the results
\ba
      T_{00}(r) &=& N_c\,m\;c_0\,j_0(\kappa r)^2 + \dots \;,
      \nonumber\\
      \rho_J(r) &=& \frac13\;c_0\;\kappa r \;j_0(\kappa r)j_1(\kappa r) 
      + \dots \;,
      \nonumber\\
      s(r) &=& \frac{N_c\kappa}{2m} \;c_0\biggl(
      -j_0^\prime(\kappa r)j_1(\kappa r)
      -\frac1r\,j_0(\kappa r)j_1(\kappa r)
      +j_0(\kappa r)j_1^\prime(\kappa r)\biggr) + \dots \;,\nonumber\\
      p(r) &=& \frac{N_c\kappa}{6m} \;c_0\biggl(
      -j_0^\prime(\kappa r)j_1(\kappa r)
      +\frac2r\,j_0(\kappa r)j_1(\kappa r)
      +j_0(\kappa r)j_1^\prime(\kappa r)\biggr) - \frac{N_c\kappa}{4mR^4}\;,
      + \dots \label{Eq:EMT-densities-mR}
\ea
where the dots indicate subleading terms.
For $r>R$ the densities are zero due to the $\Theta_V $ not
shown here for brevity. Notice that $T_{00}(r)={\cal O}(\varepsilon^{-1})$ 
and the dots indicate terms of ${\cal O}(\varepsilon^0)$, 
$\rho_J(r)={\cal O}(\varepsilon^0)$ and the dots indicate 
terms of ${\cal O}(\varepsilon)$, while $p(r)$ and $s(r)$ are 
both of ${\cal O}(\varepsilon)$ with dots indicating
terms of ${\cal O}(\varepsilon^2)$.

Integrating $T_{00}(r)$ in (\ref{Eq:EMT-densities-mR}) over the
volume yields 
\be\label{Eq:MN-mR}
      M_N=\frac{N_c}{R}\biggl(\varepsilon^{-1}
      +\frac56\,\pi^2\varepsilon+{\cal O}(\varepsilon^2)\biggr)
      = N_c\,m\biggl(1+\frac56\,\pi^2\varepsilon^2+\dots\biggr)\,.
\ee 
The term of ${\cal O}(\varepsilon^0)$ contributing to $T_{00}(r)$ 
in (\ref{Eq:EMT-densities-mR}) integrates exactly to zero,
and the limit $M_N=N_cm$ is approached from above, i.e.\ with 
positive  ${\cal O}(\varepsilon)$ corrections.
Integrating $\rho_J(r)$ in (\ref{Eq:EMT-densities-mR}) over the
volume yields the nucleon spin $\int\di^3\rho_J(r)=\frac12$ up to 
the order at which the expansion (\ref{Eq:w-mR}) of $\omega_0$
is truncated (if one does not expand the exact expression 
for $\rho_J(r)$ integrates of course to $\frac12$ ``to all orders'').
The pressure and shear forces in (\ref{Eq:EMT-densities-mR}) 
comply with the von Laue condition and the lower dimensional
conditions in (\ref{Eq:stability}) also up to the order at which 
the expansion of $\omega_0$ in (\ref{Eq:w-mR}) is truncated
(and are of course also valid to all orders if we do not expand).

Notice that the virial theorem is always valid as long as $\varepsilon\neq0$.
In the expansions in (\ref{Eq:EMT-densities-mR}) the connection to the 
virial theorem is not visible. The leading term in $M_N=N_cm+\dots$ is
irrelevant for the virial theorem and drops out from $M_N^\prime(R)$.
Stability, pressure and the von Laue condition are all encoded in 
the subsubleading terms of ${\cal O}(\varepsilon)$ in $T_{00}(r)$ 
in (\ref{Eq:EMT-densities-mR}).
This explains why the energy density is of ${\cal O}(\varepsilon^{-1})$
but the pressure and shear forces are of ${\cal O}(\varepsilon)$. 

\begin{figure}[b!]
\begin{centering}
\includegraphics[width=5.5cm]{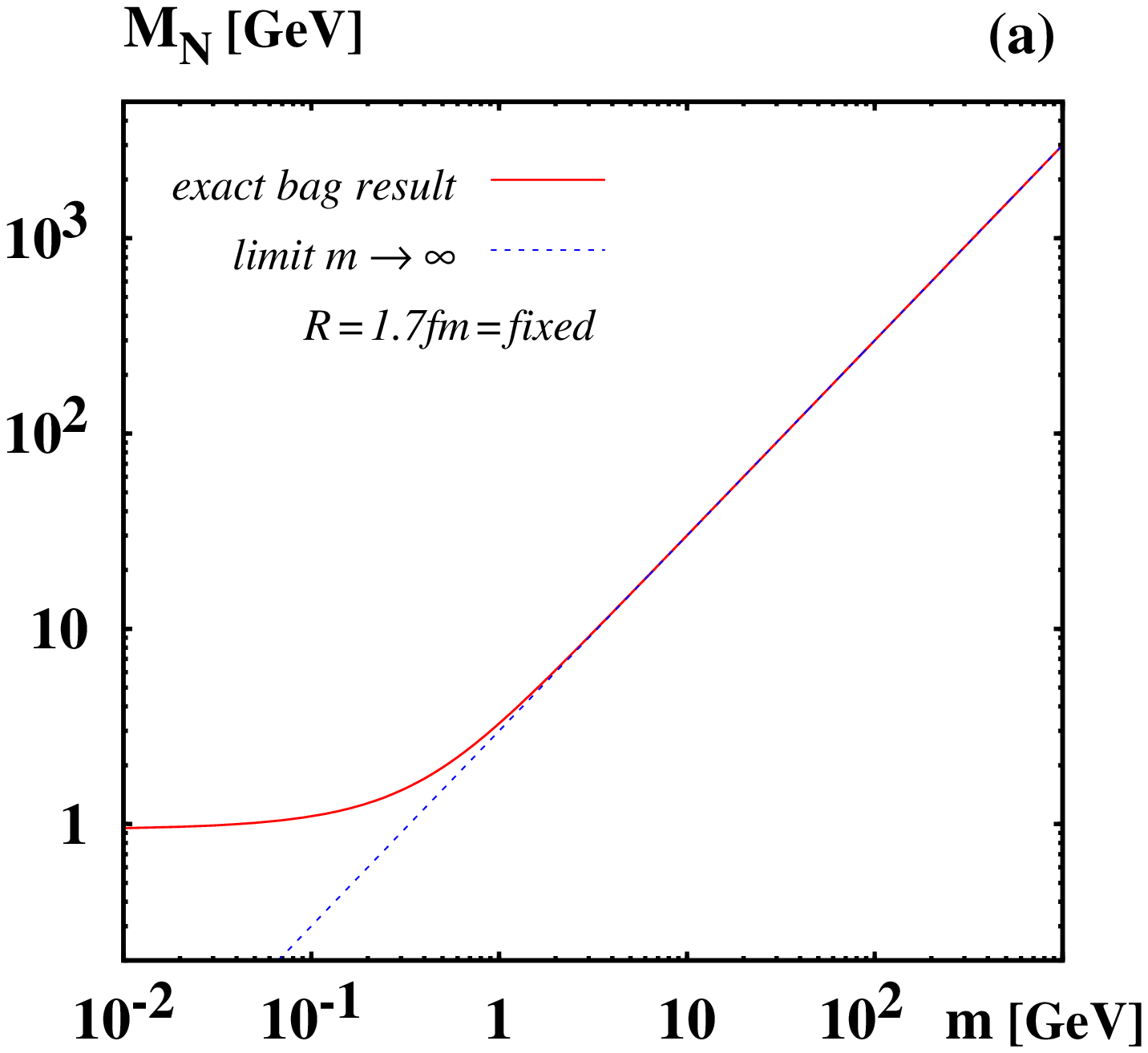}
\includegraphics[width=5.5cm]{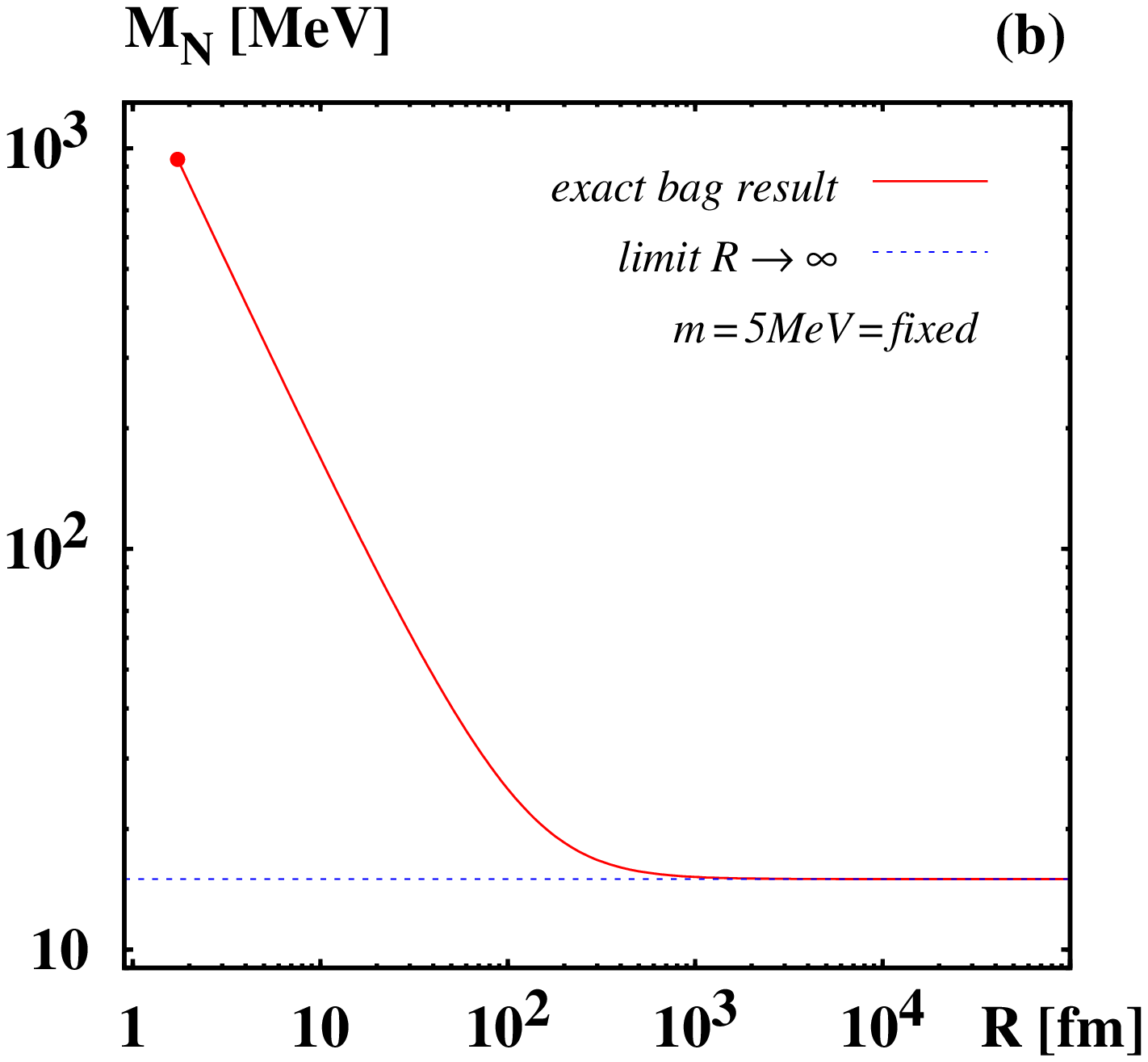}
\includegraphics[width=5.5cm]{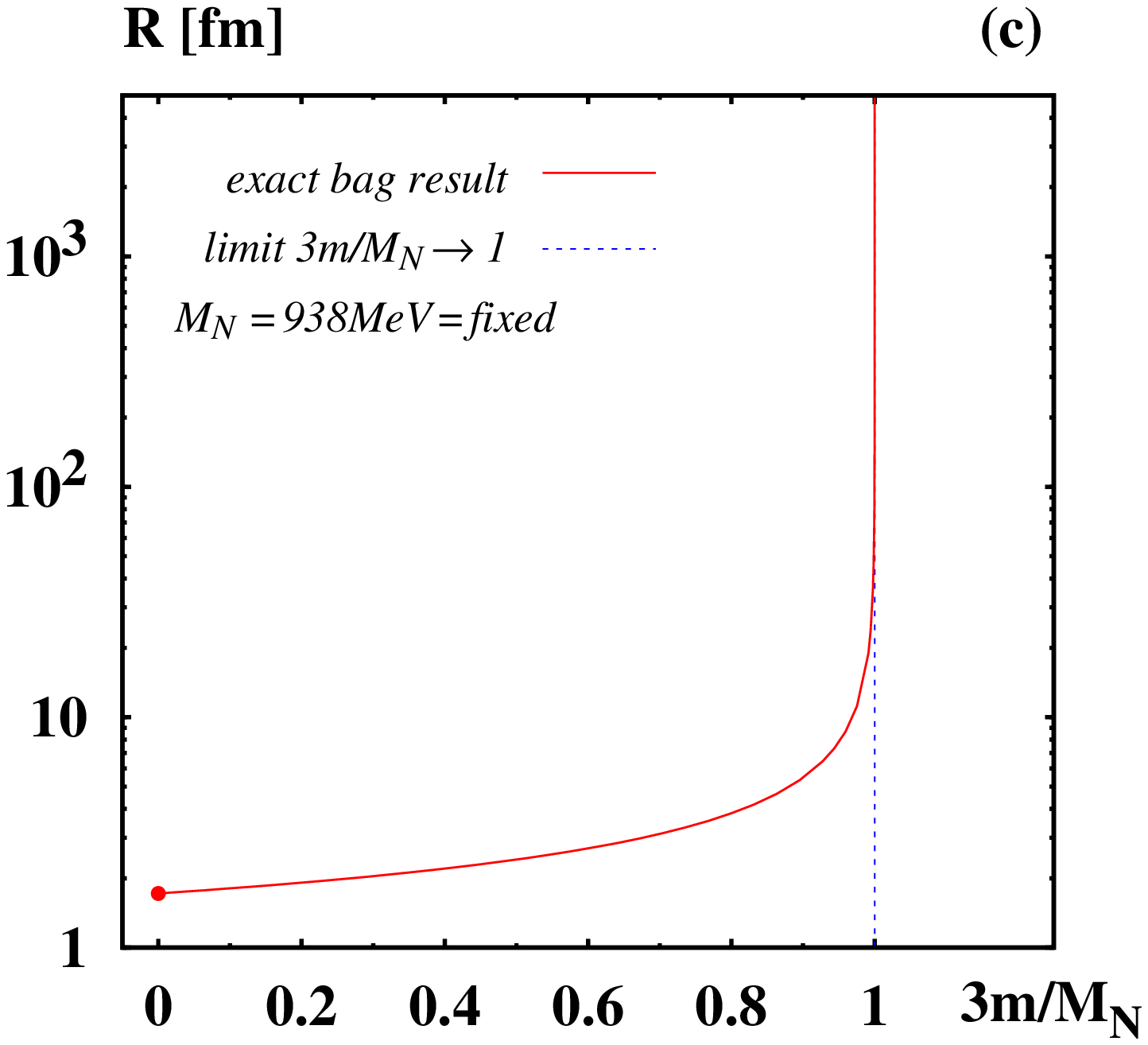}

\includegraphics[width=5.5cm]{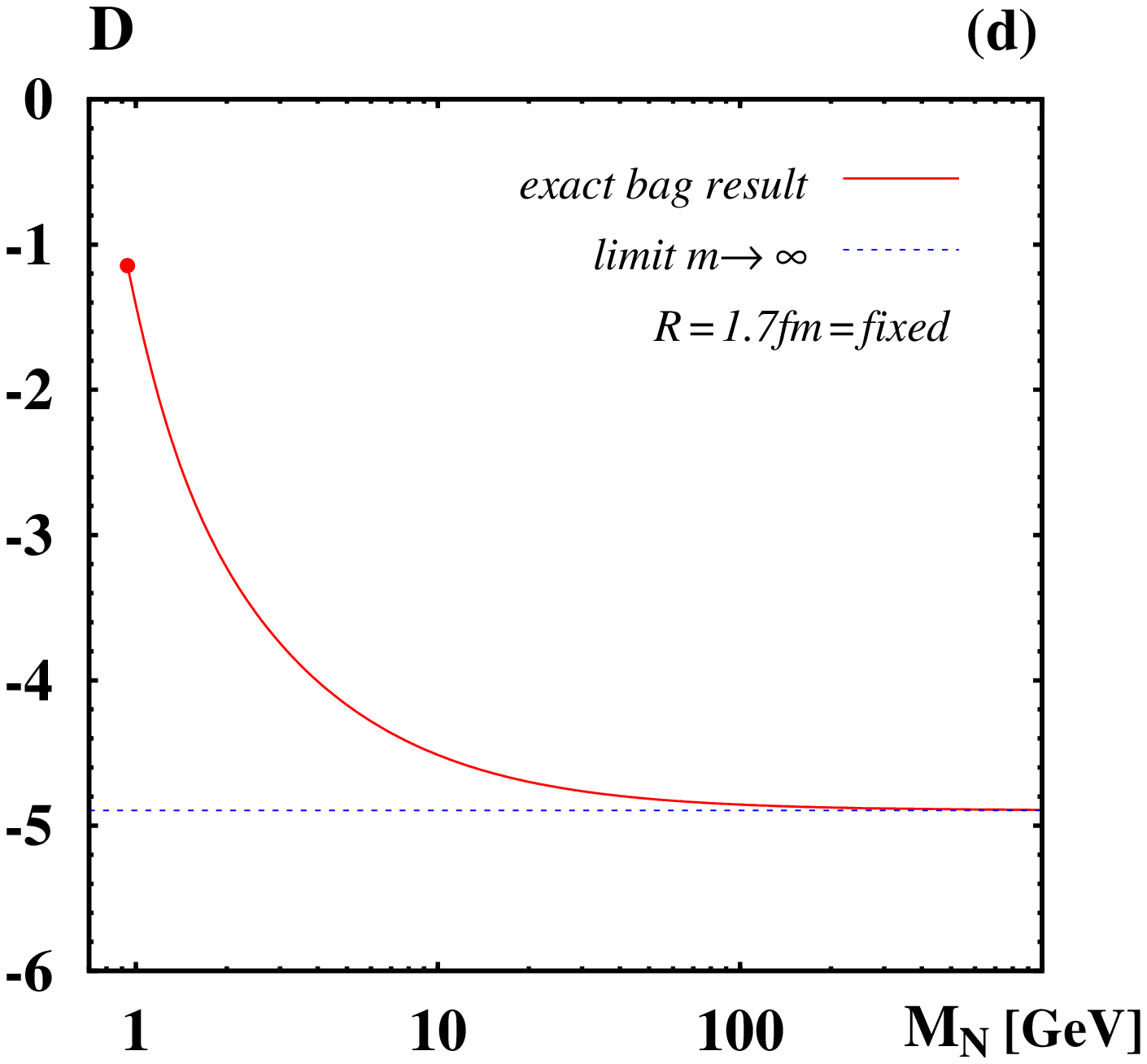}
\includegraphics[width=5.5cm]{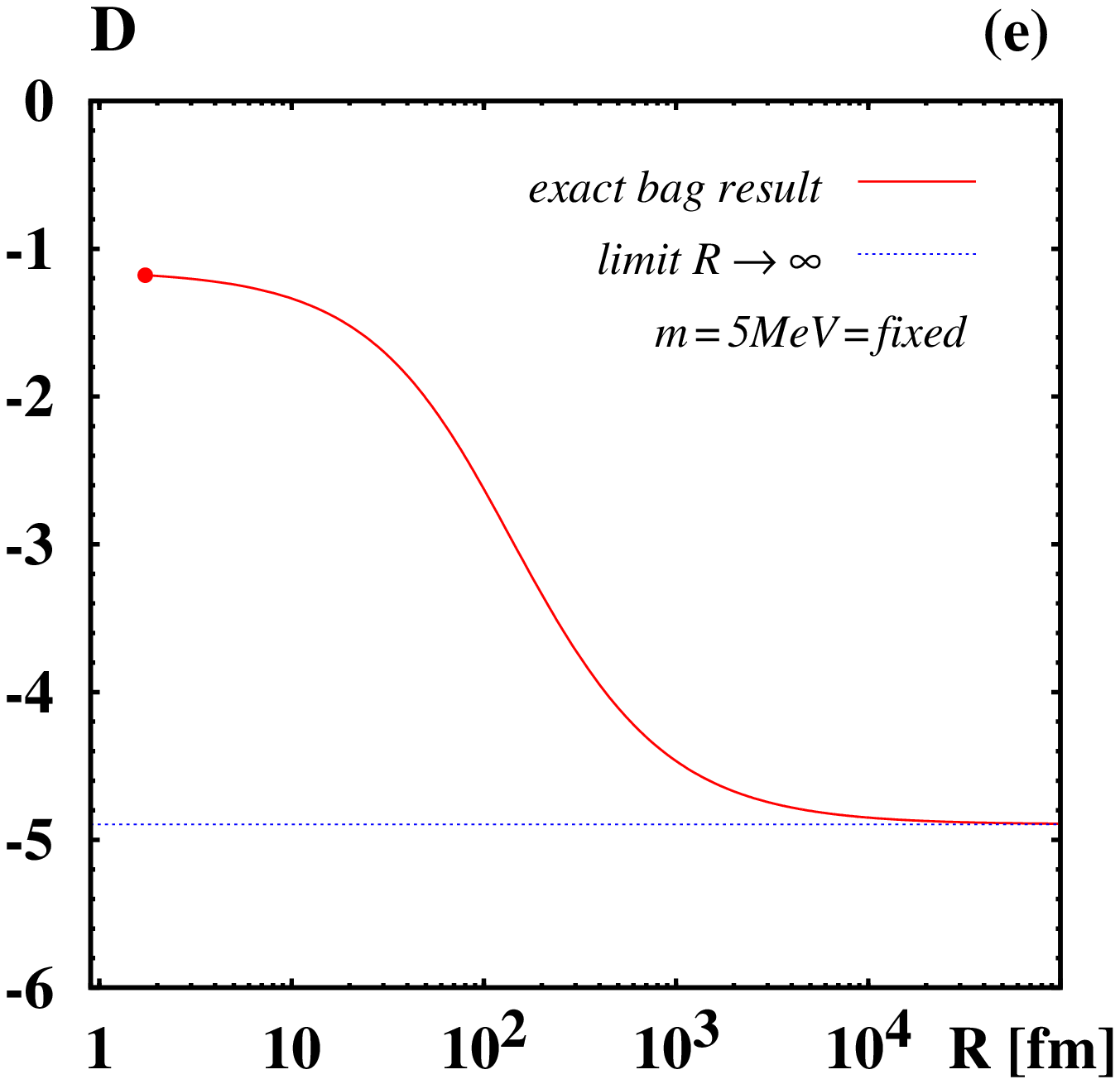}
\includegraphics[width=5.5cm]{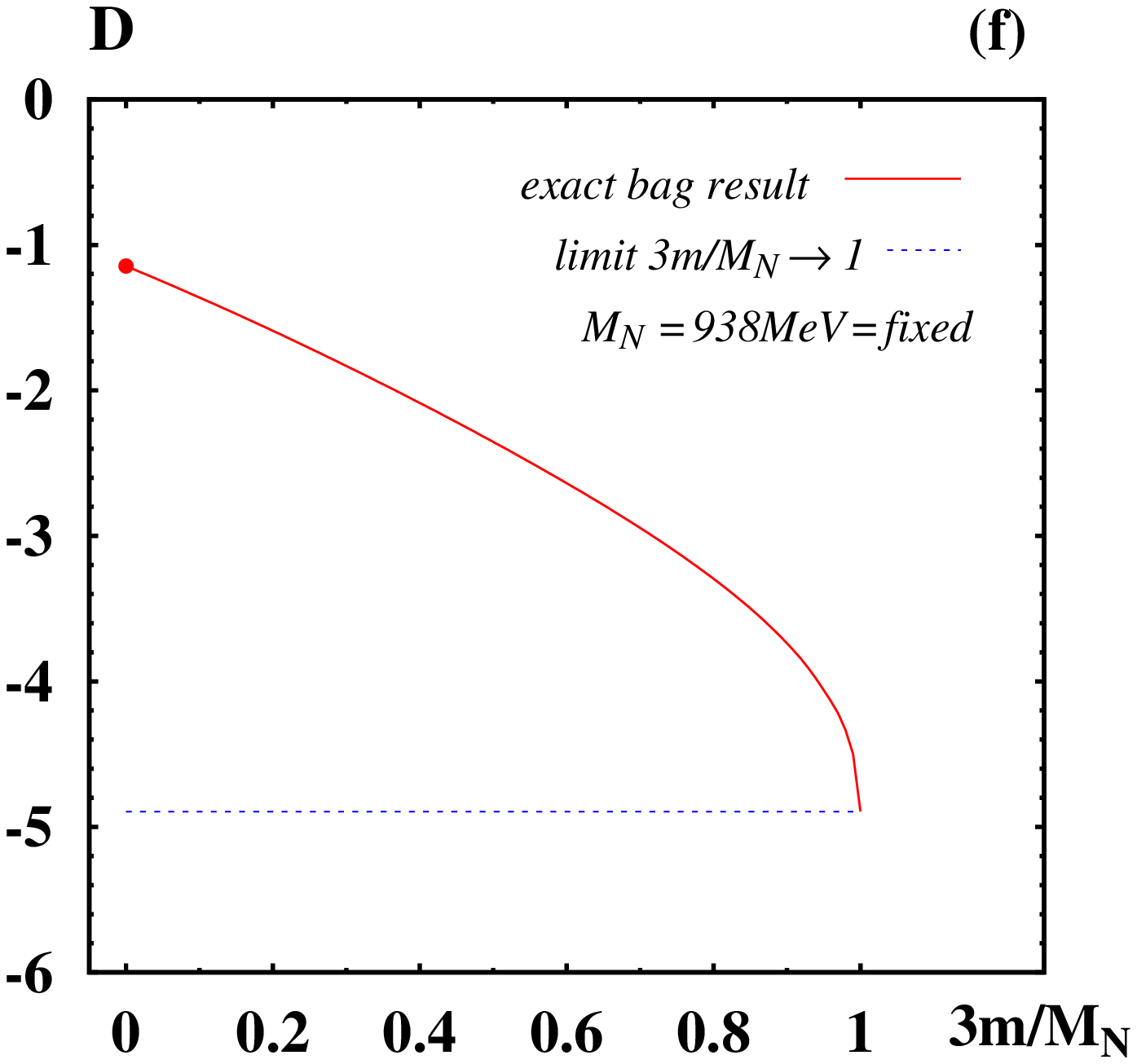}
\end{centering}
\caption{\label{Fig-7:D-limits-mR}
(a)  $M_N$ as function of quark mass $m$ for fixed $R=1.7\,{\rm fm}$. 
(b)  $M_N$ as function of bag radius $R$ for fixed $m=5\,{\rm MeV}$. 
(c)  $R$ vs.\ $m$ (in units of $\frac13M_N$) for fixed
     $M_N=938\,{\rm MeV}$.
(d)  The $D$-term vs $M_N$ for fixed $R=1.7\,{\rm fm}$. 
(e)  The $D$-term vs $R$ for a fixed $m=5\,{\rm MeV}$.
(f)  The $D$-term vs $m$ (in units of $\frac13M_N$) for 
     fixed $M_N=938\,{\rm MeV}$.
     The ``physical point'' with $M_N=938\,{\rm MeV}$, $R=1.7\,{\rm fm}$
     is marked (this point corresponds to $m=5\,{\rm MeV}$ 
     in (b), (d) and zero else).}
\end{figure}

Using the expansion for $p(r)$ and $s(r)$ in (\ref{Eq:EMT-densities-mR})
we obtain from (\ref{Eq:d1-from-s(r)-and-p(r)})  the result
\be\label{Eq:D-term-bag-non-rel}
	D = -\;N_c^2\,\biggl(\frac{4\,\pi^2-15}{45} 
        - \frac{2\pi^2}{15}\,\varepsilon+{\cal O}(\varepsilon^2)\biggr)
        \,.
\ee
The limit of the $D$-term in Eq.~(\ref{Eq:D-term-bag-non-rel}) 
applies to three different situations:
\ba
\mbox{(i)}    && \mbox{$\;R=\mbox{fixed}$, \ \ $m\to\infty$}, \nonumber\\
\mbox{(ii)}   && \mbox{$\,m=\mbox{fixed}$, \ \ $R\to\infty$}, \nonumber\\
\mbox{(iii)}  && \mbox{$m\to\frac13\,M_N$, $\,R\to\infty$, \ 
                                          $M_N=\mbox{fixed}$},\label{Eq:limits}
\ea
to be discussed below. The limits (i) and (ii) 
were briefly discussed in \cite{Hudson:2017oul}.
The Figs.~\ref{Fig-7:D-limits-mR}a--c show how $m$, $R$, $M_N$ 
are correlated in those limits. The Figs.~\ref{Fig-7:D-limits-mR}d--f 
show the behavior of the $D$-term.

%=============== BEGIN FIGURE 7: D LIMITING CASES ==================
%================= END FIGURE 7 ====================================

The case (i) in (\ref{Eq:limits}) corresponds to the ``heavy quark limit'' 
where the nucleon mass $M_N\to N_c\,m$ becomes large, 
see Fig.~\ref{Fig-7:D-limits-mR}a. 
For $m\gtrsim 1\,{\rm GeV}$ we have $M_N\approx N_c\,m$ with a $10\,\%$ 
or better accuracy. The asymptotics $M_N=N_c\,m$ is shown as dashed line
in  Fig.~\ref{Fig-7:D-limits-mR}a. 
This is intuitively expected: in the heavy quark limit one expects 
that hadron masses are largely due to the heavy quark mass. 
In this limit the quarks become ``non-relativistic:'' it is 
$\alpha_+={\cal O}(\varepsilon^0)$ while $\alpha_-={\cal O}(\varepsilon)$
such that the upper component of the spinor (\ref{Eq:bag-wave-function}) 
dominates and the lower component goes to zero. Interestingly, the $D$-term 
is proportional to $\alpha_+\alpha_-$, see Eq.~(\ref{Eq:D-term-bag}), 
but does not vanish because $M_N\propto\varepsilon^{-1}$ also
enters its definition, see Eq.~(\ref{Eq:d1-from-s(r)-and-p(r)}).
Thus $D\propto M_N\alpha_+\alpha_-$ has a non-zero limit, 
see (\ref{Eq:D-term-bag-non-rel}).
In Fig.~\ref{Fig-7:D-limits-mR}d we show how the $D$-term changes 
as one varies the quark mass from $m=0$ up to $1\,{\rm TeV}$,
with the asymptotic result (\ref{Eq:D-term-bag-non-rel}) shown as 
dashed line. 

In the limit (ii) in (\ref{Eq:limits}) the boundary is moved to infinity 
for fixed $m$ chosen to be $5\,{\rm MeV}$ in Fig.~\ref{Fig-7:D-limits-mR}b.
Intuitively one would expect to recover ``free quarks'' as the 
boundary is moved further and further away and the system becomes
more and more loosely bound. Indeed, also here $M_N\to N_c\,m$ 
(though in contrast to limit (i) the quarks may still be relativistic 
since $m$ does not need to be large as long as it is non-zero).
This limit is approached from above according to Eq.~(\ref{Eq:MN-mR}) 
as shown in Fig.~\ref{Fig-7:D-limits-mR}b where $R$ is varied from 
$1.7\,{\rm fm}$ up to {$1\,$\AA} with the asymptotic result $M_N=N_c\,m$ 
shown as dashed line.
Also in this limit the $D$-term approaches the asymptotic value
(\ref{Eq:D-term-bag-non-rel}) as shown in Fig.~\ref{Fig-7:D-limits-mR}e.
Remarkably, the $D$-term of a free fermion is zero \cite{Hudson:2017oul}, 
but here we do {\it not} recover this result, even though we deal with 
a more and more loosely bound system.
The reason is as follows. As $R$ becomes large the ``confinement''
of the fermions inside an increasingly large cavity becomes less and
less important, and the mass of the bound state approaches 
$M_N\to N_c\,m$. But no matter how small the ``residual interactions'' 
in an increasingly large cavity are: they remain non-zero, 
enter the description of the internal shear and pressure forces, and 
generate a non-zero $D$-term. How this happens can be traced back
on the technical level through, for instance, the virial theorem, 
see Appendix~\ref{App}.
To recover a free theory one has to take the limit $R\to\infty$
much earlier, on the Lagrangian level in Eq.~(\ref{Eq:Lagrangian-bag})
\cite{Hudson:2017oul}.

The limit (iii) in (\ref{Eq:limits}) is also very interesting. Here 
we assume throughout a system with the fixed (physical) value of the 
nucleon mass, but we allow the model parameters $m$, $R$ to vary such 
that the internal model dynamics interpolates all the way from highly 
relativistic ($m=0$) to highly non-relativistic ($m\to\frac13M_N$).
In the bag model the physical situation is of course more realistically 
reproduced for highly relativistic quarks rather than for non-relativistic
ones. But it is insightful to investigate such a transition from 
highly to non-relativistic system within a quark model.
A convenient measure for this transition is $m$ expressed in units of
$\frac13\,M_N$, i.e.\ the variable $3m/M_N$ whose range is $0\le 3m/M_N\le 1$. 
When $3m/M_N\to0$ we deal with highly relativistic (massless) quarks 
in a relatively small system of radius $R=1.7\,{\rm fm}$ which 
corresponds to the ``physical situation'' in this model.
When $3m/M_N\to 1$ we deal with a trully non-relativistic model of the 
nucleon: in this limit the nucleon mass is $100\,\%$ due to the 
``constituent quark mass.'' In order to maintain in this limit
the fixed (physical) value of the nucleon mass (in a system where 
the mass of the bound state is nearly entirely due to the mass of 
its constituents), it is necessary that the system becomes more 
loosely bound which implies that the size of the system must increase. 
In the strict limit $m\to\frac13\,M_N$ the bag radius diverges. 
The connection of $m$ (in units of $\frac13\,M_N$) and $R$ for fixed 
$M_N=938\,{\rm MeV}$ is shown in Fig.~\ref{Fig-7:D-limits-mR}c.
For instance, if we wanted $99.999\,\%$ of nucleon mass to be due to
the constituent quark masses, then $R=0.57\,\mu{\rm m}$ would be required.
It should be stressed that, while the system becomes more loosely bound in 
the sense that the binding energy decreases, we nevertheless still have 
confinement (in the specific way it is modelled in the bag model; it 
should be kept in mind that the binding energy is positive in a 
confining system).
Since in the limit (iii) it is $m\to \frac13\,M_N$ while $R\to\infty$
the $D$-term is again given by the limit $mR\to\infty$ quoted in
Eq.~(\ref{Eq:D-term-bag-non-rel}). How the $D$-term behaves 
during the transition from a highly relativistic ($3m/M_N=0$) 
to a highly non-relativistic ($3m/M_N\to1$) system with fixed $M_N$ 
is shown in Fig.~\ref{Fig-7:D-limits-mR}f.
For the last point included in this figure it is $M_N-3m=10\,\rm eV$ and 
$R=4\,\mbox{\AA}$, which are numbers natural for systems in atomic physics.

%=============== BEGIN FIGURE 8: D vs mR UNIVERSAL CURVE ===========
\begin{figure}[b!]
%\vspace{-5mm}
\begin{centering}
\includegraphics[width=7.5cm]{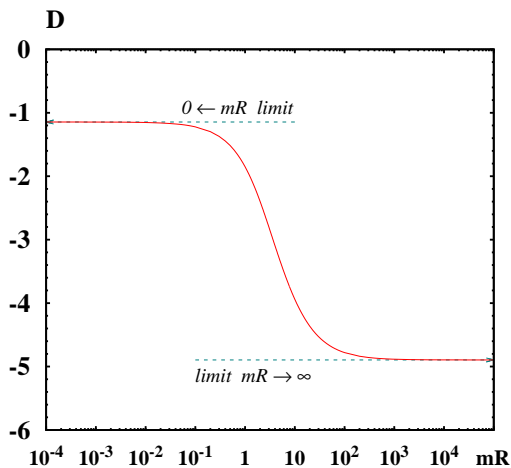}
\end{centering}
\caption{\label{Fig-8:D-vs-mR}
The $D$-term vs $mR$ in the bag model. As a dimensionless quantity 
the $D$-term only depends on the bag model parameters $m$ and $R$ 
through the dimensionless variable $mR$.}
\end{figure}
%================= END FIGURE 8 ====================================

The way the limiting value (\ref{Eq:D-term-bag-non-rel}) of the $D$-term
is approached in Figs.~\ref{Fig-7:D-limits-mR}d--f is characteristic for
the three different limits in (\ref{Eq:limits}). 
When we plot $D$ as function $mR$, the results from 
Figs.~\ref{Fig-7:D-limits-mR}d--f are in all 3 cases on 
a single universal curve shown in Fig.~\ref{Fig-8:D-vs-mR}.
Since $D$ is dimensionless. it can only depend on the bag model
parameters $m$ and $R$ in terms of the dimensionless variable $mR$. 
It is shown in Fig.~\ref{Fig-8:D-vs-mR} how the $D$-term depends on this
dimensionless variable $mR$. The ``physical situation'' for the proton
corresponds to the limit $mR\to0$ (fixed $R=1.7\,{\rm fm}$ and light
or massless up- and down-quarks). The limit $mR\to\infty$ can refer to 
the 3 different limiting cases in (\ref{Eq:limits}) discussed above.

One limiting case remains to be mentioned: fixed $m$ and $R\to0$.
In this limit one obtains a ``point-like'' particle whose mass diverges 
as $M_N\propto\frac1R$. This divergence is analogous to the difficulties 
associated with the description of point-like particles or point-like 
electric charges in classical physics. The description of the ``internal 
structure'' in a ``point-like particle'' is of no immediate interest.
We therefore refrain from discussing this limit further. 
The result for the $D$-term in this peculiar limit is, however, also 
shown in Fig.~\ref{Fig-8:D-vs-mR} in the direction $mR\to0$.

%====== SECTION 8: BOGOLIUBOV MODEL ================================
\section{Counter-example Bogoliubov model}
\label{Sec-8:Bogoliubov}

In all theoretical approaches so far the $D$-terms of particles
were found negative, 
except for free fermion fields where $D=0$ \cite{Hudson:2017oul}. 
It is an interesting question whether positive $D$-terms can be
realized at all in a physical system.

In fact, positive $D$-terms were found for {\it unphysical} states 
with spin and isospin $S=I\ge\frac52$ in the rigid rotator approach 
in the Skyrme model \cite{Perevalova:2016dln}. Hadronic states with
such (``exotic'') quantum numbers are artifacts of the rigid rotator 
approach and not realized in nature. When computing masses and other 
properties of such states one notices nothing unusual. But a more 
careful investigation of the EMT densities reveals why these states 
are unphysical: they do not obey the basic mechanical stability criterion, 
namely the positivity of normal forces $\frac23\,s(r)+p(r)>0$. So the rigid 
rotator states with $S=I\ge\frac52$ have positive $D$-terms, but they are 
also unphysical \cite{Perevalova:2016dln}.

Despite its simplicity and drawbacks the bag model is from the point of 
view of mechanical stability a perfectly reasonable and theoretically 
consistent framework with negative $D$-term.
However, a model which in some sense may be viewed as a 
predecessor of the bag model \cite{Thomas:2001kw}, 
the model of Bogoliubov \cite{Bogo:1967},
is insightful in this respect. 
In a certain limit the Bogoliubov model basically corresponds 
to the bag model except that the bag constant $B$ is absent. 
The nucleon mass is given by $M_{N,\rm Bogo}=3\frac{\omega_0}{R}$ and for 
$R=1.29\,{\rm fm}$ the physical value of the nucleon mass is reproduced.
An interesting parameter-free prediction of the Bogoliubov model
is that for massless quarks the ratio of Roper and nucleon masses is 
$M_{\rm Roper}/M_N=(2\omega_0+\omega_1)/(3\omega_0)=1.55$ is close
to the experimental value $1.53$ although in retrospective this has 
to be considered a ``happy coincidence'' \cite{Thomas:2001kw},
because the model is actually ill-defined.

One way to understand this is to notice that the nucleon mass
$M_{N,\rm Bogo}=3\frac{\omega_0}{R}$ is determined by fixing the
bag radius by hand and not by a dynamical calculation \cite{Thomas:2001kw}, 
unlike the minimization procedure in the bag model underlying the
virial theorem, see Sec.~\ref{Sec-5a:T00} and App.~\ref{App}.
[In the bag model we have {\it two} free parameters, $B$ and $R$,
one of which is dynamically determined by the virial theorem, and the 
other can then be fixed to reproduce a chosen hadron mass.]
In fact, it is not possible to minimize $M_{N,\rm Bogo}=3\frac{\omega_0}{R}$ 
whose minimum occurs for $R\to\infty$ \cite{Thomas:2001kw}.

\begin{figure}[t!]
\begin{centering}
\includegraphics[height=3.95cm]{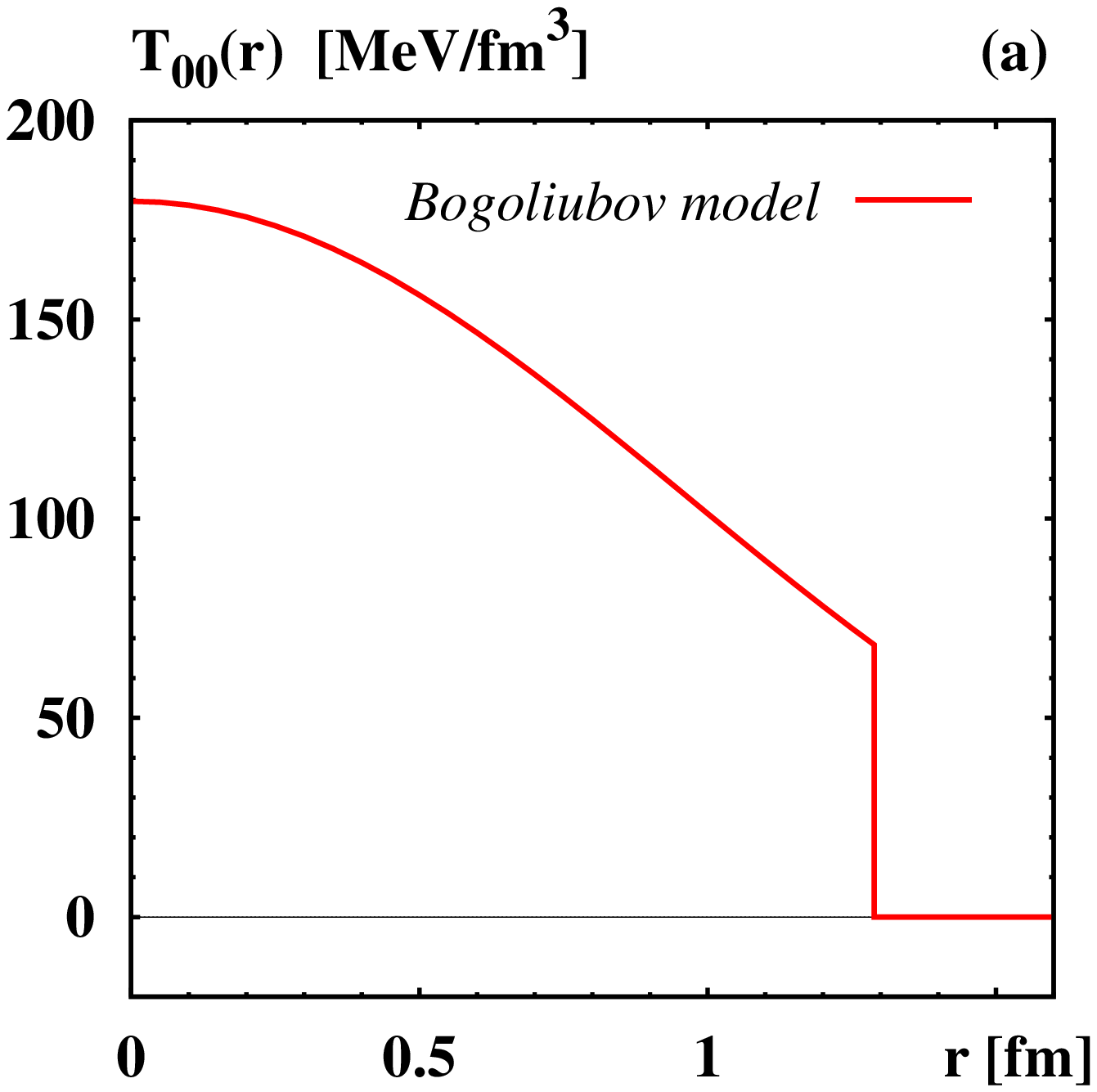} \
\includegraphics[height=3.95cm]{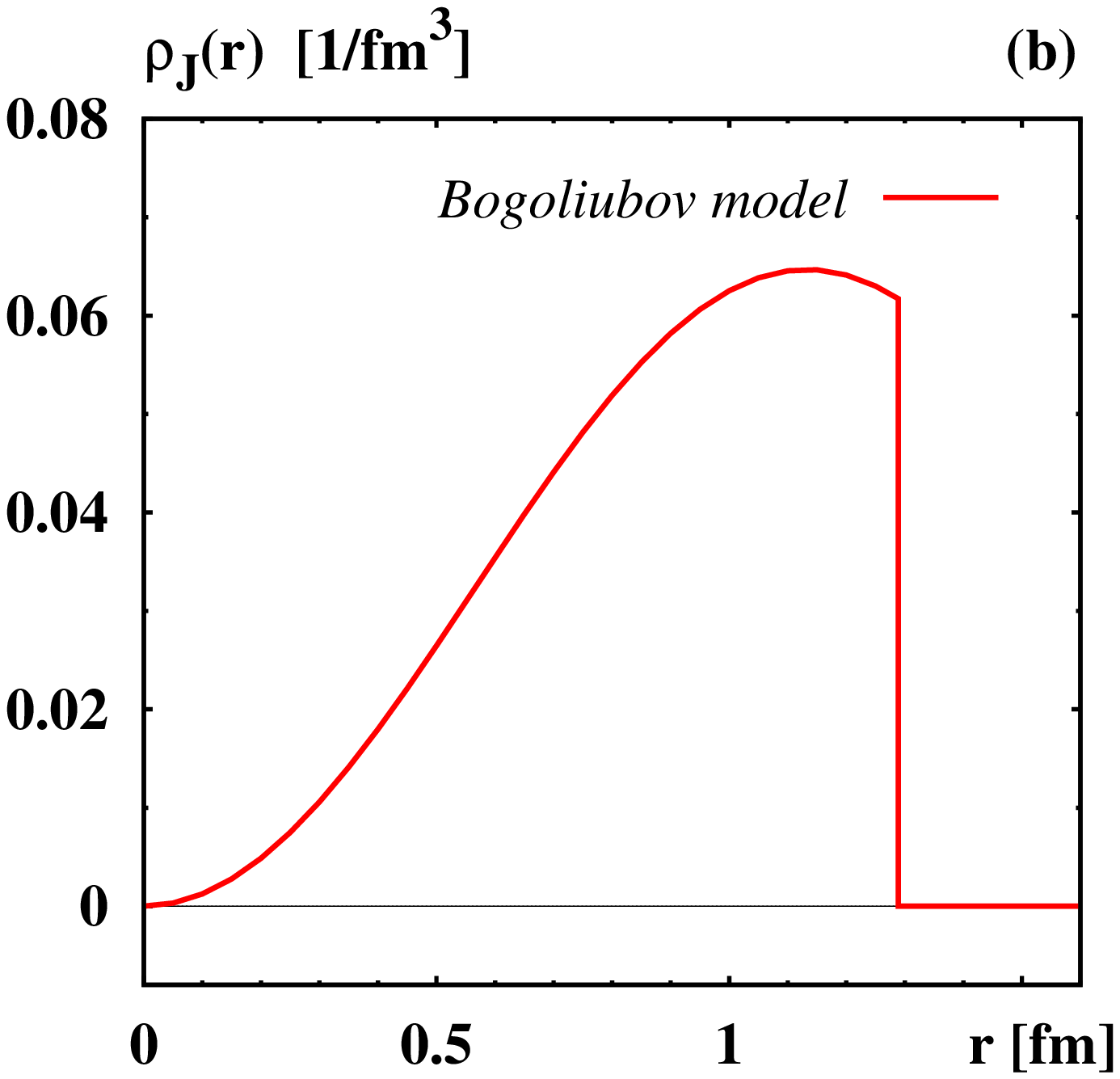} \
\includegraphics[height=3.95cm]{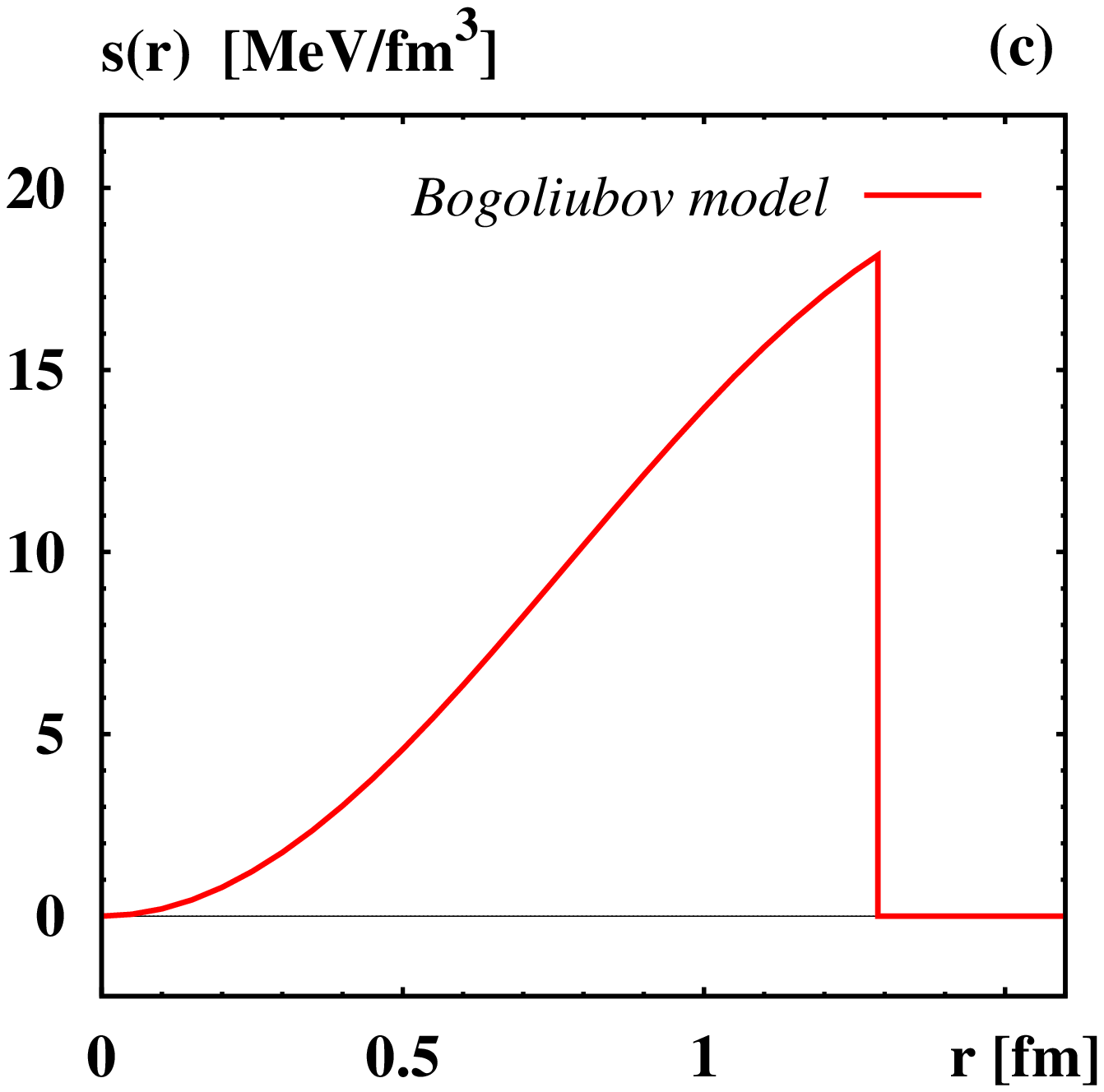} \ 
\includegraphics[height=3.95cm]{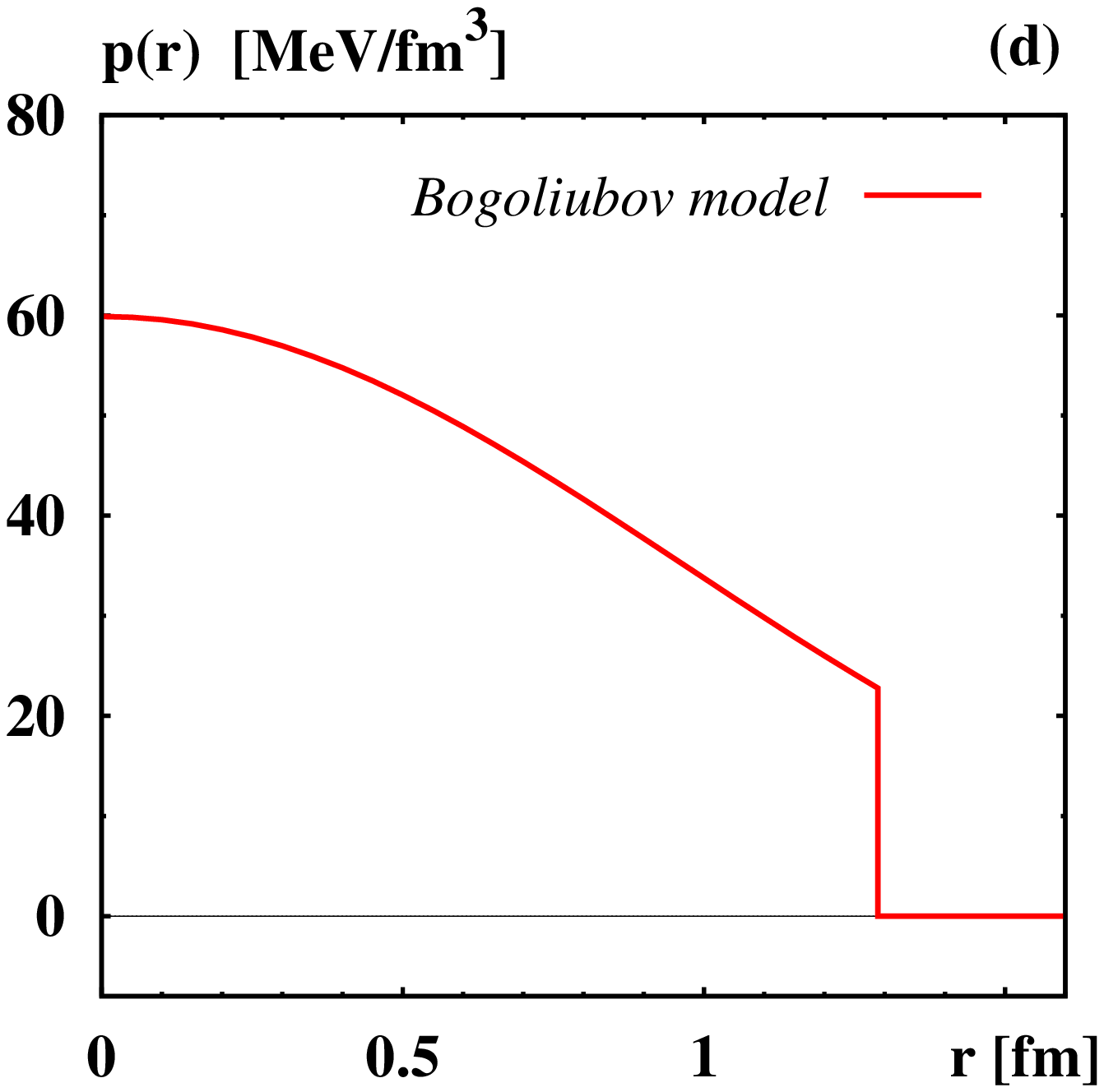}
\par\end{centering}
\caption{\label{Fig-9:Bogo} 
EMT densities as functions of $r$ in the Bogoliubov model: 
(a) energy density $T_{00}(r)$, 
(b) angular momentum density $\rho_J(r)$, 
(c) shear forces $s(r)$, and 
(d) pressure $p(r)$. This version of the Bogoliubov model
corresponds to the bag model with the bag constant $B$ absent.
The EMT densities are similar to the bag model except for the
pressure which exhibits no node and does not comply with the
von Laue condition, which means this is an inconsistent, 
unphysical solution.}
\end{figure}

The EMT densities shown in Fig.~\ref{Fig-9:Bogo} 
illustrate what goes wrong in this model. The results for 
$T_{00}(r)$, $\rho_J(r)$, $s(r)$ in Figs.~\ref{Fig-9:Bogo}a--c look 
very similar to the bag model results in Figs.~\ref{Fig-2:T00-p-s}a--c
and do not hint at anything unusual. They could in principle describe 
a consistent system: e.g.\ $\int\di^3r\,T_{00}(r)$ yields the physical
nucleon mass, and $\int\di^3r\,\rho_J(r)$ yields the nucleon spin $\frac12$.
The inconsistency of the Bogoliubov model becomes apparent when
we inspect the pressure distribution in Fig.~\ref{Fig-9:Bogo}d:
$p(r)$ exhibits no node(!), and hence cannot comply with the von 
Laue condition %(and neither with its lower dimensional analogs) 
in Eq.~(\ref{Eq:stability}).
Clearly, $\int_0^\infty\di r\,r^2p(r)>0$ means
that the internal forces are not compensated, and this solution
actually ``explodes.'' This is a consequence of fixing in this
model the bag radius by hand \cite{Thomas:2001kw}. In other words,
there are no attractive forces in this model that would stabilize 
the solution at some finite radius (as it occurs in the bag model). 
Since the positive (repulsive) forces in the center of the nucleon 
are not compensated, the solution ``explodes:'' matter is dispersed all 
over the space. This corresponds to the observation that the ``minimum'' 
of $M_{N,\rm Bogo}$ occurs only for $R\to\infty$ \cite{Thomas:2001kw}.

From the pressure distribution in Fig.~\ref{Fig-9:Bogo}d we would obtain 
a positive $D$-term by means of the Eq.~(\ref{Eq:d1-from-s(r)-and-p(r)}).
It is interesting to remark that using the shear forces in 
Fig.~\ref{Fig-9:Bogo}c we however would obtain a negative $D$-term
from Eq.~(\ref{Eq:d1-from-s(r)-and-p(r)}).
This mismatch persists even in the limit $R\to\infty$ and
reflects the fact that the EMT is not conserved in this model.
This is not surprising: the ``by-hand-fixing'' of the bag
radius corresponds to ``external forces'' which are imposed 
on the system, but are not present in the Lagrangian. As
a consequence the dynamics is incomplete, and the EMT 
not conserved. Equivalently one may notice that, due to the
absence of the bag constant, there is no form factor $\bar{c}^G(t)$ 
and the constraint $\sum_i\bar{c}^i(t)=0$ is not satisfied.

To conclude this section, we notice that so far no consistent 
physical system has been found where the $D$-term would be positive. 
The excursion to the Bogoliubov model, which is nicely presented in the
historical context in \cite{Thomas:2001kw}, has only revealed an example
where a positive $D$-term is encountered due to an incomplete dynamical
description of a system. One way to cure the inconsistencies of this
model consists in introducing a bag constant. We have seen in the
previous sections how, from the point of view of mechanical
stability, this yields to a consistent description.

%=============== BEGIN FIGURE 9: BOGOLIUBOV MODEL ==================
%================= END FIGURE 9 ====================================

\newpage
%====== SECTION 9: CONCLUSIONS =====================================
\section{Conclusions}
\label{Sec-9:conlusions}

In this work we have explored the bag model to study the EMT form 
factors $A^a(t)$, $J^a(t)$, $D^a(t)$, $\bar{c}^a(t)$ and the EMT densities. 
The quark contributions ($a=u,\,d$) to the EMT form factors 
are defined in terms of the single-quark wave-functions and the SU(4) 
spin-flavor factors needed to construct the nucleon wave-functions.
The form factors factors $A^a(t)$, $J^a(t)$, $D^a(t)$ receive only quark
contributions, i.e.\ in these cases the total form factors are given by 
$A(t)=A^u(t)+A^d(t)$ and analogous for $J(t)$, $D(t)$. 
In principle, also the bag makes contributions to form factors which can 
be interpreted as ``gluonic'' contributions. Only the form factor 
$\bar{c}^a(t)$ receives such a gluonic contribution.

It is crucial to check that all relations derived from 
$\partial_\mu T^{\mu\nu}=0$ are valid, and to demonstrate the 
mechanical stability of the model. The theoretical consistency is 
reflected in various ways. 
We have shown that the bag model description of the EMT form factors 
is consistent in the large-$N_c$ limit. The constraints $A(0)=1$ and 
$J(0)=\frac12$ are satisfied, and $\sum_a\bar{c}^a(t)=0$ holds for all $t$. 
Since the bag contribution 
is not described in terms of a wave function, it was necessary to determine 
the gluonic form factor $\bar{c}^G(t)$ using a different method by resorting 
to the EMT density formalism. The large-$N_c$ formulation of the bag model 
correctly reproduces the general large-$N_c$ counting rules for the EMT 
form factors \cite{Goeke:2001tz}. The usage of the large-$N_c$ limit has 
moreover the advantage of resolving technical problems associated with 
form factor calculations in ``independent particle models'' like 
the bag model. When considering the large-$N_c$ limit our expressions 
for the EMT form factors agree with those from Ref.~\cite{Ji:1997gm}. 
We have shown that the $1/N_c$ corrections associated with our large-$N_c$
treatment of the EMT form factors are relatively small for $|t|\ll M_N$.

The large-$N_c$ limit automatically provides a rigorous justification 
for the concept of 3D densities. 
We studied the energy density $T^{00}(r)$, the angular momentum density 
$J^i(\vec{r}) = \epsilon^{ijk} r^j T^{0k}(\vec{r})$, and the distributions
of shear forces and pressure related to the stress tensor $T^{ij}(\vec{r})$.
We have shown that the bag model EMT densities comply with all general
requirements including the von Laue condition which is a necessary
condition for stability. The bag model also complies with analogous
lower-dimensional stability conditions. Another important result
is that the angular momentum density $J^i(\vec{r})$ in the bag model 
can be decomposed in monopole and quadrupole terms which are 
model-independently related to each other.

We presented an extensive study of the $D$-term in the bag model,
not only for the nucleon but also for other hadrons 
including $N^\star$-resonances, vector mesons, $\Delta$-resonances,
and hypothetical highly excited bag model states. We have shown that
in all cases the $D$-term is negative. We made the interesting observation 
that asymptotically the $D$-terms grow as $D=-\,{\rm const}\times M^{8/3}$ 
with the mass $M$ of the excitation. Interestingly, the
same asymptotic dependence was found for high excitations in the $Q$-ball
system \cite{Mai:2012cx} even though the internal structure of the excited
states in the two systems is much different: for instance, the pressure in
the $N^{\rm th}$ excited state exhibits $(2N+1)$-nodes in the $Q$-ball
system, but one and only one node in the bag model. We are not aware
whether the growth $D=-\,{\rm const}\times M^{8/3}$ of the $D$-term 
with the mass $M$ of the excitation is a general result, or a 
common peculiarity of these two (very different) systems. 
It will be interesting to investigate this result in other 
theoretical systems. At this point it is not known how to
access information on the EMT form factors of $N^\ast$ states,
but information on transition form factors can in principle 
be deduced from studies of hard exclusive reactions. 
This field has a lot of potential. 

The study of excited states has brought very interesting insights.
For instance, while the mass increases by about $50\,\%$ as one goes 
from the ground state (nucleon) to the first excited state (Roper),
the internal pressure in the center and the $D$-term increase by 
factor 7. This finding supports the observations made in other systems 
that the $D$-term is a quantity which most strongly reflects the
internal dynamics of the system and exhibits the strongest variations
as one for instance considers higher excited states. The ground state
and all excited states correspond to mininima of the action, and
comply therefore with the necessary stability condition provided
by the von Laue relation, and the $D$-terms are always negative.
However, only the ground state is the global minimum of the action, 
and hence absolutely stable. The excited states correspond
to local minima and can decay into the ground state.

We studied the $D$-term in three different limits:
heavy quark limit, large bag-radius limit, and non-relativistic limit. 
The $D$-term assumes the same well-defined finite value in these three 
limits which can be computed analytically. This shows that the $D$-term
is a property of all systems including non-relativistic systems.
Since $D=0$ for a free fermion \cite{Hudson:2017oul}, this also provides 
an illustration how e.g.\ even very small interactions in the bag model 
(in the limit of a very large bag radius) generate a non-zero $D$-term.

The bag model is at variance with chiral symmetry, and its 
oversimplified description cannot be expected to give accurate
predictions. But one main goal of this work was to shed light on 
the interpretation of EMT form factors in terms of 3D densities. 
For this it is crucial to use a consistent theoretical framework, 
and the bag model provides this. The simplicity of this model is 
a crucial advantage when elucidating the concepts. For instance,
it was observed in several models that the von Laue condition
$\int_0^\infty dr\,r^2p(r)=0$ is related to the virial theorem.
This is also the case in the bag model, and we were able to
show that not only this but also the lower-dimensional analogs
of the von Laue condition are satisfied provided one works 
with a solution satisfying the virial theorem. 
Another interesting observation is related to the mechanical 
stability requirement that the normal force per unit area 
$\frac23s(r)+p(r)\ge0$.
This quantity is positive inside the bag, and the point where 
it drops to zero marks the ``edge of the system,'' i.e.\
the bag boundary in our case. Such an observation can only be 
obtained in a finite size system.

Finally, we studied the EMT densities in the Bogoliubov model 
\cite{Bogo:1967}, a predecessor of the bag model in which the
bag contribution is absent and the bag radius needs to be
fixed by hand.
This model provides a counter-example for a framework where the 
nucleon is not consistently described. Fixing the bag radius by
hand (rather than by means of a dynamical equation) corresponds
to ``external forces'' which are not included in the Lagrangian.
This implies an unphysical situation in which the EMT is not
conserved and where the pressure has no node and the von Laue 
condition is not satisfied. From such a positive 
pressure one would obtain an unphysical positive $D$-term. This 
problem is solved in the bag model by introducing a non-zero bag 
constant $B$ in the Lagrangian.

It will be interesting to study the EMT form factors and the 
associated densities
in other models whose nature is classical, quantum mechanical, or field
theoretical. Such studies deepen our understanding of the hadron
structure.

\ \\
\noindent{\bf Acknowledgments.} The authors are indebted
to Cedric Lorc\'e, Maxim Polyakov and Leonard Schweitzer 
for valuable discussions.
This work was supported by NSF grant no.\ 1812423 and
DOE grant no.\ DE-FG02-04ER41309.

%\newpage
\appendix
%====== APPENDIX ===================================================
\section{Technical details and proofs}
\label{App}

This Appendix contains technical details. Let us quote first 
the expressions for the first 3 spherical Bessel functions
\be\label{App-Eq:spherical-Bessel-functions}
j_0(x) = \frac{\sin x}{x}\, , \quad
j_1(x) = \frac{\sin x}{x^2}-\frac{\cos x}{x} , \quad
j_2(x) = 3\,\frac{\sin x}{x^3}-3\,\frac{\cos x}{x^2} - \frac{\sin x}{x}\,.
\ee
Below we shall also made use of the expansion of a plane wave 
$e^{i\vec{\Delta}\,\vec{r}}$ in terms of spherical Bessel functions 
and Legendre polynomials $P_l(x)$ as well as the orthogonality 
relation of the latter
\be\label{App-Eq:expansion-plane-wave}
	e^{i\vec{q}\,\vec{r}} = \sum_{l=0}^\infty i^l(2l+1)\,
	j_l(qr)\,P_l(\cos\theta)\;, \quad
	\int_{-1}^1\di\cos\theta\;P_l(\cos\theta)P_k(\cos\theta)
	=\frac{2}{2l+1}\,\delta_{lk}\, .
\ee
In order to abbreviate the expressions below let us define the integrals 
over the combinations of spherical Bessel functions entering respectively
the expressions for $p(r)$ and $s(r)$, namely
\begin{align}
         I_n^p(\omega)=\int_0^\omega\di x\;x^n\biggl(
         j_0(x)j_1^\prime(x)-j_0^\prime(x)j_1(x)+\frac{2}{x}j_0(x)j_1(x)\biggr)\,,
         \nonumber\\
         I_n^s(\omega)=\int_0^\omega\di x\;x^n\biggl(
         j_0(x)j_1^\prime(x)-j_0^\prime(x)j_1(x)-\frac{1}{x}j_0(x)j_1(x)\biggr)\,.
         \label{App-Eq:Ipn-Isn}
\end{align}

\subsection{Virial theorem in general case}
\label{App:virial-massive}

Let us generalize the virial theorem (\ref{Eq:M-virial}) to 
general (including excited) states with $m\neq0$. In the general
case the mass of a hadron is obtained by occupying $N_{\rm const}$ 
energy levels $\varepsilon_i=\Omega_i/R$ and adding the energy 
due to the bag,
\be\label{App:virial-massive-0}
       M(R) = \frac1R\sum_i \Omega_i+\frac{4\pi}{3}\,BR^3
\ee
where the sum goes over the occupied levels $i=1,\;\dots\,,\;N_{\rm const}$ 
and $N_{\rm const}$ denotes the number of constituents with $N_{\rm const}=2$ 
for mesons and $N_{\rm const}=N_c$ for baryons.

The $\Omega_i=\sqrt{\omega_i^2+m^2R^2}$ depend on $R$ explicitly, 
and the $\omega_i$ implicitly through the transcendental equation 
(\ref{Eq:omega-transcendental-eq}). The derivative of $\omega_i$
with respect to $R$ is determined by differentiating 
Eq.~(\ref{Eq:omega-transcendental-eq}) with respect to $R$ which,
upon exploring (\ref{Eq:omega-transcendental-eq}) to eliminate 
trigonometric functions, yields
\be\label{App:virial-massive-I}
	\frac{\partial\omega_i}{\partial R} 
        = \frac{m \,\omega_i}{2\Omega_i(\Omega_i-1)+mR}\,.
\ee
Using the result (\ref{App:virial-massive-I}) we obtain the virial theorem 
valid for $m\neq0$ and excited states which is given by
\be\label{App:virial-massive-II}
        M_N^\prime(R) = -\,\frac{1}{R^2}\;
        \sum_i\frac{2(\Omega_i-1)\omega_i^2}{2\Omega_i(\Omega_i-1)+m R}
        +4\pi\,R^2 B \stackrel{!}{=} 0 
	\quad \Leftrightarrow \quad 
        4\pi R^4 B = \sum_i\frac{2(\Omega_i-1)\omega_i^2}
                                {2\Omega_i(\Omega_i-1)+m R}
\ee
If one takes $m\to0$ the derivative (\ref{App:virial-massive-I}) 
vanishes and the virial theorem (\ref{App:virial-massive-II}) 
reduces to Eq.~(\ref{Eq:M-virial}) for the nucleon.

\subsection{Proof of von Laue condition}
\label{App:von-Laue}

For notational convenience we present the proof for the nucleon case. 
The generalization to other bag states is straight forward. Integrating 
$p(r)$ in Eq.~(\ref{Eq:p+s-bag}) over $\di^3r$ and using the substitution 
$r\to x = \omega r/R$ yields
\ba\label{App:Laue-I}
  \int\di^3r\,p(r) 
  & = & N_c\,A^{2}\;\alpha_{+}\alpha_{-}\,\frac{R^2}{3\omega_0^2}\;I_2^p(\omega_0)
  -\frac{4\pi}{3}\,BR^3\,.
\ea
The integral over the Bessel functions $I_2^p(\omega_0)$ is defined 
in Eq.~(\ref{App-Eq:Ipn-Isn}) and yields
\be\label{App:Laue-II}
  I_2^p(\omega_0) = \frac{\omega_0^2-\sin^2\omega_0}{\omega_0}\,.
\ee
Inserting (\ref{App:Laue-II}) into Eq.~(\ref{App:Laue-I}) we find
\be\label{App:Laue-III}
       \int\di^3r\,p(r) 
       = N_c\,A^{2}\;\alpha_{+}\alpha_{-}\,\frac{R^2}{3\omega_0^2}\;
       \frac{\omega_0^2-\sin^2\omega_0}{\omega_0}-\frac{4\pi}{3}\,BR^3
       \stackrel{!}{=}0\,,
\ee
That Eq.~(\ref{App:Laue-III}) is zero becomes apparent after inserting 
the expressions for $A$ and $\alpha_\pm$ defined in the context of 
Eq.~(\ref{Eq:bag-wave-function}), exploring the transcendental
equation (\ref{Eq:omega-transcendental-eq}) to eliminate trigonometric 
functions and some tedious algebra, which yields
\be\label{App:Laue-IV}
  \int\di^3r\,p(r) 
    = \frac{N_c}{3R}
       \frac{2\omega_0^2(\Omega_0-1)}{2\Omega_0(\Omega_0-1)+m R}-
       \frac{4\pi}{3}\,BR^3
    = -\;\frac{1}{3}\;R\,M_N^\prime(R)
  \stackrel{!}{=}0\,
\ee
where in the second step we made use of the virial theorem
(\ref{App:virial-massive-II}) for the nucleon case.

To prove the 2D analog of the von Laue condition we consider
\be
    \int_0^\infty\di r\;r\biggl(-\frac13\,s(r)+p(r)\biggr)
       = N_c\,\frac{A^{2}}{4\pi}\;\alpha_{+}\alpha_{-}\,\frac{R}{3\omega}
           \biggl[-I_1^s(\omega_0)+I_1^p(\omega_0)\;\biggr]
           -\frac{1}{2}\,BR^2 
           = -\;\frac{1}{8\pi}\;M_N^\prime(R)
           \stackrel{!}{=}0\,,
           \label{App:Laue-V}
\ee
where in the last step we used 
Eqs.~(\ref{App:virial-massive-II},~\ref{App:Laue-III}). Similarly 
for the 1D-version of the von Laue condition we find
\be
    \int_0^\infty\di r\,\biggl(-\frac43\,s(r)+p(r)\biggr)
           = N_c\,\frac{A^{2}}{4\pi}\;\alpha_{+}\alpha_{-}\, \frac{1}{3}
           \biggl[-4I_0^s(\omega_0)+I_0^p(\omega_0)\biggr]
           -BR 
           = -\;\frac{1}{4\pi R}\;M_N^\prime(R)
           \stackrel{!}{=}0\,.
           \label{App:Laue-VI}
\ee
Notice that the integrals $I_0^s(\omega_0)$, $I_0^p(\omega_0)$,
$I_1^s(\omega_0)$, $I_1^p(\omega_0)$ are well-defined but contain 
sine- and cosine-integral terms which cancel out in 
the linear combinations in the square brackets in 
(\ref{App:Laue-V},~\ref{App:Laue-VI}).
The results (\ref{App:Laue-IV},~\ref{App:Laue-V},~\ref{App:Laue-VI})
show that the von Laue condition and its lower-dimensional analogs
are all satisfied if the virial theorem is satisfied.

\subsection{\boldmath Equivalence of $D$-term expressions}
\label{App:D-equivalence}

In this Section let us distinguish the expressions $D_p$ and $D_s$ 
for the $D$-term in terms of pressure and shear forces as defined
in Eq.~(\ref{Eq:d1-from-s(r)-and-p(r)}). For $D_p$ we have
\be\label{App:D-equiv-01}
  D_p = M_N\int\di^3r\,r^2p(r) 
      = M_N\biggl(N_c\,A^{2}\;\alpha_{+}\alpha_{-}\,\frac{R^4}{3\omega_0^4}
        I_4^p(\omega_0)  - \frac{4\pi}{5}\,BR^5\biggr)\,,
\ee
where the integral over Bessel functions yields
\be\label{App:D-equiv-02}
  I_4^p(\omega_0) = \frac{\omega_0^3}{3} + \omega_0 - \omega_0\,\sin^2\omega_0 
  - \sin\omega_0\,\cos\omega_0
\ee
Exploring the expression (\ref{App:Laue-III}) for the von Laue condition 
to eliminate $B$ yields
\begin{align}
        D_p = 
        M_N N_c\,A^{2}\;\alpha_{+}\alpha_{-}\,\frac{R^4}{3\omega_0^4}
        \biggl(-\frac{4}{15}\,\omega_0^3
        + \omega_0 - \frac{2}{5}\omega_0\,\sin^2\omega_0
        - \sin\omega_0\,\cos\omega_0\biggr)
\label{App:D-equiv-03}
\end{align}
and corresponds to the expression quoted in Eq.~(\ref{Eq:D-term-bag}).

To show that the expression in terms of shear forces yields the same
result we consider
\be\label{App:D-equiv-04}
  D_s = -\frac{4}{15}\,M_N\int\di^3r\,r^2s(r) 
    = -\frac{4}{5}\,M_NN_c\,A^{2}\;\alpha_{+}\alpha_{-}\,\frac{R^4}{3\omega_0^4}
      I_4^s(\omega_0)
\ee
with the integral over Bessel functions given by
\be\label{App:D-equiv-05}
  I_4^s(\omega_0) = \frac{\omega_0^3}{3} -\frac54\,\omega_0 
  +\frac54\,\sin\omega_0\,\cos\omega_0
  +\frac12\omega_0\,\sin^2\omega_0 \,.
\ee
The difference of the two expressions for the $D$-term is
\begin{align}
         D_p-D_s 
         & = M_NN_c\,A^{2}\;\alpha_{+}\alpha_{-}\,\frac{R^4}{3\omega_0^4}
         \biggl(I_4^p(\omega_0)+\frac{4}{5}\,I_4^s(\omega_0)\biggr)
         -\frac45\,\pi\,M_NBR^5  =
         -\;\frac15\;R^3M_N^{ } M_N^\prime(R)
         \stackrel{!}{=}0
\label{App:D-equiv-06}
\end{align}
where in the last step we once more made use of 
Eqs.~(\ref{App:virial-massive-II},~\ref{App:Laue-III}).
This proves that the expressions for the $D$-term in terms
of $p(r)$ and $s(r)$ are equivalent.

\subsection{\boldmath Proof that $\bar{c}^Q(t)+\bar{c}^G(t)=0$}

In the main text it was shown that at $t=0$ it is $\bar{c}^Q(0)+\bar{c}^G(0)=0$.
We now wish to generalize this proof to $t\neq0$. The proof is elementary
but tedious such that it is worth showing it in some more detail.
The starting point is $\bar{c}^Q(t)$ in (\ref{Eq:EMT-bag-T33}).
We recall that $\vec{k}^{\,\prime}=\vec{k}+\vec{\Delta}$ with 
$\vec{\Delta}=(0,0,\Delta^3)$ in our kinematics. The right-hand-side 
of (\ref{Eq:EMT-bag-T33}) is an even function of $\Delta^3$.
To show this we replace $\Delta^3\to(-\Delta^3)$ and subsequently 
substitute $k^3\to(-k^3)$ which restores the starting expression. 
This proves that $\bar{c}^Q(t)$ can be understood as a function 
of $t=-\vec{\Delta}{ }^2$ as it must for a form factor. 
In the next step we explore this to simplify the expression for 
$\bar{c}^Q(t)$ as follows. In the first term in the square brackets 
of (\ref{Eq:EMT-bag-T33}) we substitute $k^3\to k^3-\Delta^3$
and subsequently we explore that the function is even under 
$\Delta^3\to(-\Delta^3)$, which restores the original 
expression but with $\vec{k}$ and $\vec{k}^{\,\prime}$ exchanged. 
This allows us to write Eq.~(\ref{Eq:EMT-bag-T33}) as
\be\label{App-Eq:cbar-01}
	{\bar c}^Q(t)	= - \;
	b \int\!\frac{\di^3k}{(2\pi)^3\!}\,(k^{\prime3}+k^3)
	\biggl[	t_0(k^\prime)t_1(k)\,e_{k}^3\biggr], \quad
        b=\frac{4\pi A^2R^6N_c}{M_N}\alpha_+\alpha_- 
\ee
where $\vec{e}_k=\vec{k}/k$. It is convenient to work in 
coordinate space. In the formulas below Bessel functions $j_l$
with no argument will denote $j_l(w_i r/R)$ for notational 
simplicity, and the primes will denote derivatives with respect to 
$r$. In order to avoid total derivatives (which in general do not 
vanish in the finite volume integrals in the bag model and cause 
a proliferation of terms) we proceed by introducing a $\delta$-function 
as follows
\be\label{App-Eq:cbar-02a}
        {\bar c}^Q(t) = -\;b
            \Biggl[\int\!\frac{\di^3k}{(2\pi)^3\!}\;t_0(k)k^3 
                   \int\!\frac{\di^3q}{(2\pi)^3\!}\,t_1(q)\,e_{q}^3 +
                   \int\!\frac{\di^3k}{(2\pi)^3\!}\,t_0(k)
                   \int\!\frac{\di^3q}{(2\pi)^3\!}\,t_1(q)\,q^3e_{q}^3
                   \Biggr]
                   \int\di^3r\;
                   e^{i\vec{r}(\vec{k}^{\,\prime}-\vec{q})}\,.
\ee
In the next step we invert the Fourier transforms, where 
$\Theta_V=\Theta(R-r)$ is used for brevity and we use identities
\be\label{App-Eq:cbar-02b}
         \int\!\frac{\di^3q}{(2\pi)^3\!}\,t_0(q)e^{i\vec{r}\vec{q}}
         =j_0\; \frac{\Theta_V}{4\pi R^3}\, , \quad
         \int\!\frac{\di^3q}{(2\pi)^3\!}\;\vec{e}_{q}t_1(q)\,e^{-i\vec{r}\vec{q}} 
         =-i\vec{e}_{r}j_1\;\frac{\Theta_V}{4\pi R^3}\,,
\ee
where (as before) $j_i=j_i(\omega r/R)$ for brevity.
This yields 
\begin{align}\label{App-Eq:cbar-02c}
     {\bar c}^Q(t) 
%        & = \frac{b }{(4\pi R^3)^2}\;\int\!\di^3r\,e^{i\vec{r}\vec{\Delta}}
%            \Biggl[
%              (-i\nabla^3j_0)(ie_{r}^3j_1)
%              + j_0 \, (\kt{i}\nabla^3ie_{r}^3j_1)
%             \Biggr]\,\Theta_V\nonumber\\
        & = \frac{b  }{(4\pi R^3)^2}\;\int\!\di^3r\,e^{i\vec{r}\vec{\Delta}}
            \Biggl[(e_r^3)^2\biggl(
                 j_0^\prime j_1 
              -  j_0 j_1^\prime 
              + \frac{j_0 j_1 }{r}\biggr)
              - \frac{j_0 j_1 }{r}
             \Biggr]\,\Theta_V\,.
\end{align}
Finally we explore that $e_r^3=\vec{e}_z\cdot\vec{e}_r=\cos\theta$
such that $(e_r^3)^2=\frac23P_2(\cos\theta)+\frac13P_0(\cos\theta)$.
Making use of the expansion of $e^{i\vec{r}\vec{\Delta}}$ and the orthogonality
of Legendre polynomials in (\ref{App-Eq:expansion-plane-wave}),
we obtain
\be
       {\bar c}^Q(t) 
         = \frac{b  }{(4\pi R^3)^2}\;\int_V\!\di^3r\,
            \Biggl[\biggl(
\underbrace{-\frac23j_2(\Delta r)+\frac13j_0(\Delta r)}_{
            \frac1\Delta\,j_1^\prime(\Delta r)}
\biggr)
              \biggl(
                 j_0^\prime j_1 
              -  j_0 j_1^\prime \biggr)
              +  \biggl(
\underbrace{-\frac13j_2(\Delta r)-\frac13j_0(\Delta r)}_{
             = -\frac{j_1(\Delta r)}{\Delta r}}\biggr)
              \frac{2j_0 j_1 }{r}
             \Biggr]
\ee
where the underbraces indicate useful identities. Another helpful identity 
is $2j_0j_1=-\frac{\partial\;}{\partial r}[r^2(j_0^\prime j_1-j_0j_1^\prime)]$.
After integrating over the solid angle we find that the $r$-integrand is a 
total derivative
\begin{align}
        {\bar c}^Q(t) 
        & = \frac{4\pi b }{(4\pi R^3)^2\Delta}\;\int_0^R\di r
            \Biggl[
              \frac{\partial\;}{\partial r}\biggl[
                j_1(\Delta r)\;
                r^2\biggl(
                 j_0^\prime j_1 
              -  j_0 j_1^\prime \biggr)\biggr]
             \Biggr]
         = -\,c_0
              \,\frac{j_1(\Delta R)}{\Delta R}\,.
\end{align}
In the massless case the prefactor $c_0$ is given by
\be
       c_0 =    \frac{b}{4\pi R^3}\frac{\omega_0}{R}\;
              \biggl(
                  j_0(\omega_0)j_1^\prime(\omega_0)
                 -j_0^\prime(\omega_0)j_1(\omega_0)\biggr)
         = \frac34
\ee
which follows from using the transcendental equation
(\ref{Eq:omega-transcendental-eq}).
In the massive case the result is a different fraction, and
the last step is lengthier and one has to use the expression for 
$B$ from the massive virial theorem to show that the constraint 
$\sum_i\bar{c}^i(t)$ holds also here.

\subsection{\boldmath EMT form factor of the anti-symmetric EMT}
\label{App:anti-symmetric-EMT}

For completeness we discuss the form factors of the canonical EMT 
defined by
$
        T^{\mu\nu,q}_{\rm can} = \frac{1}{2}\overline{\psi}_q(
        -i\gamma^\mu\overset{ \leftarrow}\partial{ }^\nu
        +i\gamma^\mu\overset{\rightarrow}\partial{ }^\nu)\psi_q
$. 
The canonical EMT can be decomposed in two
parts: a symmetric part whose form factors $A^q(t)$, $J^q(t)$, 
$D^q(t)$, $\bar{c}^q(t)$ were discussed in the main text,
and an antisymmetric part which is characterized by a single
form factor \cite{Bakker:2004ib}
\ba
    \la p^\prime| 
    \frac12\biggl(\hat{T}^{\mu\nu,q}_{\rm can}(0)-\hat{T}^{\nu\mu,q}_{\rm can}(0)\biggr)
    |p\rangle
    = F_{\rm can}^q(t)\ 
    \bar u(p^\prime)\,\frac{i(P^{\mu}\sigma^{\nu\rho}-P^{\nu}\sigma^{\mu\rho})
    \Delta_\rho}{4M_N}u(p)\,,
\ea
which in the bag model is given by
\be\label{Eq:App-Fcan}
    F_{\rm can}^q(t) =
    4\pi A^2R^6\int\!\frac{\di^3k}{(2\pi)^3\!}\,
    \biggl[2\alpha_+\alpha_-\frac{\varepsilon_0}{\Delta^3} \biggl(
            t_0(k^\prime)t_1(k)\,\frac{k^3}{k}
            -  t_0(k)t_1(k^\prime)\,\frac{k^{\prime3}}{k^\prime}\biggr)
    + k_\perp^2 \alpha_-^2
    \frac{t_1(k)}{k}\,\frac{t_1(k^\prime)}{k^\prime}\biggr] \, .
\ee
The expression (\ref{Eq:App-Fcan}) agrees up to the sign with
the bag model results for the axial form factor $G_A^q(t)$.
Thus, we recover $ F_{\rm can}^q(t) = -G_A^q(t)$ which is a
model independent result \cite{Bakker:2004ib}. This shows 
that also the canonical EMT is consistently described 
within the bag model.

\newpage

%====== REFERENCES =========================================================

\end{document}